\documentclass[11pt]{article}

\usepackage{epsfig,psfrag}
\usepackage{color}
\usepackage{array}

\usepackage{amsmath,amsfonts,amssymb}

\renewcommand{\vec}[1]{\boldsymbol{#1}}

\newtheorem{theorem}{Theorem}[section]
\newtheorem{proposition}{Proposition}[section]

\newcommand{\R}{\mathbb{R}}
\newcommand{\T}{\mathbb{T}}
\newcommand{\cP}{\mathcal{P}}
\newcommand{\cH}{\mathcal{H}}
\newcommand{\del}{\partial}

\title{A kinetic model for the sedimentation of rod--like particles}
\author{
Christiane Helzel\thanks{Institute of Mathematics,
  Heinrich-Heine-University D\"usseldorf, D\"usseldorf, Germany \hfill\break Email:  {\tt christiane.helzel@hhu.de}},
\and 
Athanasios E. Tzavaras
\thanks{Computer, Electrical, Mathematical Sciences \& Engineering Division, King Abdullah University of Science and Technology (KAUST), Thuwal, Saudi Arabia.
Email: {\tt athanasios.tzavaras@kaust.edu.sa} }
\thanks{Institute of Applied and Computational Mathematics,
FORTH, Heraklion, Greece.}
}

\date{}

\begin{document}
\maketitle

\begin{abstract}
We consider a coupled system consisting of a kinetic equation  coupled  to a macroscopic 
Stokes (or Navier-Stokes) equation and describing the motion of a suspension of rigid rods
in gravity. A reciprocal coupling  leads to the formation of clusters:  The buoyancy force creates a 
macroscopic velocity gradient that causes the microscopic particles to align 
so that their sedimentation reinforces the formation of clusters 
of higher particle density. We provide a quantitative analysis of cluster formation.
We derive a nonlinear moment closure model, which consists of
evolution equations for the density and second order moments and that uses the structure of
spherical harmonics to suggest a closure strategy.
For a rectilinear flow we employ the moment closure together with a quasi-dynamic approximation
to derive an effective equation,  The effective equation is an advection-diffusion equation with nonisotropic
diffusion coupled to a Poisson equation, and belongs to the class of the so-called flux-limited Keller-Segel models.
For shear flows, we provide an argument for the
validity of the effective equation and perform numerical comparisons that indicate good agreement between the original system 
and  the effective theory. Finally, a linear stability analysis on the moment system shows that
linear theory predicts a wavelength selection mechanism for the
cluster width, provided that the Reynolds number is larger than zero. 
%
%
%
\end{abstract}

%
%
%

\pagestyle{myheadings}
\thispagestyle{plain}

%
%
%
\section{Introduction}

Complex fluids with suspended microstructure might include diverse 
materials such as   polymeric solutions
with a long chain structure, or  suspensions of non-deformable particles with
an orientation (for example rod-like particles). 
Such systems might be modeled via continuum modeling of viscoelastic fluids, or
via mesoscopic models within the realm of kinetic theory, or even as particles in a fluid.
Multi-scale interactions cause interesting phenomena in the dynamics of complex flows, and the area 
forms an interesting testing ground for examining the effect of change of scale and
the passage from microscopic to mesoscopic to macroscopic theories.

An interesting system is offered by a suspension of rod-like particles in a dilute solution
under the influence of gravity. For such a system, the macroscopic flow 
can cause a change of orientation for  the suspended microstructure, which
 in turn,  produces an elastic stress that interacts with and modifies the macroscopic flow.   
The sedimentation of dilute suspensions of rod-like 
particles can lead  to  cluster formations and  has been studied
by several authors in theoretical, numerical and experimental work,
see the recent review paper by Guazzelli and Hinch \cite{GH2011} and 
references therein.

Experimental works by Guazzelli and coworkers \cite{HG96, HG99, MBG07a}
reveal the following scenarios:
Starting from a well-stirred suspension packets of particles form after some time. 
These packets seem to have a mesoscopic equilibrium width, suggesting 
that the density of particles acquires variations of a characteristic length 
scale. Within a cluster, individual particles are aligned with the
direction of gravity during most of the time; occasionally they flip. 
The average settling speed in a suspension is larger than the sedimentation 
speed of a single particle oriented in the direction of gravity. 
The mechanism of cluster formation was described in a
fundamental paper of Koch and Shaqfeh \cite{KS89}.
In recent years the sedimentation of rod-like  (and other
orientable) particles has also been studied via numerical simulations
of multi-scale models. Gustavsson and Tornberg \cite{TG06,GT09}
used a very detailed description of rod-like particles in a dilute
suspension based on a slender body approximation. They were able to
simulate suspensions with up to a few hundred particles and a domain
size of the order of a few particle length. Butler and Shaqfeh
\cite{BS2002} used a lower order slender body description. Saintillan
et al.\ \cite{SDS05} accelerated this algorithm using fast summation
techniques. This allowed them to simulate several thousand
particles. Wang and Layton \cite{WL09} used the immersed boundary method
for their two-dimensional numerical studies. All numerical studies
confirm the basic experimental findings: Packet formation and
alignment in gravity direction. Note that the models used in those
simulations are of more microscopic nature than the model considered
here. Instead of a number density function for the rod orientation in
every point of the domain, those authors model a large number of individual particles. 
Moreover, Brownian effects are not considered in those models.

Our objective here is to provide a quantitative desciption of cluster formation by deriving
an effective theory via asymptotic methods.
We focus on a multi-scale model, described in Section \ref{section:model}, 
an extension of models for dilute suspensions of rod-like particles
described in Doi and Edwards \cite{book:DE86} and studied in \cite{OT08,HO06}. In non-dimensional form the
resulting model reads
\begin{equation}\label{intro-model}
\begin{split}
\partial_t f  + \nabla_{\vec{x}} \cdot \left(\vec{u} f \right)
& +\nabla_{\vec{n}} \cdot \left(P_{\vec{n}^\bot} \nabla_{\vec{x}} \vec{u} \vec{n} f \right) - \nabla_{\vec{x}} \cdot \left(\left(I + \vec{n} \otimes \vec{n} \right)
\vec{e}_3 f\right)\\
&  =  D_r \Delta_{\vec{n}}f + \gamma \nabla_{\vec{x}} \cdot \left(I+\vec{n} \otimes \vec{n} \right) \nabla_{\vec{x}} f\\
\sigma  & = \int_{S^{d-1}} \left( d \vec{n} \otimes \vec{n} - I \right) f\,  d \vec{n}
\\
Re \left(\partial_{t} \vec{u} + \left( \vec{u} \cdot \nabla_{\vec{x}} \right) \vec{u}\right) & = \Delta_{\vec{x}} \vec{u} - \nabla_{\vec{x}} p
+ \delta \gamma \nabla_{\vec{x}} \cdot \sigma + \delta 
\left(m -  \int_{S^{d-1}} f \, d \vec{n} \right) \vec{e}_3
\\
\nabla_{\vec{x}} \cdot \vec{u} & = 0
\end{split}
\end{equation}
where $f (t, \vec{x}, \vec{n})$ describes the distribution function of particles 
as a function of time $t$, space $\vec{x} \in \R^d$ and the orientation
of the rod-like particles 
$\vec{n} \in S^{d-1}$, $\vec{u} (t, \vec{x})$
stands for  the velocity of the solvent, $p(t, \vec{x})$ is the pressure,  
and $D_r$, $\gamma$, $\delta$ and $Re$ stand for non-dimensional numbers, and $m$ stands for the total mass of the
suspended rods.
We note that the processes, which lead
according to Koch and Shaqfeh \cite{KS89} to the
instability of the sedimentation process, are included in this model. The model and
its properties is described in Section \ref{section:model}, while the cluster formation mechanism 
is detailed in section \ref{section:mechanism}. 

Due to the high-dimensionality of the problem, numerical computations of the full five dimensional problem
(for $d=3$) are cumbersome. We are therefore interested in
the  derivation of partial differential equations (pdes), which describe the macroscopic flow without
resolving the microscopic structure. Such nonlinear models should also
provide further insight into the basic nonlinear mechanism which leads to 
cluster formation. Sections \ref{section:moment-closure} and 
\ref{section:quasi-dynamic} are devoted to the derivation of such simpler
pde models. In Section \ref{section:moment-closure}, we perform a moment closure from the full
multi-scale model, based on deriving equations for the moments, and using the structure of
spherical harmonics to suggest a closure strategy. We adapt the results to various special
flows of interest, including rectilinear flows and shear flows. The obtained models 
\eqref{eqn:mc-rectilinear} and \eqref{eqn:mc-shear} have some analogies to
the familiar Oldroyd-B models  in continuum viscoelasticity  (see Renardy \cite{Ren08}
for a description and further references).

The moment closure leads to a $d$-dimensional system of pdes
($1$-dimensional for shear flow,
$2$-dimensional for rectilinear flow) and thus reduces the dimension
of the problem.  A second approximation,  explained in Section \ref{section:quasi-dynamic}, 
leads to a scalar evolution equation for the particle density coupled to the
equation of the floe and provides an even  simpler
description of the process of cluster formation.  
We call this quasi-dynamic approximation and it 
consists of setting the higher order moments to their local equilibrium  and evaluating them in terms 
of the zero-th moment and its spatial derivatives.
It leads to an emerging effective equation for the particle density.
For the case of rectilinear flow along the $z$-direction of the
three-dimensional space  $\vec{x} = (x,y,z)$,
with ansatz
$$
 f = f (t, x, y, \vec{n}) \, , \quad \vec{u} =  \big ( 0, 0, w(t,x,y)  \big )^T  \, ,
$$
the resulting effective system reads
\begin{equation}
\label{intro-effsys}
\begin{split}
\del_t \rho &= \nabla \cdot \left (  
\frac{1}{\kappa + 10 |\nabla w|^2} \left [ I - \frac{36 |\nabla w|^2 }{\kappa + 46 |\nabla w|^2} \; \frac{\nabla w}{|\nabla w|} \otimes \frac{\nabla w}{|\nabla w|} \right ]
 \tfrac{\kappa}{30} \rho  \nabla \big ( w +\tfrac{1}{3} \ln \rho \big ) \right )
\\
 Re \; \partial_t w &= \triangle_{(x,y)}  w + \delta \left( \bar \rho - \rho \right)
\end{split}
\end{equation}
where $\rho = \int f d\vec{n}$ and $\kappa = 42^2$.

The system \eqref{intro-effsys}  should be compared to the Keller-Segel model ({\it e.g.} \cite{JL92,HV96,BDP06}) used
as a model for chemotaxis. There are two differences: (a) that the diffusion in \eqref{intro-effsys}$_1$ is anisotropic; 
more important (b) that both convection and diffusion are flux-limited, thus making the system belong to the general type
of flux-limited chemotaxis systems  ({\it e.g} \cite{DS06,VSMS13,Per13}).

We discuss in sections \ref{sec:remarks} and \ref{sec:justify}  the nature of the quasi-dynamic approximation 
and provide some justifying arguments. For the case of shear flows, an eigenvalues analysis  of a linearized system 
suggests that the approximation should be valid for moderate strain rates. 
Moreover, extensive numerical simulations for shear flows confirm that the solution
structure obtained from the scalar evolution equation agrees very well with the solution structure of the full model. 
In upcoming work, we plan to provide numerical studies for the computationally more challenging case of rectilinear flows. 

In appendices \ref{section:appendix2} and \ref{app3}, we summarize properties of the
differential operator on the sphere as well as properties of the spherical harmonics, 
which are extensively used in this paper. In appendix
\ref{appendix:linstab}, we present results of a linear stability
analysis for shear flows. In particular,  we show that linear stability
theory predicts a wavelength selection only if the Reynolds number of
the macroscopic flow is larger than zero.

%
%
%
%

\section{The mathematical model} \label{section:model}
We describe a kinetic model for sedimentation in dilute suspensions of 
rod-like Brownian particles. Models of this type were introduced by Doi and Edwards, see
\cite[Ch. 8]{book:DE86}. 
In \cite{OT08} and \cite{HO06}  a related model for suspensions of rod-like 
particles was considered. This model is extended here to account for the effects of 
gravity in an effort to describe sedimentation of rod-like particles in a solvent.
%
%
\begin{figure}[hbt]
\centerline{
\input{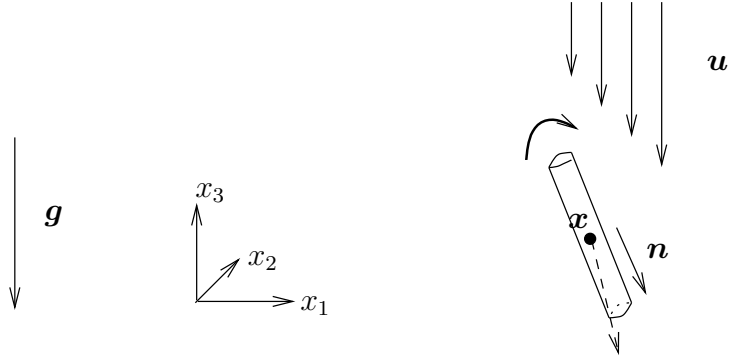}}
\caption{Basic notation for rod-like molecule which is falling sidewards.
\label{figure:sed_pic}}
\end{figure}

We consider inflexible rod-like particles of thickness $b$ which is
much smaller than the particles length $l$.
Our considerations are restricted to the 
dilute regime which is characterized by the relation $\nu \ll l^{-3}$ where
$\nu$ is the number density. Note that in contrast to the model considered
in \cite{OT08}, the number density is not constant here. The orientation 
of a rod-like particle is characterized by $\vec{n} \in S^{d-1}$ where
$d$ is the dimension of the macroscopic physical space $\Omega \subset \mathbb{R}^d$. 

We denote by $\vec{e}_3$ the unit vector in the upward direction and by
$\vec{g}= - g \vec{e}_3$ the acceleration of gravity with gravitational constant $g$.
Furthermore, let $m_0$ denote the mass of an individual rod-like 
particle and $\vec{G} = -m_0 g \vec{e}_3$ be the force of gravity on a single 
particle. Some of our basic notation is depicted in Figure 
\ref{figure:sed_pic}.

In a quiescent suspension (i.e.\ macroscopic velocity $\vec{u}=0$) each
particle sediments at a speed depending on its orientation $\vec{n}$ according 
to
\begin{equation*}
\begin{split}
\frac{d \vec{x}}{dt} & =  \frac{1}{\zeta_{||}} \left(\vec{G} \cdot \vec{n}\right) \vec{n} + \frac{1}{\zeta_\bot}\left(\vec{G} - \left(\vec{G}\cdot \vec{n} \right) \vec{n} \right)\\
& = \left( \frac{1}{\zeta_{||}} \vec{n} \otimes \vec{n} + 
\frac{1}{\zeta_\bot}\left(I - \vec{n}\otimes \vec{n} \right) \right) \vec{G}\\
& = \frac{1}{\zeta_\bot} \left( \vec{n}\otimes\vec{n} + I\right) \vec{G}
\end{split}
\end{equation*} 
where $\zeta_{||}$ and $\zeta_\bot$ are the frictional coefficients in the 
tangential and the normal direction. On the basis of the Kirkwood theory $\zeta_{||}$, $\zeta_\bot$  satisfy 
$\zeta_\bot = 2 \zeta_{||}$,  see \cite[App 8.I]{book:DE86}.  
In particular, a particle with a horizontal 
orientation sediments slower than a particle with a vertical orientation and
a particle of oblique orientation moves also sideways.

In addition a macroscopic velocity gradient causes a rotation according to
the equation
\begin{equation*}
\begin{split}
\frac{d\vec{n}}{dt} = P_{\vec{n}^\bot} \nabla_{\vec{x}} \vec{u} \vec{n}
\end{split}
\end{equation*}
where $P_{\vec{n}^\bot} \nabla_{\vec{x}} \vec{u} \vec{n} := \nabla_{\vec{x}}\vec{u} \vec{n} - \left( \nabla_{\vec{x}} \vec{u} \vec{n} \cdot \vec{n} \right) \vec{n}$ is the projection of the vector $\nabla_{\vec{x}} \vec{u} \vec{n}$ on the tangent space at $\vec{n}$.

We next include the effects of rotational and translational Brownian motion and account
for the macroscopic mean flow $\vec{u}(\vec{x},t)$. The model then becomes the system of
stochastic differential equations
\begin{equation}
\begin{aligned}
d \vec{x} & = \vec{u } dt +  \left( \frac{1}{\zeta_{||}} \vec{n} \otimes \vec{n} + 
\frac{1}{\zeta_\bot}\left(I - \vec{n}\otimes \vec{n} \right) \right) \vec{G} \, dt
\\
&\qquad +  \sqrt{ \frac{2 k_B \theta}{\zeta_{||}} \vec{n} \otimes \vec{n} + 
  \frac{2 k_B \theta}{\zeta_{\perp}} (I - \vec{n}\otimes \vec{n})}  \; dW 
\\
d\vec{n} &= P_{\vec{n}^\bot} \nabla_{\vec{x}} \vec{u} \vec{n} \, dt + 
 \sqrt{\frac{2 k_B \theta}{\zeta_r} }  \;  dB
\end{aligned}
\end{equation}
where $W$ is the translational Brownian motion and $B$ is the rotational Brownian motion,
$\zeta_r$ is the rotational friction coefficient, $k_B$ is the Boltzmann constant and
 $\theta$  the absolute temperature.

The above stochastic equations may be equivalently expressed via the Smoluchowski
equation for the evolution of the  {\em local orientational distribution function}: 
\begin{equation}\label{eqn:smol_a}
\begin{split}
\partial_t f & +  \nabla_{\vec{x}} \cdot \left[ \left( \vec{u} +  \frac{1}{\zeta_{\bot}} \left( \vec{n} \otimes \vec{n} + I \right)
\left( - m_0 g \vec{e}_3 \right) \right) f \right]
+ \nabla_{\vec{n}} \cdot \left( P_{\vec{n}^\bot} \nabla_{\vec{x}} \vec{u} \vec{n} f \right) \\
& = \frac{k_B \theta}{\zeta_r} \Delta_{\vec{n}} f + \frac{k_B \theta}{\zeta_\bot} \, 
\nabla_{\vec{x}} \cdot
\left( \vec{n} \otimes \vec{n} + I \right) 
\nabla_{\vec{x}} f. 
\end{split}
\end{equation}
Here $f(\vec{x},t,\vec{n}) \, d\vec{n}$ describes the number of particles
per unit volume at macroscopic position $\vec{x}$ and time $t$ with 
orientations in the element centered at $\vec{n}$ and of volume $d\vec{n}$. 
The second term on the left hand side of (\ref{eqn:smol_a}) models transport
of the center of mass of the particles due to the macroscopic flow velocity 
and due to gravity. The last term on the left hand side models the 
rotation of the axis due to a macroscopic velocity gradient 
$\nabla_{\vec{x}} \vec{u}$. 
The terms on the right hand side describe  rotational as well as translational diffusion.  
They together amount to non-isotropic spatial diffusion, which is one of the main
features of the model at hand.
The gradient, divergence and Laplacian on the sphere are denoted by
$\nabla_{\vec{n}}$, $\nabla_{\vec{n}} \cdot$ and $\Delta_{\vec{n}}$, while
the gradient and divergence in the macroscopic flow domain are denoted by
$\nabla_{\vec{x}}$ and $\nabla_{\vec{x}} \cdot$.
The total number of rod-like particles is
\begin{equation*}
\int_\Omega \int_{S^{d-1}} f(\vec{x},t,\vec{n})\, d\vec{n} \, d\vec{x} =
\int_\Omega \int_{S^{d-1}} f(\vec{x},0,\vec{n})\, d\vec{n} \, d\vec{x} = N,
\end{equation*}
i.e.\ $f$ has dimensions of number density.
It is convenient to rewrite the Smoluchowski equation in the form
\begin{equation}\label{eqn:smol0}
\begin{split}
\partial_t f & + \nabla_{\vec{x}} \cdot \left(\vec{u} f \right)
+ \nabla_{\vec{n}} \cdot \left( P_{\vec{n}^\bot} \nabla_{\vec{x}} \vec{u} \vec{n} f \right) \\
& = D_r  \Delta_{\vec{n}} f + 
D_\bot  \nabla_{\vec{x}} \cdot 
 \left( I + \vec{n} \otimes \vec{n} \right) 
\Big ( \nabla_{\vec{x}} f + \frac{1}{k_B \theta} \, 
 f  \, \nabla_{\vec{x}} U \Big), 
\end{split}
\end{equation} 
where $D_r := \frac{k_B \theta}{\zeta_r}$ and $D_\bot := \frac{k_B \theta}{\zeta_\bot}$ stand 
for the  rotational and translational diffusion coefficients and $U(x) = m_0 g (\vec{x} \cdot \vec{e}_3)$ is the potential of the  gravity force
$\vec{G} = -\nabla U = - m_0 g \vec{e}_3$.

As can be seen from (\ref{eqn:smol0}), a velocity gradient $\nabla_{\vec{x}} \vec{u}$ distorts an isotropic distribution $f$ which leads to an increase in entropy. 
Thermodynamic consistency requires 
 that this is balanced by a stress tensor $\sigma(\vec{x},t)$ given by
\begin{equation}\label{eqn_sigma}
\sigma(\vec{x},t) := k_B \theta \int_{S^{d-1}} \left( d \vec{n} \otimes \vec{n} - I \right) f(\vec{x},t, \vec{n}) d \vec{n}.
\end{equation}
(see \cite[Sec 8.6]{book:DE86} and compare with Section \ref{consistency}).

Local variations in the density $m_0 \int_{S^{d-1}}f d\vec{n}$ 
lead to spatial variations 
of the specific weight of the suspension that generally can not be compensated 
by a hydrostatic pressure and thus trigger a fluid motion (buoyancy).
The macroscopic flow is described by the  Navier-Stokes equation.
Let $\rho_f$ be the density of the 
fluid which is assumed to be constant. 
The balance laws of mass and momentum have the
form
\begin{equation}\label{eqn:stokes1}
\begin{split}
\rho_f \Big [  \partial_t \vec{u} + \left(\vec{u} \cdot \nabla_{\vec{x}}\right )
\vec{u} \Big ] & 
= \mu \triangle_{\vec{x}}  \vec{u} -  \nabla_{\vec{x}} p + \nabla_{\vec{x}} \cdot \sigma 
\\
&\quad  - \rho_f g \vec{e}_3 - 
\Big ( \int_{S^{d-1}} f d\vec{n} \Big ) m_0 g \vec{e}_3
\\
\nabla_{\vec{x}} \cdot \vec{u} & = 0 
\end{split}
\end{equation}

The term $\rho_f g \vec{e}_3$ can be incorporated to the pressure. 
For the linear stability analysis it will be convenient to express the momentum
equation in the equivalent form
\begin{equation}
\begin{split}
\rho_f \left(\partial_t \vec{u}+ \left(\vec{u} \cdot \nabla_{\vec{x}}\right) 
\vec{u}  \right) & =    \mu \Delta_{\vec{x}}  \vec{u} 
- \nabla_{\vec{x}} p' + \nabla_{\vec{x}} \cdot \sigma  
 + \left(\frac{N}{V} -\int_{S^{d-1}} f \, d \vec{n} \right) m_0 g \vec{e}_3\\
  \nabla_{\vec{x}} \cdot \vec{u}  &= 0  
\label{eqn_stokes_2}
\end{split}
\end{equation} 
by redefining the pressure,
$$p' = p +  \rho_f g \vec{e}_3 \cdot \vec{x} + 
\frac{m_0  N g}{V} \vec{e}_3 \cdot \vec{x} \, ,
$$
to account for the hydrostatic pressures,
where $V$ is the volume occupied by the 
suspension and $N$ the total number of rod-like particles.

We summarize the final model :
\begin{align}
\partial_t f & =  -\nabla_{\vec{x}} \cdot \left(\vec{u} f \right)
- \nabla_{\vec{n}} \cdot \left( P_{\vec{n}^\bot} \nabla_{\vec{x}} \vec{u} \vec{n} f \right)  +  D_r \Delta_{\vec{n}} f 
\nonumber \\
&  + D_\bot 
\nabla_{\vec{x}} \cdot \left( I + \vec{n} \otimes \vec{n} \right) 
\left( \nabla_{\vec{x}} f + \frac{1}{k_B \theta} 
m_0 g \vec{e}_3  f \right) \label{eqn:smol_physdim}
\\
\sigma(\vec{x},t) & =  k_B \theta \int_{S^{d-1}} \left( d \vec{n} \otimes \vec{n} - I \right) f(\vec{x},t, \vec{n}) d \vec{n}\label{eqn:sigma_physdim}
\\
\nabla_{\vec{x}} \cdot \vec{u} &= 0 \label{eqn:divu}
\\
\rho_f \left(\partial_t \vec{u} + \left(\vec{u} \cdot \nabla_{\vec{x}}\right) 
\vec{u} \right) & =   \mu \Delta_{\vec{x}}  \vec{u} 
- \nabla_{\vec{x}} p + \nabla_{\vec{x}} \cdot \sigma   - 
\Big ( \int_{S^{d-1}} f \, d \vec{n} \Big ) m_0 g \vec{e}_3
\label{eqn:stokes_physdim}
\end{align}

\subsection{Thermodynamic consistency of the model}
\label{consistency}

To show thermodynamic consistency of the model we use the free energy functional
\begin{equation}\label{eqn:entropy}
A[f] := \int_\Omega \int_{S^2} \left( k_B \theta f \ln f + f U (\vec x) \right) d \vec{n} d \vec{x},
\end{equation}
where $U(\vec x) = m_0 g \vec{x} \cdot \vec{e}_3$ is the gravitational potential.

\begin{proposition}
For $f$ satisfying the Smoluchowski equation \eqref{eqn:smol_physdim},
the free energy $A[f]$ defined in (\ref{eqn:entropy}) satisfies the identity
\begin{equation} 
\label{freen}
\begin{split}
&\partial_t A[f] + D_r k_B \theta \int_{\Omega} \int_{S^2} f |\nabla_{\vec{n}} \ln f|^2 
d\vec{n} d\vec{x}
\\
&+ D_\perp  k_B \theta \int_{\Omega} \int_{S^2}
\nabla_{\vec{x}} \big ( \ln f + \frac{1}{k_B \theta} U \big ) \cdot 
\big (  I + \vec{n} \otimes \vec{n} \big) 
\nabla_{\vec{x}} \big ( \ln f + \frac{1}{k_B \theta} U \big ) d\vec{n} d\vec{x}
\\
&\quad = \int_{\Omega} \nabla_{\vec{x}} \vec{u} : \sigma d\vec{x}
+ \int_\Omega m_0 g \vec{e}_3 \left( \int_{S^2} f d\vec{n}\right)  \cdot \vec{u} d\vec{x}
\end{split}
\end{equation}
Moreover,  the total energy
\begin{equation}
E[\vec{u}, f] = \int_{\Omega}  \left ( \frac{1}{2} \rho_f |\vec{u}|^2
+ \int_{S^2} \Big ( (k_B \theta) f \ln f + f U (\vec x) \Big ) d \vec{n}\right ) d\vec{x}
\end{equation}
of the system \eqref{eqn:smol_physdim}-\eqref{eqn:stokes_physdim} dissipates.
\end{proposition}

{\bf Proof:} We differentiate (\ref{eqn:entropy}) with respect to $t$,
\begin{equation}
\partial_t A[f]  = 
\int_\Omega \int_{S^2} \Big ( k_B \theta ( 1 + \ln f ) + U(\vec{x}) \Big ) f_t \, d\vec{n} d \vec{x} \label{eqn_dtA}
\end{equation}
and use (\ref{eqn:smol_physdim}) to express the various contributions.

The  contribution of the transport term $-\nabla_{\vec{x}} \cdot (\vec{u}f)$ gives
\begin{equation*}
\begin{split}
I_{tr} &= 
- \int_\Omega \int_{S^2} \Big ( k_B \theta ( 1 + \ln f ) + U(\vec{x}) \Big ) 
\nabla_{\vec x} \cdot (\vec{u} f ) \, d\vec{n} d \vec{x} 
\\
&= \; \int_\Omega \int_{S^2} \Big ( k_B \theta  \nabla_{\vec x} f +  f \nabla_{\vec x} U \Big ) 
\cdot \vec{u} d\vec{n} d \vec{x} 
\\
& \stackrel{(\ref{eqn:divu})}{=}   \int_\Omega  m_0 g\vec{e}_3 \left(
\int_{S^2} f d \vec{n} \right) \cdot \vec{u} \, d \vec{x}
\end{split}
\end{equation*} 
The contribution of the drift term 
$-\nabla_{\vec{n}} \cdot \left( P_{\vec{n}^\bot} \nabla_{\vec{x}}\vec{u} \vec{n} f \right)$ is :
\begin{equation*}
\begin{split}
I_{dr} &=  
- \int_\Omega \int_{S^2} \Big ( k_B \theta ( 1 + \ln f ) + U(\vec{x}) \Big ) 
\nabla_{\vec n} \cdot 
\left( P_{\vec{n}^\bot} \nabla_{\vec{x}} \vec{u} \vec{n} f \right)\, d\vec{n} d \vec{x} 
\\
& \stackrel{(\ref{A.1})}{=} 
\int_\Omega \int_{S^2} k_B \theta \nabla_{\vec n} \ln f \cdot 
 P_{\vec{n}^\bot} \left( \nabla_{\vec{x}} \vec{u} \vec{n}  f \right) \, d\vec{n} d \vec{x} 
\\
&= \int_\Omega \nabla_{\vec{x}} \vec{u} : k_B \theta \int_{S^2} 
\vec{n} \otimes \nabla_{\vec{n}} f \, d\vec{n} \, d\vec{x}
\\
& \stackrel{(\ref{A.3}), \eqref{eqn:sigma_physdim}}{=}
  \int_\Omega \nabla_{\vec{x}} \vec{u} : \sigma \, d \vec{x}.
\end{split}
\end{equation*}

The contribution of rotational diffusion leads to
\begin{equation*}
\begin{split}
I_{rd} &=
 \int_\Omega \int_{S^2} \Big ( k_B \theta ( 1 + \ln f ) + U(\vec{x}) \Big ) 
       D_r \Delta_{\vec{n}} f \, d \vec{n} d \vec{x}
\\
&=
\int_\Omega \int_{S^2} \Big ( k_B \theta ( 1 + \ln f ) + U(\vec{x}) \Big ) 
       D_r  \nabla_{\vec n} \cdot ( f \nabla_{\vec n} \ln f ) \, d \vec{n} d \vec{x}
\\
& = - \frac{(k_B \theta)^2}{\zeta_r} \int_\Omega \int_{S^2} | \nabla_{\vec{n}} \ln f |^2 f \, d \vec{n} d \vec{x} \, .
\end{split}
\end{equation*}
Finally, the contribution of the last term in
(\ref{eqn:smol_physdim}), modeling  the effect of translational friction and translational diffusion,
reads
\begin{equation*}
\begin{split}
I_{tdf} &=  
\int_\Omega \int_{S^2} \Big ( k_B \theta ( 1 + \ln f ) + U(\vec{x}) \Big ) 
   D_\bot \nabla_{\vec{x}} \cdot \left( I + \vec{n} \otimes \vec{n} \right) 
   \Big( \nabla_{\vec{x}} f +  \frac{f}{k_B\theta} \nabla_{\vec x} U \Big)
\\
& = - \frac{(k_B \theta)^2}{\zeta_\bot}\int_{S^2} \int_\Omega
\nabla_{\vec{x}} \left( \ln f + \frac{1}{k_B\theta} U  \right) \cdot \left( I + \vec{n} \otimes \vec{n} \right) f \nabla_{\vec{x}} 
\left( \ln f + \frac{1}{k_B\theta} U \right) 
\end{split}
\end{equation*}
Combining all these contributions together yields \eqref{freen}.

Next, we multiply the Navier-Stokes equation \eqref{eqn:stokes_physdim} by $\vec{u}$
and integrate over $\Omega$ to obtain the balance of the kinetic energy
\begin{equation}\label{kinen}
\begin{split}
\frac{d}{dt} \int_\Omega \frac{1}{2} \rho_f |\vec{u}|^2 d\vec{x} 
&+ \mu \int_\Omega \nabla_{\vec{x}} \vec{u} : \nabla_{\vec{x}} \vec{u} \, d \vec{x} 
\\
& = - \int_\Omega \nabla_{\vec{x}} \vec{u} : \sigma \, d \vec{x} 
 - \int_\Omega 
\vec{u} \cdot  m_0 g \vec{e}_3 \left(\int_{S^2} f d \vec{n} \right) d \vec{x} \, .
\end{split}
\end{equation}
Combining \eqref{freen} and \eqref{kinen} leads to the balance of total energy.
In particular, it follows that the total energy dissipates.

\subsection{Non-dimensionalization}
We first list the dimensions of the terms that appear in the equations. 
The units of mass, length and time are denoted by $M$, $L$ and $T$. We also
monitor the dependence on the number of particles $N$.

\begin{itemize}
\item $v$: velocity $\left[\frac{L}{T}\right]$
\item $m_0 g$: mass $\times$ acceleration 
$\left[ \frac{ML}{T^2} \right]$; 
\item $k_B \theta$: energy = force $\times$ length $\left[ \frac{M L^2}{T^2}\right]$
\item $\zeta_\bot$: translational friction orthogonal to rod = force / velocity $\left[ \frac{M}{T} \right]$ \\
$D_\bot = \frac{k_B \theta}{\zeta_\bot}$ $\left[\frac{L^2}{T}\right]$ 
\item $\zeta_r$: rotational friction = torque / rotatational velocity $\left[ \frac{M L^2}{T} \right]$ \\
\quad $D_r = \frac{k_B \theta}{\zeta_r}$ $\left[\frac{1}{T}\right]$
\item $\mu = \frac{\mbox{force/area}}{\mbox{velocity gradient}}$ $\left[ \frac{M}{LT}\right]$
\item $p$: pressure = force/area $\left[\frac{M}{LT^2}\right]$
\item $f$: number of particles / volume $\left[ \frac{N}{L^3}\right]$
\item $\sigma \sim (k_B \theta) f$ $\left[\frac{M N}{L T^2} \right]$
\end{itemize}

Now we consider a change of scale of the form
\begin{equation}\label{nondim}
t = T \hat{t}, \, \vec{x} = X \hat{\vec{x}}, \, \vec{u} = \frac{X}{T} \hat{\vec{u}}, \,
f = \frac{N}{V} \hat{f}, \, p = \frac{\mu}{T} \hat{p}, \, \sigma = (k_B \theta) \frac{N}{V} \hat{\sigma}.
\end{equation}
%
In these new units the Smoluchowski equation takes the form
\begin{equation}\label{eqn:smol-scale1}
\begin{split}
& \partial_{\hat{t}} \hat{f}  
+ \nabla_{\hat{\vec{x}}} \cdot \left( \hat{\vec{u}} \hat{f} \right) 
+ \nabla_{\vec{n}} \cdot \left( P_{\vec{n}^\bot} \nabla_{\hat{\vec{x}}} 
\hat{\vec{u}} \vec{n} \hat{f} \right)
= T D_r \Delta_{\vec{n}} \hat{f} 
\\
&+ \frac{T}{X^2} D_\bot \nabla_{\hat{\vec{x}}} \cdot \left( I + \vec{n} \otimes \vec{n} \right) \nabla_{\hat{\vec{x}}} 
\hat{f}
+ \frac{D_\bot}{X} T \frac{m_0 g}{k_B \theta} \nabla_{\hat{\vec{x}}} \cdot
\left( \left( I + \vec{n} \otimes \vec{n} \right) \vec{e}_3 \hat{f} \right)
\end{split}
\end{equation} 
The non-dimensionalization of the elastic stress tensor \eqref{eqn:sigma_physdim} leads to the expression
\begin{equation}
\hat{\sigma} = \int_{S^{d-1}} \left( d \vec{n}\otimes \vec{n} - I \right) \hat{f} d\vec{n} \, , 
\end{equation}
while the conservation of momentum for the flow leads to
\begin{equation}
\label{eqn:nsdim}
\begin{split}
& \frac{X^2}{T \mu} \rho_f  \Big [ \partial_{\hat{t}} \hat{\vec{u}}
+ \left( \hat{\vec{u}} \cdot \nabla_{\hat{\vec{x}}}\right) \hat{\vec{u}}  \Big ]
\\
& = \Delta_{\hat{\vec{x}}} \hat{\vec{u}} - \nabla_{\hat{\vec{x}}} \hat{p}
+ \frac{XT}{\mu} \frac{k_B \theta}{X} \frac{N}{V} \nabla_{\hat{\vec{x}}} \cdot \hat{\sigma}  - \frac{N}{V} m_0 g \frac{XT}{\mu} \left(  \int_{S^{d-1}} 
\hat{f} \, d\vec{n} \right) \vec{e}_3.
\end{split}
\end{equation}

We have used two length scales in \eqref{nondim}:   a length scale $X$ that is microscopic in nature 
 and a length scale $L$  standing for the size of the macroscopic domain and 
entering only through the  volume  $V$ occupied by the suspension $V = O( L^3 )$. $T$ is an observational time scale and its role
will be clarified later. The ratio $X/T$ is fixed at this point to be the velocity of sedimentation
\begin{equation}\label{eqn:v-scale}
\frac{X}{T} = \frac{m_0 g}{\zeta_\bot} =: v_{sed},
\end{equation}
i.e.\ the velocity scale is proportional to the motion of a single rod falling due to gravity in a friction dominated flow. 

A review of the Navier-Stokes equation indicates that there are three dimensionless numbers at play: 
First, a Reynolds number based on the velocity $X/T = v_{sed}$
\begin{equation}
\label{defRe}
Re:=  \frac{ X v_{sed}}{\tfrac{\mu}{\rho_f}} = \rho_f \frac{X^2}{T \mu} \, .
\end{equation}
Second, a dimensionless number $\Gamma$ describing the ratio of elastic versus viscous stresses at the fluid,
\begin{align}
\label{defGamma}
\Gamma :=  \frac{XT}{\mu} \frac{k_B \theta}{X} \frac{N}{V} 
&= \left ( \frac{ X^2 \rho_f}{T \mu} \right ) \left ( \frac{N}{V} \frac{ k_B \theta}{\rho_f  \left ( \tfrac{X}{T} \right )^2 } \right )
\\
\nonumber
&=  \; \;  \frac{ {\rm inertial}}{ {\rm viscous}}  \quad \; \;  \frac{ {\rm elastic} }{ {\rm inertial} }
\end{align}
Third, a dimensionless number $\delta$ describing the ratio between boyancy forces and viscous stresses,
\begin{align}
\label{defdelta}
\delta := \frac{N}{V} m_0 g \frac{XT}{\mu} &= \left ( \frac{ X^2 \rho_f}{T \mu} \right ) \left ( \frac{ N m_0 g}{ V \rho_f \tfrac{X}{T^2} } \right )
\\
\nonumber
&= \; \;  \frac{ {\rm inertial}}{ {\rm viscous}}  \quad  \frac{ {\rm boyancy} }{ {\rm inertial} }  
\end{align}

In interpreting the above definitions one observes that $V \rho_f v_{sed}^2 = V \rho_f  \left( \tfrac{X}{T} \right)^2$ is the total kinetic energy of
the sedimenting solution, while $N k_B \theta$ stands for the total elastic energy of entropic origin of the microstructure. Hence, the last term in
\eqref{defGamma} is the ratio of elastic over inertial forces.  The term $V \rho_f \tfrac{X}{T^2}$ is the inertial force of the solution at the selected
length and time scales, while $N m_0 g$ stands for the total boyancy force. Hence, the last term in \eqref{defdelta} stands for the
ratio of boyancy over inertial forces, and in the usual practice of fluid mechanics it will be denoted as $\frac{1}{Fr}$ where $Fr$ is a Froude number.
Hence, we have
$$
\delta = Re \frac{1}{Fr}
$$
We also define
\begin{equation}
\label{defgamma}
\gamma := \frac{\Gamma}{\delta} = \frac{ k_B \theta}{X m_0 g}
\end{equation}
and note that $\gamma$ stands for the ratio of elastic over boyancy forces, while $\delta$ is the ratio of the boyancy over viscous stresses.
The non-dimensional form of the Navier-Stokes equation then becomes
\begin{equation}
Re \left(\partial_{\hat{t}} \hat{\vec{u}} + 
\left(\hat{\vec{u}} \cdot \nabla_{\hat{\vec{x}}}\right) \hat{\vec{u}}\right) 
= \Delta_{\hat{\vec{x}}} \hat{\vec{u}} - \nabla_{\hat{\vec{x}}} \hat{p}
+ \delta \gamma \nabla_{\hat{\vec{x}}} \cdot \sigma - \delta
\left(  \int_{S^{d-1}} \hat{f} \, d \vec{n} \right) \vec{e}_3. 
\end{equation}

We turn now to the transport equation \eqref{eqn:smol-scale1}.  We introduce the Deborah number
$$
De := \frac{1}{D_r T}
$$
which as usual expresses the ratio of a stress relaxation time $\tfrac{1}{D_r}$ over the observational time scale $T$.
We also note that by virtue of \eqref{defgamma} and \eqref{eqn:v-scale} 
$$
\frac{T}{X^2} D_\bot \frac{1}{\gamma} = \frac{T}{X^2} \frac{k_B \theta}{\zeta_\bot} \frac{ X m_0 g}{k_B \theta} =1
$$
Hence, the kinetic equation \eqref{eqn:smol-scale1} may be expressed in the dimensionless form
\begin{equation*}
\begin{split}
& \partial_{\hat{t}} \hat{f}  
+ \nabla_{\hat{\vec{x}}} \cdot \left( \hat{\vec{u}} \hat{f} \right) 
+ \nabla_{\vec{n}} \cdot \left( P_{\vec{n}^\bot} \nabla_{\hat{\vec{x}}} 
\hat{\vec{u}} \vec{n} \hat{f} \right)
= \frac{1}{De}  \Delta_{\vec{n}} \hat{f} 
+ \nabla_{\hat{\vec{x}}} \cdot \left( I + \vec{n} \otimes \vec{n} \right)  
\left ( \gamma \nabla_{\hat{\vec{x}}} \hat{f}  + \vec{e}_3 \hat{f} \right)
\end{split}
\end{equation*} 
dependent on two dimensionless numbers: $\gamma$ defined in \eqref{defgamma}  and the Deborah number $De$ (or equivalently the observational time scale $T$).
In the sequel we will use the notation $D_r$  (of the rotational diffusion coefficient) in the place of  $\frac{1}{De}$ in order to simplify the notation.

We summarize the non-dimensional form of the equations 
(dropping the hats)
\begin{equation}\label{eqn:system_nondim}
\begin{split}
\partial_t f  + \nabla_{\vec{x}} \cdot \left(\vec{u} f \right)
& +\nabla_{\vec{n}} \cdot \left(P_{\vec{n}^\bot} \nabla_{\vec{x}} \vec{u} \vec{n} f \right) - \nabla_{\vec{x}} \cdot \left(\left(I + \vec{n} \otimes \vec{n} \right)
\vec{e}_3 f\right)\\
&  =  D_r \Delta_{\vec{n}}f + \gamma \nabla_{\vec{x}} \cdot \left(I+\vec{n} \otimes \vec{n} \right) \nabla_{\vec{x}} f\\
\sigma  & = \int_{S^{d-1}} \left( d \vec{n} \otimes \vec{n} - I \right) f\,  d \vec{n}\\
Re \left(\partial_{t} \vec{u} + \left( \vec{u} \cdot \nabla_{\vec{x}} \right) \vec{u}\right) & = \Delta_{\vec{x}} \vec{u} - \nabla_{\vec{x}} p
+ \delta \gamma \nabla_{\vec{x}} \cdot \sigma - \delta
\left(  \int_{S^{d-1}} f \, d \vec{n} \right) \vec{e}_3\\
\nabla_{\vec{x}} \cdot \vec{u} & = 0
\end{split}
\end{equation}
If we express the Navier-Stokes equation in the equivalent form (\ref{eqn_stokes_2}), then the associated non-dimensional form is given by
\begin{equation}\label{eqn_stokes_3}
\begin{split}
Re \left(\partial_{t} \vec{u} + \left( \vec{u} \cdot \nabla_{\vec{x}} \right) \vec{u}\right) & = \Delta_{\vec{x}} \vec{u} - \nabla_{\vec{x}} p
+ \delta \gamma \nabla_{\vec{x}} \cdot \sigma + \delta 
\left(1-  \int_{S^{d-1}} f \, d \vec{n} \right) \vec{e}_3\\
\nabla_{\vec{x}} \cdot \vec{u} & = 0
\end{split}
\end{equation}

\subsection{Multi-scale mechanism for instability and cluster
  formation}\label{section:mechanism}
The multi-scale mechanism that leads to the instability and the
formation of clusters was first explained by Koch and Shaqfeh, see
\cite{KS89}. 
%
%
\begin{figure}[htb]
(a)\includegraphics[width=0.2\textwidth]{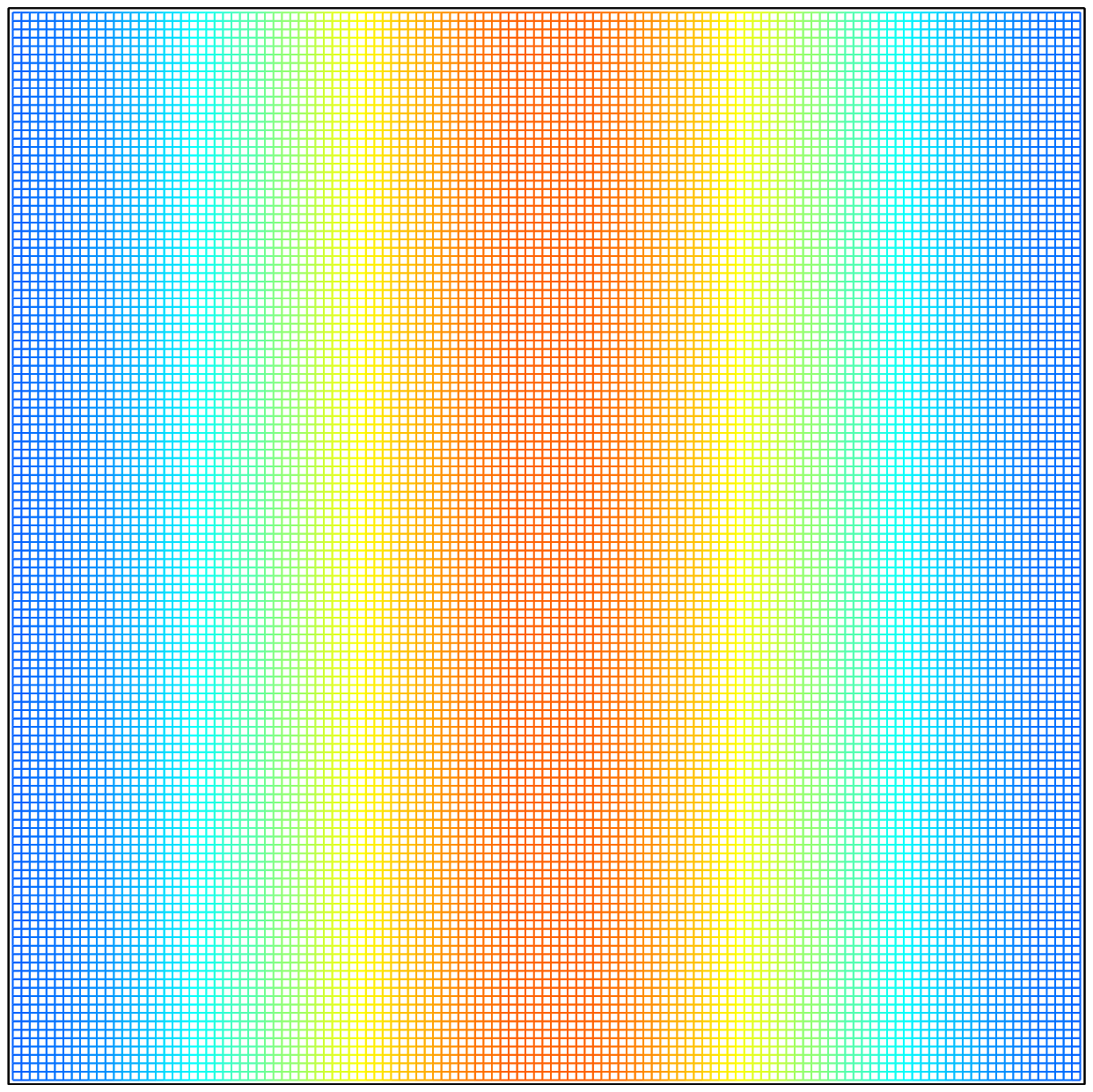}\hfil
(b)\includegraphics[width=0.2\textwidth]{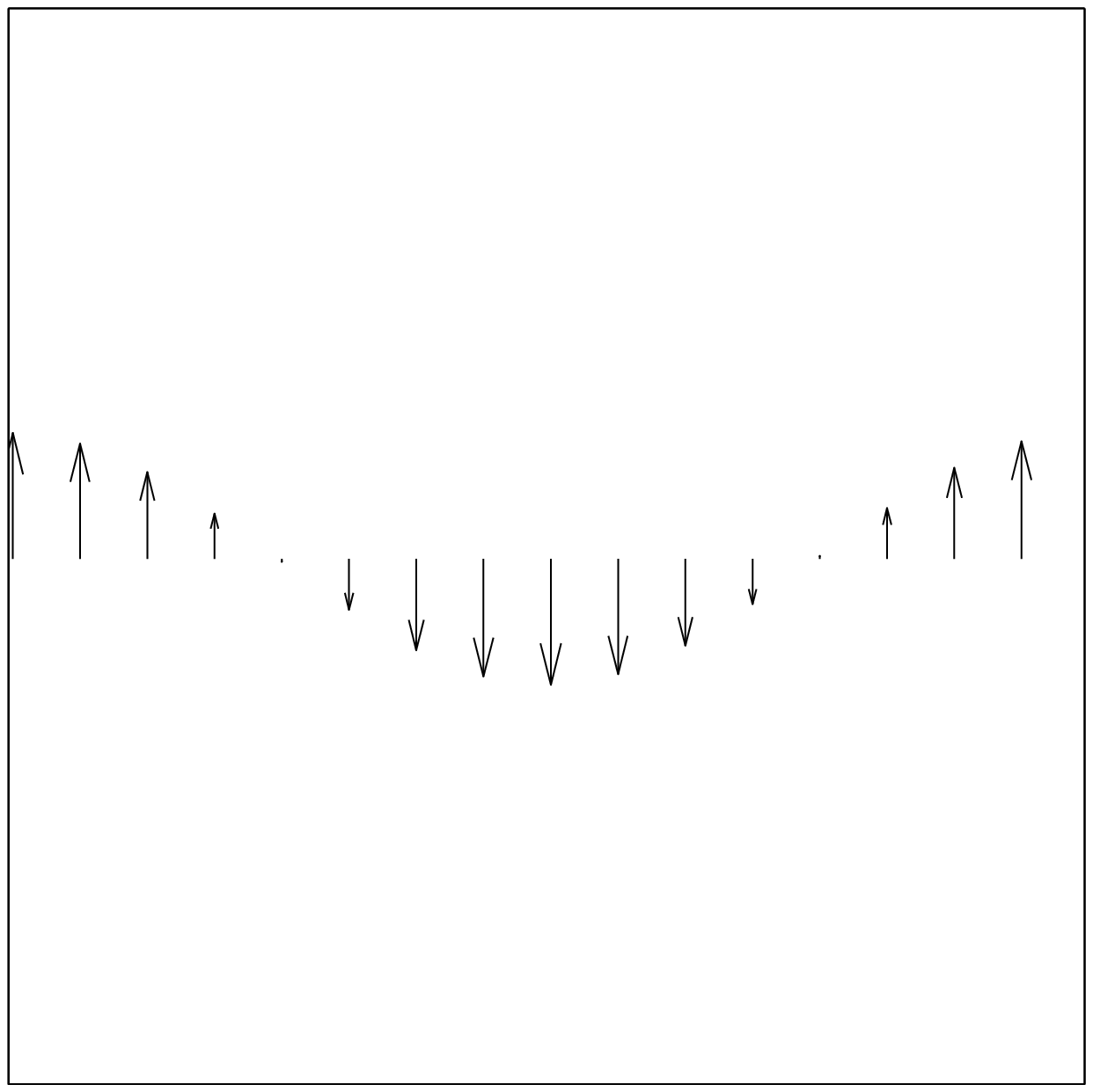}\hfil
(c)\includegraphics[width=0.2\textwidth]{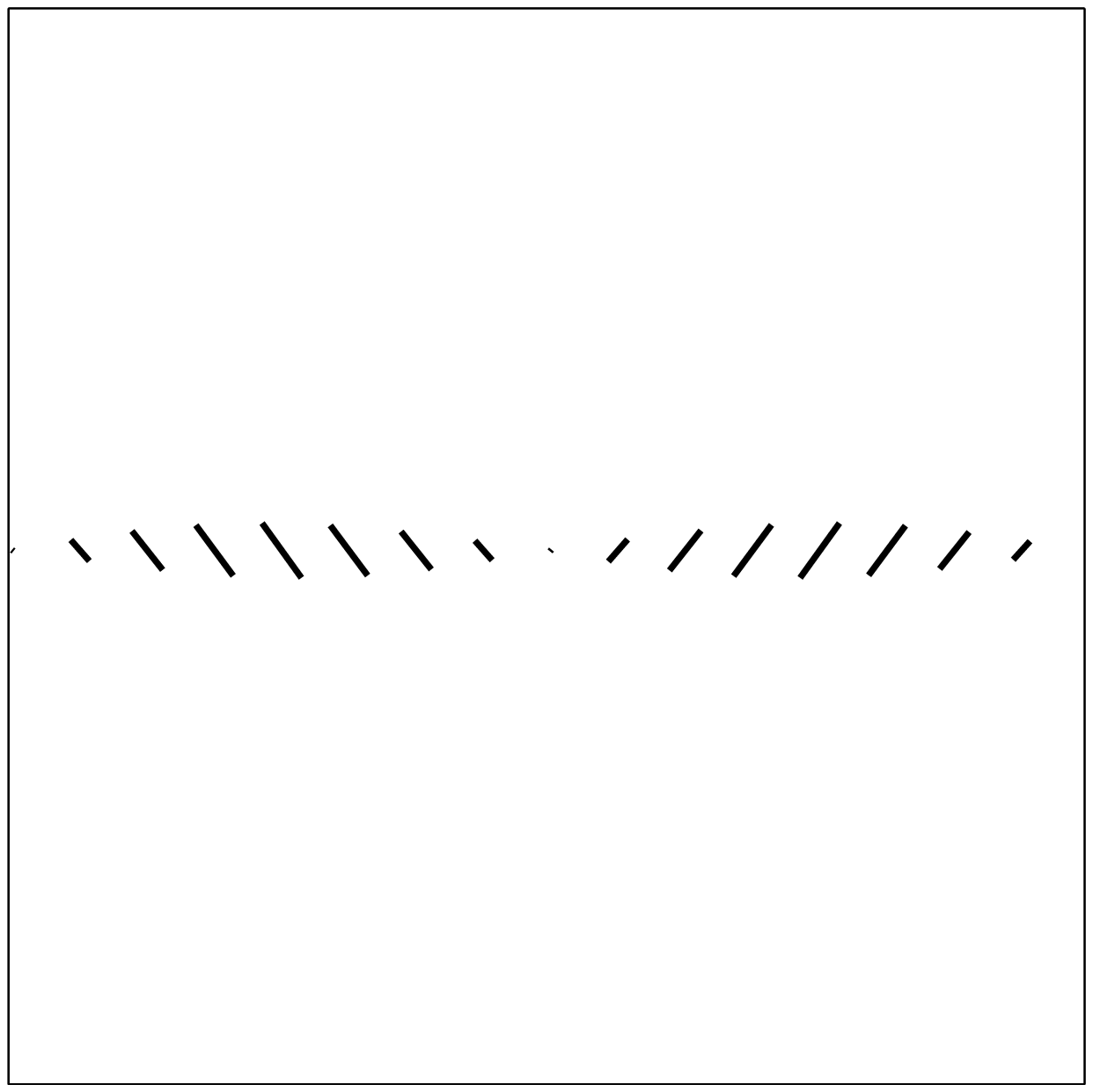}\hfil
(d)\includegraphics[width=0.2\textwidth]{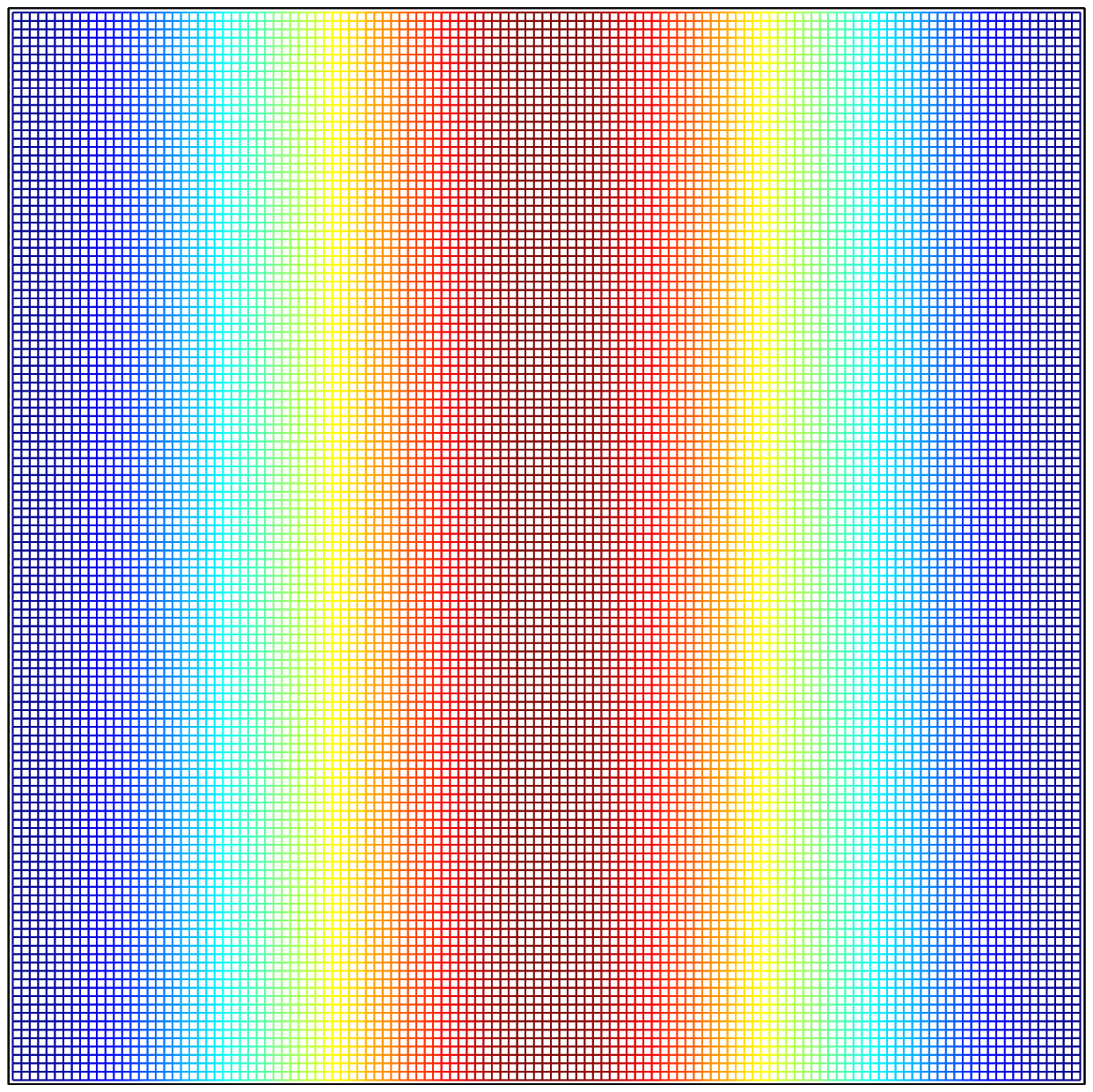}
\caption{\label{figure:concentration}{Illustration of the concentration mechanism (a) initial 
density modulation, (b) velocity field, (c) microscopic orientation: plot of $\lambda \vec{r}$, where $\lambda$ is the largest eigenvalue of $\sigma$ and $\vec{r}$ is the corresponding eigenvector, 
(d) increased density modulation at later time.}}
\end{figure}

In our kinetic model, the function 
$$
\rho(\vec{x},t) = \int_{S{^{d-1}}} f(\vec{x},t,\vec{n}) d \vec{n}
$$
measures the density of rod like particles.
By virtue of the buoyancy term in the Stokes equation, a density 
modulation (as indicated in Figure \ref{figure:concentration} (a)) triggers a modulated 
shear flow with flow direction $\vec{e}_3$ (see Fig.\ \ref{figure:concentration} (b)).

By virtue of the microscopic drift term on the sphere 
$\nabla_{\vec{n}} \cdot \left(P_{\vec{n}^\bot} 
\nabla_{\vec{x}}\vec{u}\vec{n} f \right)$ this shear destroys the uniform
distribution $f$ in $\vec{n}$. For moderate local Deborah numbers the 
distribution $f$ slightly concentrates in a direction at 45 degrees 
between the flow direction and the shear direction  as shown in Fig.\ \ref{figure:concentration} (c).
For larger shear rates the distribution $f$ concentrates more pronounced 
in direction of gravity.
In this figure we plotted the average microscopic orientation $\lambda
\vec{r}$, where $\lambda$ is the largest eigenvalue of $\sigma$ and
$\vec{r}$ is the corresponding eigenvector.

By virtue of the term $- \nabla_{\vec{x}}\cdot \left(
\left( I + \vec{n} \otimes \vec{n} \right) \vec{e}_3 f \right)$
this nonuniform distribution in $\vec{n}$ implies that particles on average 
fall in a direction which is at a non-vanishing angle between flow
direction and shear direction. 
Hence this term acts as a horizontal drift term for the  
modulated density $\rho$ . In fact, it reinforces the original horizontal 
modulation of the density $\rho$ since particles with an orientation as shown
in Fig.\ \ref{figure:concentration} (c) move towards the center.

The goal of this article is to provide a quantitative analysis of the mechanism of cluster formation. 
In the process we will use the model  \eqref{eqn:system_nondim} as well as some
simplified problems derived from that system.

\subsection{Perturbation of the quiescent flow - the linearized problem}

The problem \eqref{eqn:system_nondim} admits a special class of trivial solutions describing
a quiescent flow. One verifies that functions  $(\bar f, \bar u, \bar \sigma, \bar p)$ given by
\begin{equation}
\bar f = f_0 \, , \quad \bar u = 0 \, , \quad \bar \sigma = \sigma_0 
 \, , \quad \bar p = p_0 (x) = - \delta f_0 a_{d-1} \vec{e}_3 \cdot \vec{x}
\end{equation}
where $f_0$ is a constant,
$$
\sigma_ 0 := f_0 \int_{S^{d-1}} \left( d \vec{n} \otimes \vec{n} - I \right)  d \vec{n}  \, ,
$$
and $a_{d-1} = \int_{S^{d-1}} 1 d \vec{n}$ the area of the (d-1)-dimensional sphere, are special
solutions for any constant $f_0$, where $p_0 (x)$ stands for hydrostatic pressure induced by
the presence of the rods.

We derive now the linearized equation obeyed for a perturbation of the quiescent flow. For simplicity
take $f_0 =1$, that is the basic flow is $\bar f = 1$, $\bar u = 0$, $\bar p = p_0 (x) = - \delta a_{d-1} \vec{e}_3 \cdot \vec{x}$.
Write
\begin{equation}
f = 1 + F  \, , \quad  u  \, , \quad \sigma = \sigma_0 + \Sigma
 \, , \quad p = p_0 (x) + P
\end{equation}
a perturbation of the flow and let
$$
m  = \int \int_{S{^{d-1}}} F_0  (\vec{x},\vec{n}) d \vec{n} d\vec{x}
$$
be the initial mass of the perturbation (which is of course conserved).

One computes easily that the linearized system satisfied by 
the perturbation $(F, u, P)$ of the flow is
\begin{equation}\label{linsystem}
\begin{split}
\partial_t F  + \nabla_{\vec{n}} \cdot ( P_{\vec{n}^\bot} \nabla_{\vec{x}} \vec{u} \vec{n}  )  
- \nabla_{\vec{x}} \cdot \left(\left(I + \vec{n} \otimes \vec{n} \right) \vec{e}_3  F \right)
 &=  D_r \Delta_{\vec{n}} F  + \gamma \nabla_{\vec{x}} \cdot \left(I+\vec{n} \otimes \vec{n} \right)  \nabla_{\vec{x}} F 
\\
\Sigma  & = \int_{S^{d-1}} \left( d \vec{n} \otimes \vec{n} - I \right) F\,  d \vec{n}
\\
Re \partial_{t} \vec{u} = \Delta_{\vec{x}} \vec{u} 
- \nabla_{\vec{x}}  ( P + \delta m \vec{e}_3 \cdot \vec{x} )
&+ \delta \gamma \nabla_{\vec{x}} \cdot \Sigma + \delta
\big ( m -  \int_{S^{d-1}} F \, d \vec{n} \big) \vec{e}_3\\
\nabla_{\vec{x}} \cdot \vec{u} & = 0
\end{split}
\end{equation}

The study of the linearized system \eqref{linsystem} is a challenging problem, even if one restricts to the case $\gamma =0$ where
the elastic effects of the microstructure decouple. Nevertheless, it is an important problem for the following reason: Statistical analysis of 
realizations of sedimenting flows in the experimental works \cite{HG96, HG99} indicates that
there exists a characteristic length associated to the cluster formation. It would be important to study if the linearized analysis can 
predict such a length selection mechanism and how it relates to the various parameters of the flow. 
We have taken a step in this direction in Appendix \ref{appendix:linstab}: for a linearized moment closure approximation
of shear flow, we perform a linear stability analysis and show that when $Re > 0$ then there exists a most unstable wavelength.

\subsection{Some special sedimenting flows}

We now present some special flows that are adapted to sedimentation of rigid rods in the vertical
direction. The flows take place in $\R^3$ with the generic point $\vec{x} = (x, y, z)^T$.
The gravity is in the direction of the negative $z$-axis, while the rigid rods take values on the sphere $\vec{n} = (n_1, n_2, n_3)^T \in S^{2}$.

\subsubsection{Rectilinear flows}
We consider first the question whether  \eqref{eqn:system_nondim} admits solutions following the 
{\it ansatz}  of  a rectilinear flow, namely 
\begin{equation}
\label{ansatz}
f= f(t,x,y , \vec{n} ) \, , \quad 
\vec{u} = 
\begin{pmatrix} u \\ v \\ w \end{pmatrix} 
=
\begin{pmatrix} 0 \\ 0  \\ w (t,x.y) \end{pmatrix} 
 \, , \quad
 p= p(t,x,y,z) \, .
\end{equation}
The incompressibility
condition is automatically satisfied,  equation \eqref{eqn:system_nondim}$_1$ gives
\begin{equation}
\label{microre}
\begin{aligned}
\partial_t f &+ \nabla_{\vec{n}} \cdot \left [ P_{\vec{n}^\bot}  (0, 0, n_1 \del_x w + n_2 \del_y w)^T  f  \right ]
- \del_x (n_1 n_3 f) -  \del_y (n_2 n_3  f  )
\\
&= D_r \Delta_{\vec{n}} f
+ \gamma  \Big ( (1+ n_1^2 ) \del_x^2 f + 2 n_1 n_2 \del_x \del_y f + (1 + n_2^2)  \del_y^2 f \Big ) \, ,
\end{aligned}
\end{equation}
while the Navier-Stokes equation \eqref{eqn:system_nondim}$_3$ reduces to
\begin{equation}
\label{press}
\begin{aligned}
\partial_x p &= \delta \gamma ( \partial_x \sigma_{11} + \partial_y \sigma_{12} )
\\
\partial_y p &= \delta \gamma ( \partial_x \sigma_{21} + \partial_y \sigma_{22} )
\\
\partial_z p &= - Re \, \frac{\partial w}{\partial t} + \triangle_{(x,y)} w +  \delta \gamma ( \partial_x \sigma_{31} + \partial_y \sigma_{32} )
- \delta \int_{S^{2}} f \, d \vec{n} 
\end{aligned}
\end{equation}
where $\triangle_{(x,y)}$ stands for the two-dimensional Laplace operator and
\begin{equation}
\label{sigmare}
\sigma  = \sigma(t,x,y) = \int_{S^{2}} \left( 3 \vec{n} \otimes \vec{n} - I \right) f\,  d \vec{n} \, .
\end{equation}

The ansatz \eqref{ansatz} implies the right hand side in \eqref{press}$_3$ depends only on $(x,y)$. We deduce
that the pressure 
$$
p = \kappa z + P(x,y) \, ,
$$
where $\kappa$ is arbitrary and reflects the effect
of an imposed pressure gradient  (or it might even be that $\kappa = \kappa (t)$ if the imposed gradient varies with time).
The functions  $(f, w)$ are selected by solving the coupled  system consisting of \eqref{microre}, \eqref{sigmare} and
\begin{equation}
\label{momentumre}
 Re \, \frac{\partial w}{\partial t} =  \triangle_{(x,y)} w +  \delta \gamma ( \partial_x \sigma_{31} + \partial_y \sigma_{32} )
+  \delta  ( m  -  \int_{S^{2}} f \, d \vec{n} ) \, .
\end{equation}
In \eqref{momentumre} we selected the constant $\kappa = \del_z p = - \delta m$ with $m$ the total mass
$$
m = \iint \int_{S^{2}}  f\,  d \vec{n} \, dx \, dy 
$$
which is conserved under appropriate (say periodic) boundary conditions in $(x,y)$. Note that there is an arbitrary 
gradient of pressure that  can be imposed from the outside. The selected constant is the one amounting to equilibration
of the mean flow.

Finally, in order to understand the solution of \eqref{microre}, \eqref{sigmare}, \eqref{momentumre} as a rectilinear flow
we have to select the pressure $P(x,y)$  by solving
\begin{equation}
\begin{aligned}
\partial_x P &= \delta \gamma ( \partial_x \sigma_{11} + \partial_y \sigma_{12} )
\\
\partial_y P &= \delta \gamma ( \partial_x \sigma_{21} + \partial_y \sigma_{22} )
\end{aligned}
\end{equation}
This system is consistent provided that $\sigma$ satisfies the consistency  condition 
$$
\del_{xy} ( \sigma_{11} - \sigma_{22} ) + ( \del_{yy} -\del_{xx} ) \sigma_{12} = 0 \, .
$$
Such conditions reflect symmetries of the initial data that give rise to the special rectilinear flow and will
not be pursued further here. In the sequel, we only consider the special case $\gamma =0$ reflecting the case that the
translational Brownian motion is negligible relative to boyancy. For the case $\gamma = 0$ we have $P(x,y) = const$ and the pressure takes
the form $p = -m z$.

\subsubsection{Shear flows}

Shear flows follow the ansatz
\begin{equation}
\label{ansatz2}
f= f(t,x, \vec{n} ) \, , \quad 
\vec{u} =
\begin{pmatrix} 0 \\ 0  \\ w (t,x) \end{pmatrix} 
 \, , \quad
 p= p(t,x,z) \, .
\end{equation}
where $x$ is the horizontal direction and $z$ is the vertical direction (as before). However, $\vec{n}$ is still allowed to take
values in $S^2$ which means that the rigid rods are allowed to move out of the plane of the shear flow. Shear flow is a special case of the rectilinear flow, and adapting the
equations from the previous section we conclude that
$(f,w)$ satisfy the coupled system
\begin{equation}
\label{microreshear}
\begin{aligned}
\partial_t f + \nabla_{\vec{n}} \cdot \left [ P_{\vec{n}^\bot}  (0, 0, n_1 \del_x w )^T  f  \right ]
- \del_x (n_1 n_3 f) &= D_r \Delta_{\vec{n}} f + \gamma   (1+ n_1^2 ) \del_x^2 f 
\\
 Re \, \frac{\partial w}{\partial t} =  \del_x^2 w +  \delta \gamma  \partial_x \sigma_{31} 
&+  \delta  \Big ( m  -  \int_{S^{2}} f \, d \vec{n} \Big )
\end{aligned}
\end{equation}


%
%
%
%
\section{Derivation of a nonlinear moment closure}
\label{section:moment-closure}
Henceforth we take for simplicity $D_r  =1$ and  consider the Smoluchowski equation in the form
\begin{equation}
\label{eqn:moment-1}
\begin{aligned}
\partial_t f  &+ \nabla_{\vec{x}} \cdot (\vec{u} f) + \nabla_{\vec{n}}
\cdot \left( P_{\vec{n}^\bot} \nabla_{\vec{x}} \vec{u} \vec{n} f
\right) 
\\
&= \Delta_{\vec{n}} f + \gamma \nabla_{\vec{x}} \cdot \left( I +
  \vec{n} \otimes \vec{n} \right) \nabla_{\vec{x}} f +
\nabla_{\vec{x}} \cdot \left( \left( I + \vec{n} \otimes \vec{n}
  \right) f \vec{e}_3 \right)
  \end{aligned}
\end{equation}

\subsection{Equations of moments}
The objective is to derive a moment system at the level of second moments. We define the quantities:
\begin{eqnarray*}
0^{th} \mbox{ moment:} \quad & \rho := \int_{S^{d-1}} f d \vec{n} & \\
2^{nd} \mbox{moment:} \quad & S := \int_{s^{d-1}} \left( \vec{n}
  \otimes \vec{n} - \frac{1}{d} I \right) f d\vec{n}, \quad & i.e.\ S_{ij} =
\int_{S^{d-1}} \left( n_i n_j - \frac{1}{d} \delta_{ij} \right) f
d\vec{n}\\
4^{th} \mbox{moment:} \quad & P := \int_{S^{d-1}} \vec{n} \otimes
\vec{n} \otimes \vec{n} \otimes \vec{n} f d\vec{n}, \quad & i.e.\
P_{\alpha \beta i j} = \int_{S^{d-1}} n_\alpha n_\beta n_i n_j f d\vec{n}
\end{eqnarray*} 

The evolution equation for $\rho$ can easily be obtained by
integrating the Smoluchowski equation (\ref{eqn:moment-1}) over the
sphere $S^{d-1}$. Note that there is no contribution from the third and fourth
term. Thus we obtain
\begin{equation*}
\begin{split}
\partial_t \rho + \vec{u} \cdot \nabla_{\vec{x}} \rho & = \gamma
\nabla_{\vec{x}} \cdot \int_{S^{d-1}} \left( I + \vec{n} \otimes
  \vec{n} \right) \nabla_{\vec{x}} f d \vec{n} + \nabla_{\vec{x}} \cdot
\int_{S^{d-1}} \left( I + \vec{n} \otimes \vec{n} \right) f \vec{e}_3
d\vec{n}\\
& = \gamma \nabla_{\vec{x}} \cdot \nabla_{\vec{x}} \cdot \int_{S^{d-1}}
\left( \left( \vec{n} \otimes \vec{n} - \frac{1}{d} I \right) +
  \frac{d+1}{d} I \right) f d\vec{n} \\
& \hspace*{2cm} + \nabla_{\vec{x}} \cdot
\int_{S^{d-1}} \left( \left( \vec{n} \otimes \vec{n} - \frac{1}{d} I
  \right) + \frac{d+1}{d} I \right) \vec{e}_3 f d\vec{n}  \\
& = \gamma \nabla_{\vec{x}} \cdot \nabla_{\vec{x}} \cdot \left( S +
  \frac{d+1}{d} I \rho \right) + \nabla_{\vec{x}} \cdot \left[ \left( S +
  \frac{d+1}{d} I \rho \right) \vec{e}_3 \right]
\end{split}
\end{equation*}

Next we derive the evolution equation for the components of $S$. Note
that in the following we make frequent use of the Einstein summation convention.
Multiplying (\ref{eqn:moment-1}) with $\left( \vec{n} \otimes \vec{n}
  - \frac{1}{d} I \right)$ and integrating over $S^{d-1}$, provides
\begin{equation*}
\begin{split}
\partial_t S_{\alpha \beta} & + \left( \vec{u} \cdot \nabla_{\vec{x}}
\right) S_{\alpha \beta} + \underbrace{\int_{S^{d-1}} \left( n_\alpha n_\beta -
  \frac{1}{d} \delta_{\alpha \beta} \right) \nabla_{\vec{n}} \cdot
\left( P_{\vec{n}^\bot} \nabla_{\vec{x}} \vec{u} \vec{n} f \right)
d\vec{n}}_{(I_1)_{\alpha\beta}} \\
& = \underbrace{\int_{S^{d-1}} \left( n_\alpha n_\beta - \frac{1}{d}
  \delta_{\alpha \beta} \right) \Delta_{\vec{n}} f d\vec{n}}_{(I_2)_{\alpha\beta}} + \gamma
\underbrace{
\int_{S^{d-1}} \left( n_\alpha n_\beta - \frac{1}{d} \delta_{\alpha
    \beta} \right) \nabla_{\vec{x}} \cdot \left( I + \vec{n} \otimes
  \vec{n} \right) \nabla_{\vec{x}} f d\vec{n}}_{(I_3)_{\alpha\beta}}\\
& + \underbrace{\int_{S^{d-1}} \left( n_\alpha n_\beta - \frac{1}{d}
  \delta_{\alpha \beta} \right) \nabla_{\vec{x}} \cdot \left( I +
  \vec{n} \otimes \vec{n} \right) f \vec{e}_3 d\vec{n}}_{(I_4)_{\alpha\beta}}.
\end{split}
\end{equation*}
We now calculate the different integrals separately.

\medskip
\noindent
{\bf Calculation of $I_1$:} 

\begin{equation*}
\begin{split}
I_1 & = \int_{S^{d-1}} \left( \vec{n} \otimes \vec{n} - \frac{1}{d} I
\right) \nabla_{\vec{n}} \cdot P_{\vec{n}^\bot} \nabla_{\vec{x}}
\vec{u} \vec{n} f d\vec{n} \\
& = - \int_{S^{d-1}}\nabla_{\vec{n}} \left( \vec{n} \otimes \vec{n} -
  \frac{1}{d} I \right) \cdot P_{\vec{n}^\bot} \nabla_{\vec{x}}
\vec{u} \vec{n} f d\vec{n} \\
& = - \int_{S^{d-1}} \nabla_{\vec{n}} \left( \vec{n} \otimes \vec{n} \right) \cdot
P_{\vec{n}^\bot} \nabla_{\vec{x}} \vec{u} \vec{n} f d\vec{n}\\
& = - \underbrace{\int_{S^{d-1}} \nabla_{\vec{n}} \left( \vec{n} \otimes \vec{n}
\right) \cdot \nabla_{\vec{x}} \vec{u} \vec{n} f d\vec{n}}_{I_1^1} +
\underbrace{\int_{S^{d-1}} \vec{n} \left( \left( \vec{n} \cdot \nabla_{\vec{n}} \right)
  \vec{n} \otimes \vec{n} \right) \cdot \nabla_{\vec{x}} \vec{u}
\vec{n} f d\vec{n}}_{I_1^2} 
\end{split}
\end{equation*}
where 
\begin{equation*}
\begin{split}
(I_1^1)_{\alpha \beta}  & = \int_{S^{d-1}} \nabla_{\vec{n}} \left(
  n_\alpha n_\beta \right) \cdot \nabla_{\vec{x}} \vec{u} \vec{n} f
d\vec{n} \\
& = \int_{S^{d-1}} \frac{\partial}{\partial n_i} \left( n_\alpha
  n_\beta \right) \frac{\partial u_i}{\partial x_j} n_j f d \vec{n}\\
& = \int_{S^{d-1}} \left( \delta_{i \alpha} n_\beta + n_\alpha
  \delta_{i\beta} \right) \frac{\partial u_i}{\partial x_j} n_j f
d\vec{n}\\
& = \int_{S^{d-1}} \frac{\partial u_\alpha}{\partial x_j} n_j n_\beta f
  + \frac{\partial u_\beta}{\partial x_j} n_\alpha n_j f d\vec{n} \\
& = \int_{S^{d-1}} \left( \frac{\partial u_\alpha}{\partial x_j}
  \left( n_j n_\beta - \frac{1}{d} \delta_{j\beta}\right) +
  \frac{\partial u_\beta}{\partial x_j} \left( n_\alpha n_j -
    \frac{1}{d} \delta_{\alpha j} \right) + \frac{1}{d} \left(
    \frac{\partial u_\alpha}{\partial x_\beta} + \frac{\partial
      u_\beta}{\partial x_\alpha} \right) \right) f d\vec{n}\\
& = \frac{\partial u_\alpha}{\partial x_j} S_{j \beta} + 
\frac{\partial u_\beta}{\partial x_j} S_{\alpha j} + \frac{1}{d} 
\left( \frac{\partial u_\alpha}{\partial x_\beta} + \frac{\partial
    u_\beta}{\partial x_\alpha} \right) \rho
\end{split}
\end{equation*}
i.e.\ 
\begin{equation*}
I_1^1 = \nabla_{\vec{x}}  \vec{u} S + S \nabla_{\vec{x}} \vec{u}^T +
\frac{1}{d} \left( \nabla_{\vec{x}} \vec{u} + \nabla_{\vec{x}}
  \vec{u}^T \right) \rho,
\end{equation*}
and 
\begin{equation*}
\begin{split}
(I_1^2)_{\alpha \beta} & = \int_{S^{d-1}} n_k \frac{\partial}{\partial
  n_k} \left( n_\alpha n_\beta \right) n_i \frac{\partial
  u_i}{\partial x_j} n_j f d\vec{n} \\
& = \int_{S^{d-1}} n_k \left( \delta_{\alpha k} n_\beta + n_\alpha
  \delta_{k \beta} \right) n_i \frac{u_i}{\partial x_j} n_j f d\vec{n}
\\
& = 2 \frac{\partial u_i}{\partial x_j}  \int_{S^{d-1}} n_\alpha n_\beta n_i n_j f d\vec{n}
\end{split}
\end{equation*}
i.e.\ 
$$
I_1^2 = 2 P : \nabla_{\vec{x}} \vec{u}.
$$
Therefore, 
$$
I_1 = - \nabla_{\vec{x}} \vec{u} S - S \nabla_{\vec{x}} \vec{u}^T -
\frac{1}{d} \left( \nabla_{\vec{x}} \vec{u} + \nabla_{\vec{x}}
  \vec{u}^T \right) \rho + 2 P : \nabla_{\vec{x}} \vec{u}.
$$

\medskip
\noindent
{\bf Calculation of $I_2$:}
\begin{equation*}
\begin{split}
I_2 & = \int_{S^{d-1}} \left( \vec{n} \otimes \vec{n} - \frac{1}{d} I
\right) \Delta_{\vec{n}} f d\vec{n} \\
& = \int_{S^{d-1}} \Delta_{\vec{n}} \left( \vec{n} \otimes \vec{n} -
  \frac{1}{d} I \right) f d\vec{n} \\
& \stackrel{(\ref{surfhar})}{=}  - \frac{\ell (\ell + d - 2)}{2d} \int_{S^{d-1}} \left( \vec{n} \otimes
\vec{n} - \frac{1}{d} I \right) f d \vec{n} \\
& = - 2 d S
\end{split}
\end{equation*}

\medskip
\noindent
{\bf Calculation of $I_3$:}
\begin{equation*}
\begin{split}
(I_3)_{\alpha\beta} & = \int_{S^{d-1}} \left( n_\alpha n_\beta -
  \frac{1}{d} \delta_{\alpha \beta} \right) \partial_{x_i} \left(
  \delta_{ij} + n_i n_j \right) \partial_{x_j} f d\vec{n} \\
& = \partial_{x_i} \partial_{x_j} \int_{S^{d-1}} \left( n_\alpha
  n_\beta - \frac{1}{d} \delta_{\alpha \beta} \right) \left(
  \delta_{ij} + n_i n_j \right) f d\vec{n} \\
& = \partial_{x_i} \partial_{x_j} \int_{S^{d-1}} \left( n_\alpha n_\beta
  - \frac{1}{d} \delta_{\alpha \beta} \right) f \delta_{ij} + \left(
  n_\alpha n_\beta - \frac{1}{d} \delta_{\alpha \beta} \right) n_i n_j
f d\vec{n} \\
& = \partial_{x_i} \partial_{x_j} \left( S_{\alpha \beta} \delta_{ij}
  + P_{\alpha \beta i j} - \frac{1}{d} \delta_{\alpha \beta}S_{ij} +
  \frac{1}{d^2} \delta_{\alpha \beta}\delta_{ij} \rho \right)
\end{split}
\end{equation*}

\medskip
\noindent
{\bf Calculation of $I_4$:}
\begin{equation*}
\begin{split}
(I_4)_{\alpha \beta} & = \int_{S^{d-1}} \left( n_\alpha n_\beta -
  \frac{1}{d} \delta_{\alpha \beta} \right) \partial_{x_i} \left( n_i
  n_3 + \delta_{i3}\right) f  d\vec{n} \\
& = \partial_{x_i}  \int_{S^{d-1}} \left( \left( n_\alpha n_\beta -
    \frac{1}{d} \delta_{\alpha \beta} \right) n_i n_3 + \left(
    n_\alpha n_\beta - \frac{1}{d} \delta_{\alpha \beta} \right)
  \delta_{i3} \right) f d\vec{n}\\
& = \partial_{x_i}  \int_{S^{d-1}} \big( n_\alpha n_\beta n_i n_3
  - \frac{1}{d} \delta_{\alpha \beta} \big( n_i n_3 - \frac{1}{d}
    \delta_{i3} \big) + \frac{1}{d^2} \delta_{\alpha \beta}
  \delta_{i3} + \big( n_\alpha n_\beta - \frac{1}{d} \delta_{\alpha
      \beta} \big) \delta_{i3} \big) f d\vec{n}\\
& = \partial_{x_i}  \Big( P_{\alpha \beta i 3} - \frac{1}{d}
  \delta_{\alpha \beta} S_{i3} + \frac{1}{d^2} \delta_{\alpha \beta}
  \delta_{i3} \rho + S_{\alpha \beta} \delta_{i3} \Big)
\end{split}
\end{equation*}

Putting these  together, we derive evolution equations for $\rho$ and $S$:
\begin{equation}\label{equation:rho-S-unclosed}
\begin{split}
\partial_t \rho & + \vec{u} \cdot \nabla_{\vec{x}} \rho  = \gamma
\nabla_{\vec{x}} \cdot \nabla_{\vec{x}} \cdot \left( S + \frac{d+1}{d}
  \rho I \right) + \nabla_{\vec{x}} \cdot \left( S + \frac{d+1}{d}
  \rho I \right) \vec{e}_3\\
\partial_t S_{\alpha \beta}  & + \left( \vec{u} \cdot \nabla_{\vec{x}} \right)
  S_{\alpha \beta} - \frac{\partial u_{\alpha}}{\partial_{x_j}}
    S_{j\beta} - \frac{\partial u_{\beta}}{\partial x_j} S_{\alpha j}
    - \frac{1}{d} \left( \frac{\partial u_{\alpha}}{\partial x_\beta}
      + \frac{\partial u_\beta}{\partial x_\alpha} \right) \rho + 2
    P_{\alpha \beta i j} \frac{\partial u_i}{\partial x_j} \\
& = - 2d  S_{\alpha \beta} + \gamma \partial_{x_i} \partial_{x_j} \left(
  S_{\alpha \beta} \delta_{ij} + P_{\alpha \beta i j} - \frac{1}{d}
  \delta_{\alpha \beta} S_{ij} + \frac{1}{d^2} \delta_{\alpha \beta}
  \delta_{ij} \rho \right) \\
& + \partial_{x_i}  \left( P_{\alpha \beta i 3} - \frac{1}{d}
  \delta_{\alpha \beta} S_{i3} + \frac{1}{d^2} \delta_{\alpha \beta}
  \delta_{i3} \rho + S_{\alpha \beta} \delta_{i3} \right) 
\end{split}
\end{equation}

\subsection{Moment closure}
The evolution equation for $\rho$ is expressed in terms of $0^{th}$ and
$2^{nd}$ moments, the evolution equation of $S$
involves $2^{nd}$ and $4^{th}$-order moments and so on. We would like to close the system at
the level of the $2^{nd}$-order moments. Motivated by the properties of harmonic polynomials 
listed in Appendix \ref{app3}  and in particular \eqref{ds2} we close the system at
the level of second moments by projecting the fourth order homogeneous polynomials to the
subspace of spherical harmonics of $2^{nd}$ and $0^{th}$-order.
To do this, we employ \eqref{ds2} and express the higher order terms of
$P_{\alpha \beta i j}$ in terms of using the harmonic polynomial basis presented
in Appendix \ref{app3}. To get a moment closure on the level of second
moments, we then  neglect all the projections to the 
$4^{th}$ order part of the basis and retain only the projections to the $2^{nd}$ and $0^{th}$-order
part of the basis.

In the sequel we restrict to special flows, shear or rectilinear flows, and implement this derivation in  Sections
\ref{section:mc-shear} and 
\ref{section:mc-rectilinear}.  For these special
flows, the only terms of the form $P_{\alpha \beta i j}$
which do not cancel are $P_{\alpha \beta 1 3} = P_{\alpha \beta 3 1}$ and 
$P_{\alpha \beta 2 3} = P_{\alpha \beta 3 2}$.  Therefore, we now restrict our
considerations to these terms.
To simplify notation, we introduce $p_{\alpha \beta i j} = n_\alpha n_\beta
n_i n_j$ and first consider these expressions. Furthermore, we
introduce the matrix $s$ with $s_{ij} = n_i n_j - \frac{1}{3}
\delta_{ij}$, $i,j = 1,2,3$.
Note that 
\begin{equation*}
\begin{split}
s & =  \begin{pmatrix}
n_1^2 - \frac{1}{3} & n_1 n_2 & n_1 n_3 \\
\cdot & n_2^2 - \frac{1}{3} & n_2 n_3 \\
\cdot & \cdot & n_3^2 - \frac{1}{3}
\end{pmatrix} 
 =  \begin{pmatrix}
\frac{1}{2} P_2^{-2} - \frac{1}{2} P_2^0 & \frac{1}{2} P_2^2 &
P_2^{-1} \\
\cdot & -\frac{1}{2} P_2^{-2} - \frac{1}{2} P_2^0 & P_2^1\\
\cdot & \cdot & P_2^0
\end{pmatrix},
\end{split}
\end{equation*}
where we used the harmonic polynomial basis of Appendix \ref{app3} and
only write out the upper part of the symmetric matrix.

Now we can verify that 
\begin{equation*}
\begin{split}
& (p_{\alpha \beta 13})_{\alpha,\beta=1,2,3}  = 
\begin{pmatrix}
n_1^3 n_3 & n_1^2 n_2 n_3 & n_1^2 n_3^2 \\
\cdot & n_1 n_2^2 n_3 & n_1 n_2 n_3^2 \\
\cdot & \cdot & n_1 n_3^3 
\end{pmatrix} \\[0.2cm]
& = 
\begin{pmatrix}
\sin^3 \theta \cos \theta \cos^3 \phi & \sin^3 \theta \cos \theta \sin
\phi \cos^2 \phi & \sin^2 \theta \cos^2 \theta \cos^2 \phi \\
\cdot & \sin^3 \theta \cos \theta \sin^2 \phi \cos \phi & \sin^2
\theta \cos^2 \theta \sin \phi \cos \phi \\
\cdot & \cdot & \sin \theta \cos^3 \theta \cos \phi
\end{pmatrix} \\[0.2cm]
& = \begin{pmatrix}
\frac{1}{4} P_4^{-3} - \frac{3}{28} P_4^{-1} + \frac{3}{7} s_{13} &
\frac{1}{4} P_4^3 - \frac{1}{28} P_4^1 + \frac{1}{7} s_{23} &
\frac{1}{14} P_4^{-2} - \frac{1}{70} P_4^0 - \frac{1}{7} s_{22} +
\frac{1}{15} \\
\cdot & -\frac{1}{4} P_4^{-3} - \frac{1}{28} P_4^{-1} + \frac{1}{7}
s_{13} & \frac{1}{14} P_4^2 + \frac{1}{7} s_{12} \\
\cdot & \cdot & \frac{1}{7} P_4^{-1} + \frac{3}{7} s_{13}
\end{pmatrix} 
\end{split}
\end{equation*} 
and
\begin{equation*}
\begin{split}
& (p_{\alpha \beta 2 3})_{\alpha,\beta=1,2,3}  = 
\begin{pmatrix}
n_1^2 n_2 n_3 & n_1 n_2^2 n_3 & n_1 n_2 n_3^2\\
\cdot & n_2^3 n_3 & n_2^2 n_3^2 \\
\cdot & \cdot & n_2 n_3^3
\end{pmatrix} \\[0.2cm]
& =  \begin{pmatrix} 
\sin^3 \theta \cos \theta \sin \phi \cos^2 \phi & \sin^3 \theta \cos
\theta \sin^2 \phi \cos \phi & \sin^2 \theta \cos^2 \theta \sin \phi
\cos \phi \\
\cdot & \sin^3 \theta \cos \theta \sin^3 \phi & \sin^2 \theta \cos^2
\theta \sin^2 \phi \\
\cdot & \cdot & \sin \theta \cos^3 \theta \sin \phi 
\end{pmatrix} \\[0.2cm]
& = \begin{pmatrix}
\frac{1}{4} P_4^3 - \frac{1}{28} P_4^1 + \frac{1}{7} s_{23} &
-\frac{1}{4} P_4^{-3} - \frac{1}{28} P_4^{-1} + \frac{1}{7} s_{13} &
\frac{1}{14} P_4^2 + \frac{1}{7} s_{12} \\
\cdot & -\frac{1}{4} P_4^3 - \frac{3}{28} P_4^1 + \frac{3}{7} s_{23} &
-\frac{1}{14} P_4^{-2} - \frac{1}{70} P_4^0 - \frac{1}{7} s_{11} +
\frac{1}{15} \\
\cdot & \cdot & \frac{1}{7} P_4^1 + \frac{3}{7} s_{23}
\end{pmatrix}.
\end{split}
\end{equation*}

Finally we drop higher order terms, i.e.\ we drop all the multiples of the
4th order basis polynomials, and obtain the approximations
\begin{equation*}
\begin{split}
(p_{\alpha \beta 1 3})_{\alpha, \beta = 1,2,3} & \approx
\begin{pmatrix}
\frac{3}{7} s_{13} & \frac{1}{7} s_{23} & -\frac{1}{7} s_{22} +
\frac{1}{15}\\
\cdot & \frac{1}{7} s_{13} & \frac{1}{7} s_{12}\\
\cdot & \cdot & \frac{3}{7} s_{13}
\end{pmatrix} \\
(p_{\alpha \beta 2 3})_{\alpha, \beta = 1,2,3} & \approx
\begin{pmatrix}
\frac{1}{7} s_{23} & \frac{1}{7} s_{13} & \frac{1}{7} s_{12} \\
\cdot & \frac{3}{7} s_{23}  & -\frac{1}{7} s_{11} + \frac{1}{15} \\
\cdot & \cdot & \frac{3}{7} s_{23}
\end{pmatrix}.
\end{split}
\end{equation*}
Now it is straight forward to see that $P_{\alpha \beta 13}$ and
$P_{\alpha \beta 2 3}$ can analogously be approximated by
\begin{equation} \label{eqn:approx-P}
\begin{split}
(P_{\alpha \beta 13})_{\alpha, \beta=1,2,3} &\approx 
 \begin{pmatrix}
\frac{3}{7} S_{13} & \frac{1}{7} S_{23} & -\frac{1}{7} S_{22} +
\frac{1}{15} \rho\\
\cdot & \frac{1}{7} S_{13} & \frac{1}{7} S_{12}\\
\cdot & \cdot & \frac{3}{7} S_{13}
\end{pmatrix} \\[0.2cm]
(P_{\alpha \beta 2 3})_{\alpha, \beta = 1,2,3} & \approx
\begin{pmatrix}
\frac{1}{7} S_{23} & \frac{1}{7} S_{13} & \frac{1}{7} S_{12} \\
\cdot & \frac{3}{7} S_{23}  & -\frac{1}{7} S_{11} + \frac{1}{15} \rho \\
\cdot & \cdot & \frac{3}{7} S_{23}
\end{pmatrix}.
\end{split}
\end{equation}
Furthermore, we note that $P_{\alpha \beta 13} =
P_{\alpha \beta 31}$ and $P_{\alpha \beta 23} = P_{\alpha \beta 3 2}$
holds for all $\alpha, \beta = 1,2,3$.

\subsection{Shear flow}\label{section:mc-shear}
In the special case of shear flow, we consider functions of the form
\begin{equation}\label{eqn:shear}
\begin{split}
f & = f(t,x,\vec{n})\\
\vec{u} & = (0,0,w(t,x))^T \\
S_{i j}  & = S_{i j} (t,x).
\end{split}
\end{equation} 
Furthermore, we restrict our considerations to the case $\gamma=0$.
Under these assumptions, the system (\ref{equation:rho-S-unclosed})
can be written in the form
\begin{equation}
\begin{split}
 \partial_t \rho &= \partial_x S_{13}
 \\
 \partial_t S_{\alpha \beta} - \frac{\partial u_\alpha}{\partial x}
S_{1\beta} &- \frac{\partial u_\beta}{\partial x} S_{\alpha 1} -
\frac{1}{d} \left( \frac{\partial u_\alpha}{\partial x_\beta} +
  \frac{\partial u_\beta}{\partial x_\alpha} \right) \rho 
  \\
&= -2 d S_{\alpha \beta} - 2P_{\alpha \beta 3 1} w_x 
 + \partial_x \left(P_{\alpha \beta 1 3} - \frac{1}{d} \delta _{\alpha \beta} S_{13} \right).
\end{split}
\end{equation}
Now we approximate $P_{\alpha \beta 3 1}$ according to (\ref{eqn:approx-P})
and obtain the nonlinear moment closure model equations
\begin{equation}\label{eqn:mc-shear}
\begin{split}
\partial_t \rho & = \partial_x S_{13}  \\
\partial_t S_{11} - \left(\frac{3}{7} - \frac{1}{d} \right) \partial_x
S_{13} & = - 2 d  S_{11} - \frac{6}{7} w_x S_{13} \\
\partial_t S_{22} - \left( \frac{1}{7} - \frac{1}{d}
\right) \partial_x S_{13} & = -2 d  S_{22} -
\frac{2}{7} w_x S_{13} \\
\partial_t S_{33} - \left(\frac{3}{7} - \frac{1}{d}\right) \partial_x
S_{13} & = -2 d  S_{33} +
\frac{8}{7} w_x S_{13} \\
\partial_t S_{13} +
\frac{1}{7} \partial_x S_{22} & = - 2 d  S_{13} + w_x \left( S_{11}
  +\frac{2}{7} S_{22}\right) + w_x \rho \left( -\frac{2}{15} +
  \frac{1}{d} \right) + \frac{1}{15} \partial_x \rho, 
\end{split}
\end{equation}
which needs to be solved together with the Stokes or Navier-Stokes equations.
The evolution equations for $S_{12}$ and $S_{23}$ are
decoupled from the system (\ref{eqn:mc-shear}) and can be neglected.

\subsection{Rectilinear flow}\label{section:mc-rectilinear}
Now we restrict our considerations to a rectilinear flow, i.e.\ taking the form 
\begin{equation}
\begin{split}
f & = f(t,x,y,\vec{n})\\
\vec{u} & = (0,0,w(t,x,y))^T\\
S & = S(t,x,y)
\end{split}
\end{equation}
Furthermore, we consider again the case $\gamma=0$. Under these assumptions,
the system (\ref{equation:rho-S-unclosed}) can be written in the form
\begin{equation}
\begin{split}
& \partial_t \rho  = \partial_x S_{13} + \partial_y S_{23}\\
& \partial_t S_{\alpha \beta} - \frac{\partial u_\alpha}{\partial x}
S_{1 \beta} - \frac{\partial u_\alpha}{\partial y} S_{2 \beta} -
\frac{\partial u_\beta}{\partial x} S_{\alpha 1} - 
\frac{\partial u_\beta}{\partial y} S_{\alpha 2} - \frac{1}{d} \left(
  \frac{\partial u_\alpha}{\partial x_\beta} + 
\frac{\partial u_\beta}{\partial x_\alpha} \right) \rho 
 \\
& = - 2d S_{\alpha \beta} - 2 w_x 
P_{\alpha \beta 31} - 2 w_y P_{\alpha \beta 32} + \partial_x \left(P_{\alpha \beta 13} -
\frac{1}{d} \delta_{\alpha \beta} S_{13} \right) + \partial_y \left(
P_{\alpha \beta 2 3} - \frac{1}{d} \delta_{\alpha \beta} S_{23} \right)\\
\end{split}
\end{equation}
Using the approximation
(\ref{eqn:approx-P}), we obtain the
nonlinear moment closure for rectilinear flow
\begin{equation}\label{eqn:mc-rectilinear}
\begin{split}
\partial_t \rho & = \partial_x S_{13} + \partial_y S_{23}\\
\partial_t S_{11} + 2 d  S_{11} & = (-2 w_x + \partial_x) \frac{3}{7}
S_{13} + (-2w_y + \partial_y) \frac{1}{7} S_{23} - \frac{1}{d}
(\partial_x S_{13} + \partial_y S_{23} ) \\
\partial_t S_{22} + 2 d  S_{22} & = (-2 w_x + \partial_x )
\frac{1}{7} S_{13} + (-2 w_y + \partial_y) \frac{3}{7} S_{23} -
\frac{1}{d} (\partial_x S_{13} + \partial_y S_{23})\\
\partial_t S_{33} + 2 d  S_{33} & = 2 w_x S_{13} + 2 w_y S_{23} + (-2
w_x + \partial_x) \frac{3}{7} S_{13} + (-2 w_y + \partial_y)
\frac{3}{7} S_{23} - \frac{1}{d} (\partial_x S_{13} + \partial_y
S_{23})\\
\partial_t S_{13} + 2 d  S_{13} & = (S_{11} + \frac{\rho}{d}) w_x +
S_{12} w_y + (-2 w_x + \partial_x) \left( -\frac{1}{7} S_{22} +
  \frac{\rho}{15} \right) + ( -2 w_y + \partial_y) \frac{1}{7}
S_{12}\\
\partial_t S_{23} + 2 d  S_{23} & = w_x S_{12} + w_y \left( S_{22} +
\frac{\rho}{d} \right)+ (-2 w_x + \partial_x) \frac{1}{7} S_{12} + (-2
w_y + \partial_y) \left( -\frac{1}{7} S_{11} + \frac{\rho}{15}
\right)\\
\partial_t S_{12} + 2 d  S_{12} & = (-2 w_x + \partial_x ) \frac{1}{7}
S_{23} + (-2 w_y + \partial_y) \frac{1}{7} S_{13},
\end{split}
\end{equation}
which should be solved together with the macroscopic Stokes or
Navier-Stokes equations.


\section{ Derivation of an effective equation via a quasi-dynamic
  approximation}
\label{section:quasi-dynamic}
In this section, we derive a scalar evolution equation for the particle
density $\rho$, which describes the cluster formation process for
intermediate and long times. We will separately consider shear flow
and rectilinear flow. The computation is much simpler for the shear flow and it
serves as a pedagogical example to explain the main approximation idea.

\subsection{Shear flow}
We consider the system (\ref{eqn:mc-shear}) describing a shear flow. The rods are allowed to
move out of the plane of the shear and  $\vec{n}$ takes values on the sphere $S^2$, hence $d=3$. The system
then is written as
\begin{equation}\label{eqn:qd-1}
\partial_t \rho = \partial_x S_{13}
\end{equation}
\begin{equation}\label{eqn:qd-2}
\partial_t S + A \partial_x S = L S + f(w_x,\rho,\rho_x),
\end{equation}
where $S = (S_{11},S_{22},S_{33},S_{13})^T$, 
$$
f(w_x,\rho,\rho_x) = 
\left( 0,0,0, \left( \tfrac{3}{15} w_x \rho
    + \tfrac{1}{15} \rho_x \right) \right)^T
$$,
\begin{equation}\label{eqn:qd-4}
A = \begin{pmatrix}
0 & 0 & 0 & \frac{1}{3}-\frac{3}{7}  \\
0 & 0 & 0 & \frac{1}{3}-\frac{1}{7} \\
0 & 0 & 0 & \frac{1}{3}-\frac{3}{7} \\
0 & \frac{1}{7}  & 0 & 0
\end{pmatrix}, \quad
L = \begin{pmatrix}
-6 & 0 & 0 & -\frac{6}{7} w_x\\
0 & -6 & 0 & -\frac{2}{7} w_x \\
0 & 0 & - 6 & \frac{8}{7} w_x \\
w_x & \frac{2}{7} w_x & 0 & -6 \end{pmatrix}.
\end{equation}
The idea of the quasi-dynamic approximation is the following: the density $\rho$ is generated by the
0-th order harmonic polynomials associated to the eigenvalue $\lambda_0 =0$ of the Laplace-Beltrami
operator; the stresses $S_{ij}$ are generated by the $2$-nd order spherical harmonics $s_{ij}$ which are eigenfunctions of Laplace-Beltrami 
associated to the common eigenvalue $\lambda= -6$. The stresses are thus expected to decay faster; similarly higher-order harmonics that
are neglected in (\ref{eqn:mc-shear}) are expected to decay even faster.  We split the system 
according to the decay rates of the modes in (\ref{eqn:qd-1}), (\ref{eqn:qd-2}) and set the faster decaying modes to
their local equilibrium; we call this approximation quasi-dynamic approximation since part of the modes
evolve dynamically while the rest of the modes are relaxed immediately to their local equilibrium values.

To accomplish that we set (\ref{eqn:qd-2}) to its
equilibrium $L S+f(w_x,\rho,\rho_x) = 0$ and the dynamics of
(\ref{eqn:mc-shear}) is approximated by 
\begin{equation}\label{eqn:qd-3}
\begin{split}
\partial_t \rho & = \partial_x S_{13}\\
S_{11} + \frac{6}{42} w_x S_{13} & = 0 \\
 S_{22} + \frac{2}{42} w_x S_{13} & = 0 \\
S_{33} - \frac{8}{42} w_x S_{13} & = 0 \\
 - \frac{7}{42} w_x S_{11} -  \frac{2}{42} w_x S_{22}  + S_{13} & = 
 \tfrac{1}{90} ( 3 \rho w_x +  \rho_x )
\end{split}
\end{equation}
As already mentioned, the underlying thinking is that $(S_{11},S_{22},S_{33},S_{13})$ relax
fast to their local equilibria, since their decay rate (at least near equilibrium) is determined
by the first nonzero eigenvalue of the Laplace-Beltrami operator,
while $\rho$ does not relax, as it corresponds to the eigenvalue
$\lambda_0 = 0$ of the Laplace Beltrami operator.

Solving the algebraic equation in  (\ref{eqn:qd-3}), we obtain
\begin{equation}
S_{13} =  \tfrac{\kappa}{90} \frac{1}{\kappa + 46 w_x^2} \big ( 3 \rho w_x + \rho_x \big) 
\qquad \mbox{where $\kappa = 42^2$} \, .
\end{equation}
The effective equation for the
evolution of $\rho$ then has the form
\begin{equation}
\label{flks}
\partial_t \rho =
\del_x  \left( \tfrac{\kappa}{90} \frac{1}{\kappa + 46 w_x^2} \big ( 3 \rho w_x + \rho_x \big)  \right )
\end{equation}
which needs to be
solved together with the Stokes or Navier-Stokes equation for shear flow
\begin{equation}
\label{shearns}
Re \, \partial_t w(t,x) = \partial_{xx} w(t,x) + \delta \left( m -
  \int_{S^{d-1}} f \, dn \right).
\end{equation} 

The system  \eqref{flks}-\eqref{shearns} should be compared to the Keller-Segel model ({\it e.g.} \cite{JL92,HV96,BDP06}) that has been extensively used
as a model for chemotaxis in biology. Compared to the Keller-Segel model, the present  system has the noteworthy difference that  
the convection and diffusion coefficients in \eqref{flks} depend nonlinearly on the gradient of the "potential"  (played here by the shear $w_x$) 
and are in fact decreasing for increasing shear. It is an example within  the general class of flux-limited systems proposed in 
\cite{DS06}, \cite{VSMS13,Per13} as models for flux-limited diffusion, although strictly speaking the present model is not included in the
models listed in the above references and has the feature to be endowed with both flux-limited
convection and diffusion. An extension that also presents anisotropic diffusion is developed in the next section to describe
cluster formation for rectilinear flows.

\subsection{Rectilinear flow}

For rectilinear flows we again apply the quasidynamic approximation, similar in spirit
as for shear flows  but now requiring more cumbersome calculations. 
We rewrite equation (\ref{eqn:mc-rectilinear}) in the form
\begin{equation}\label{eqn:rectilinear-mc-short}
\begin{split}
\partial_t \rho & = \partial_x S_{13} + \partial_y S_{23} \\
\partial_t S + A_x \partial_x S + B_y \partial_y S & = L S + f(w_x, w_y, \rho, \rho_x, \rho_y),
\end{split}
\end{equation}
with $S = (S_{11}, S_{22},S_{33},S_{13}, S_{23}, S_{12})^T$, $d=3$ and $\vec{n}$ taking values in $S^2$,
\begin{equation}
\label{matrixa}
A_x = \begin{pmatrix}
0 & 0 & 0 & \frac{1}{3} - \frac{3}{7} & 0 & 0\\
0 & 0 & 0 & \frac{1}{3} - \frac{1}{7} & 0 & 0\\
0 & 0 & 0 & \frac{1}{3} - \frac{3}{7} & 0 & 0\\
0 & \frac{1}{7} & 0 & 0 & 0 & 0\\
0 & 0 & 0 & 0 & 0 & - \frac{1}{7}\\
0 & 0 & 0 & 0 & - \frac{1}{7} & 0\end{pmatrix}, \quad
B_y = \begin{pmatrix}
0 & 0 & 0 & 0 & \frac{1}{3} - \frac{1}{7} & 0\\
0 & 0 & 0 & 0 & \frac{1}{3} - \frac{3}{7} & 0\\
0 & 0 & 0 & 0 & \frac{1}{3} - \frac{3}{7} & 0\\
0 & 0 & 0 & 0 & 0 & -\frac{1}{7}\\
\frac{1}{7} & 0 & 0 & 0 & 0 & 0\\
0 & 0 & 0 & -\frac{1}{7} & 0 & 0
\end{pmatrix},
\end{equation}
\begin{equation}
\label{matrixl}
L = \begin{pmatrix}
-6  & 0 & 0 & -\frac{6}{7} w_x & - \frac{2}{7} w_y & 0\\
0 & - 6  & 0 & - \frac{2}{7} w_x & - \frac{6}{7} w_y & 0\\
0 & 0 & -6  & \frac{8}{7} w_x & \frac{8}{7} w_y & 0 \\
w_x & \frac{2}{7} w_x & 0 & -6  & 0 & \frac{5}{7} w_y\\
\frac{2}{7} w_y & w_y & 0 & 0 & - 6  & \frac{5}{7} w_x\\
0 & 0 & 0 & -\frac{2}{7} w_y & -\frac{2}{7} w_x & -6 
\end{pmatrix}, \; \; 
f = \begin{pmatrix}
0 \\
0 \\
0 \\
\frac{3}{15} w_x  \rho +
\frac{1}{15} \rho_x\\
\frac{3}{15} w_y  \rho +
\frac{1}{15} \rho_y\\
0 \end{pmatrix}. 
\end{equation}

The quasi--dynamic approximation is obtained by setting the second
equation of (\ref{eqn:rectilinear-mc-short}) to its local equilibrium,
i.e.\   the solution of
\begin{equation}\label{eqn:equi}
L S + f(\nabla w, \rho, \nabla \rho)=0 \, .
\end{equation}
After some manipulations this leads to solving for $(S_{13}, S_{23})$ the algebraic system
\begin{equation}
\label{algsys}
\begin{aligned}
(\kappa + 10(w_x^2 + w_y^2) + 36 w_x^2 ) S_{13} + 36 w_x w_y S_{23} &= \tfrac{\kappa}{30} ( w_x \rho + \tfrac{1}{3} \rho_x)
\\
36 w_x w_y S_{13}  + (\kappa + 10(w_x^2 + w_y^2) + 36 w_y^2 ) S_{23} &= \tfrac{\kappa}{30} ( w_y \rho + \tfrac{1}{3} \rho_y)  \, , 
\end{aligned}
\end{equation}
where $\kappa = 42^2$,  and evaluating the remaining stresses via
\begin{equation}
\begin{aligned}
S_{11} &= -\tfrac{6}{42} w_x S_{13} - \tfrac{2}{42} w_y S_{23}
\\
S_{22} &= -\tfrac{2}{42} w_x S_{13} - \tfrac{6}{42} w_y S_{23}
\\
S_{23} &= \; \; \tfrac{8}{42} w_x S_{13} + \tfrac{8}{42} w_y S_{23}
\\
S_{12} &= -\tfrac{2}{42} w_y S_{13} - \tfrac{2}{42} w_x S_{23} \, .
\end{aligned}
\end{equation}

The solution of the algebraic system \eqref{algsys} is expressed in the form
\begin{equation}
\label{effstress}
\begin{pmatrix} S_{13} \\ S_{23} \end{pmatrix}
= 
\, D (\nabla w) \,  \tfrac{\kappa}{30} \rho  \nabla \Big ( w +\tfrac{1}{3} \ln \rho \Big )  \, , 
\end{equation}
where $\nabla = (\del_x, \del_y)$, and $D(\nabla w)$ is a matrix obtained by inverting \eqref{algsys} and given in
the successive forms
\begin{equation}
\label{diffmatrix}
\begin{aligned}
D (\nabla w) &:= \frac{1}{ (\kappa + 10 |\nabla w|^2)(\kappa + 46 |\nabla w|^2)}
\begin{pmatrix}
\kappa + 10 |\nabla w|^2 + 36 w_x^2  &  -36 w_x w_y 
\\
-36 w_x w_y  & \kappa + 10 |\nabla w|^2 + 36 w_y^2 
\end{pmatrix}
\\[2pt]
&= \frac{1}{\kappa + 10 |\nabla w|^2} \left [ I - \frac{36}{\kappa + 46 |\nabla w|^2} \; \nabla w \otimes \nabla w \right ]
\\[2pt]
&= \frac{1}{\kappa + 46 |\nabla w|^2} \Big [ I + \frac{36}{\kappa + 10 |\nabla w|^2 } \big ( |\nabla w |^2 I - \nabla w \otimes \nabla w \big ) \Big ]
\end{aligned}
\end{equation}

When \eqref{effstress} is introduced into \eqref{eqn:rectilinear-mc-short}$_1$ we obtain the non-isotropic diffusion equation
\begin{equation}
\label{effeq}
\del_t \rho = \nabla \cdot \left (  
D(\nabla w)   \tfrac{\kappa}{30} \rho  \nabla \big ( w +\tfrac{1}{3} \ln \rho \big ) \right ) \, ,
\end{equation}
 which is conjectured to describe the effective response of the system.

\subsection{The effective equation for the rectilinear flow}

Combining \eqref{eqn:rectilinear-mc-short}$_1$ with \eqref{effstress}, \eqref{diffmatrix} and \eqref{momentumre} (for $\gamma =0$), we obtain
 the effective equation describing the dynamics of the rectilinear sedimenting flow. 
This takes the form
\begin{equation}
\label{effsys}
\begin{split}
\del_t \rho &= \nabla \cdot \left (  
\frac{1}{\kappa + 10 |\nabla w|^2} \Big [ I - \frac{36}{\kappa + 46 |\nabla w|^2} \; \nabla w \otimes \nabla w \Big ]
 \tfrac{\kappa}{30} \rho  \nabla \big ( w +\tfrac{1}{3} \ln \rho \big ) \right )
\\
 Re \; \partial_t w &= \triangle_{(x,y)}  w + \delta \left( \bar \rho - \rho \right)
\end{split}
\end{equation}
The constant $\bar \rho$ is selected to be either the initial mass over a period (if the problem is periodic) or
the total mass $\bar \rho = \int \rho (x,t) dx$ (for the Cauchy problem). In either case the mass is conserved and the selection of
$\bar \rho$ amounts to a change of Galilean frame for observing the flow.

Equation \eqref{effsys}$_1$ describes anisotropic diffusion. The diffusion matrix $D(\nabla w)$ in \eqref{diffmatrix} 
is symmetric and positive definite. The system \eqref{effsys} is invariant under rotations in the $x,y$-plane.
Indeed, for $x' = Q x$ with $Q \in SO (2)$,  we have $\nabla_{x'} = Q^T \nabla_x$ and $|\nabla_x w| = |Q \nabla_{x'} w| = |\nabla_{x'} w|$.
Then \eqref{diffmatrix} implies
$$
D ( \nabla_x w) = D ( Q \nabla_{x'} w) = Q D( \nabla_{x'} w) Q^T
$$
and \eqref{effsys}$_1$ is invariant,
$$
\begin{aligned}
\del_t \rho &= Q \nabla_{x'} \cdot Q D( \nabla_{x'} w) Q^T \tfrac{\kappa}{30} \rho Q \nabla_{x'} \big ( w +\tfrac{1}{3} \ln \rho \big ) 
\\
&= \nabla_{x'} \cdot D( \nabla_{x'} w)  \tfrac{\kappa}{30} \rho \nabla_{x'} \big ( w +\tfrac{1}{3} \ln \rho \big ) 
\end{aligned}
$$
The same is true for \eqref{effsys}$_2$ due to the invariance of the Laplacian under rotations.

Finally, we show that \eqref{effsys} is endowed with an entropy-dissipation structure. For concreteness, we assume periodic boundary conditions
over a domain of periodicity $\T$. The total density is then conserved, and we select $\bar \rho$ in \eqref{effsys}$_2$ as
$$
\bar \rho = \int_{\T} \rho (x,t) dx = \int_{\T} \rho_0 (x) dx \, .
$$
Note that this choice can be always assured by changing Galilean frame of reference.
Moreover,
$$
 \int_{\T} w_t dx = 0 \, , \quad \int_{\T} w(x,t) dx = \int_{T} w_0 (x) dx
$$
Next, we multiply \eqref{effsys}$_1$ by $\tfrac{1}{3} (1 + \ln \rho) + w$ and \eqref{effsys}$_2$ by $w_t$ ; after some integrations by part
we respectively obtain
$$
\begin{aligned}
\del_t \big ( \tfrac{1}{3}  \rho  \ln \rho +  \rho w \big )  - \rho w_t &= 
\nabla \cdot \big ( \tfrac{1}{3} (1 + \ln \rho) + w \big ) \tfrac{\kappa}{30} \rho  D (\nabla w)  \nabla \big ( w +\tfrac{1}{3} \ln \rho \big )
\\
&\quad - \nabla \big ( w +\tfrac{1}{3} \ln \rho \big ) \cdot \tfrac{\kappa}{30} \rho  D (\nabla w)  \; \nabla \big ( w +\tfrac{1}{3} \ln \rho \big )
\\
\rho w_t + Re \, w_t^2 + \del_t \tfrac{1}{2} |\nabla w|^2 &= \nabla \cdot w_t \nabla w + \delta \bar \rho w_t
\end{aligned}
$$
and the entropy (free-energy)  dissipation identity
\begin{equation}
\label{entropyiden}
\begin{aligned}
\frac{d}{dt} \int_{\T} &\big (  \tfrac{1}{3}  \rho  \ln \rho +  \rho w + \tfrac{1}{2} |\nabla w|^2 \big ) dx 
\\
&+ \int_{\T} \left [ Re \, w_t^2 + \nabla \big ( w +\tfrac{1}{3} \ln \rho \big ) \cdot \tfrac{\kappa}{30} \rho  D (\nabla w)  \; \nabla \big ( w +\tfrac{1}{3} \ln \rho \big ) \right ]\, dx
=0 \, .
\end{aligned}
\end{equation}

\subsection{Remarks concerning the validity of the quasidynamic approximation}
\label{sec:remarks}

The validity of the conjecture that the long time dynamics of \eqref{eqn:rectilinear-mc-short} together with Stokes is described by the system
\eqref{effsys} is at present an open problem. 
In the sequel, we take up the case of rectilinear flows,  and  give some partial arguments highlighting the idea 
and the analytical difficulties for justidying the quasidynamic approximation.
Then, in the following section, we will restrict to the case of shear flows, and give numerical evidence that shows that
the quasidynamic approximation is a good representation of the full dynamics of the kinetic model for long times, 
and will present  a heuristic argument towards justifying the approximation for shear flows.

The idea behind the approximation is that the transient dynamics of $S$ is replaced  by its equilibrium  response.
This amounts to considering the simplified linear non-homogeneous problem
$$
\partial_t S + A_x \partial_x S + B_y \partial_y S  = L S + f(x)
$$
and postulating that its solution is well approximated for long-times by its local equilibrium 
$$
 L S + f(x) = 0 \, .
 $$
This would be the case provided the solutions of the homogeneous system 
\begin{equation}
\label{homosys}
\partial_t S + A_x \partial_x S + B_y  \partial_y S - L S = 0
\end{equation}
decay to zero for large times,  {\it i.e.} $S(x,t) \to 0$ as $t \to \infty$.

The solution of \eqref{homosys} can be visualized as the Trotter product of the semigroup generated by the system
of ordinary differential equations
\begin{equation}
\label{homoode}
\partial_t S  - L S = 0
\end{equation}
and  the semigroup generated by the linear hyperbolic system
\begin{equation}
\label{homohyp}
\partial_t S + A_x \partial_x S + B_y  \partial_y S = 0 \, .
\end{equation}

The eigenvalues of \eqref{homoode} are computed by finding the roots of the characteristic polynomial
 $\det (-\lambda I + L) = 0$ where
\begin{equation}
L  - \lambda I = \begin{pmatrix}
-(\lambda + 6)  & 0 & 0 & -\frac{6}{7} w_x & - \frac{2}{7} w_y & 0\\
0 & -(\lambda + 6)  & 0 & - \frac{2}{7} w_x & - \frac{6}{7} w_y & 0\\
0 & 0 & -(\lambda + 6)  & \frac{8}{7} w_x & \frac{8}{7} w_y & 0 \\
w_x & \frac{2}{7} w_x & 0 & -(\lambda + 6)  & 0 & \frac{5}{7} w_y\\
\frac{2}{7} w_y & w_y & 0 & 0 & -(\lambda + 6)  & \frac{5}{7} w_x\\
0 & 0 & 0 & -\frac{2}{7} w_y & -\frac{2}{7} w_x & -(\lambda + 6)
\end{pmatrix} \, .
\end{equation}
They can be computed via the following  formula: If $A$, $B$, $C$, $D$ are square matrices of the same size and  the matrix  $A$  is invertible, then 
$$
K :=
\begin{pmatrix}
A & B
\\
C & D
\end{pmatrix}
= 
\begin{pmatrix}
A & 0
\\
C & I
\end{pmatrix}
\;
\begin{pmatrix}
I &  A^{-1} B
\\
0  & D - C A^{-1} B
\end{pmatrix}
$$
and
\begin{equation}
\label{form}
\det K = \det A \det ( D - C A^{-1} B )
\end{equation}
Using this formula, a lengthy but straightforward calculation yields that the six eigenvalues of \eqref{homoode} are
$$
\begin{cases}
\lambda_{1,2} = -6  \quad \mbox{ with multiplicity $2$}  & \\
\lambda_{3,4} = - 6 \pm \imath \tfrac{ \sqrt{46} }{7} \sqrt{ w_x^2 + w_y^2} & \\
\lambda_{5,6} = - 6 \pm \imath \tfrac{ \sqrt{10} }{7} \sqrt{ w_x^2 + w_y^2} & \\
\end{cases}
$$
All the eigenvalues have strictly negative real parts and the solution of \eqref{homoode} converges to zero as $t \to \infty$.

The same formula can be used to calculate the eigenvalues of \eqref{homohyp} and establish that this system is hyperbolic.
Indeed, for a vector $\nu = (\nu_x , \nu_y) \in \R^2$, $\nu \ne 0$, we use formula \eqref{form} and compute that the eigenvalues
of $(-\lambda I + \nu_x A_x + \nu_y B_y)$ are
$$
\begin{cases}
\lambda_{1,2} = 0 \quad \mbox{ of multiplicity $2$}  & \\
\lambda_{3,4} = \pm \tfrac{2}{7 \sqrt{3}}  \sqrt{ \nu_x^2 + \nu_y^2} & \\
\lambda_{5,6} = \pm  \tfrac{1}{7} \sqrt{ \nu_x^2 + \nu_y^2} & \\
\end{cases}
$$
One easily checks that the eigenspace corresponding to the zero-eigenvalue is two-dimensional. The problem \eqref{homohyp}
is hence hyperbolic. It is tempting to conclude that $S$ converges to zero as time tends to infinity. Such results are
available when the matrices $A_x$ and $B_y$ are symmetric (see \cite{SK85}), but we are not aware of a corresponding theory covering
the case that the matrices $A_x$ and $B_y$ are not symmetric.

\bigskip
\bigskip

%
%
%
%

\subsection{Justification for shear  flows}
\label{sec:justify}

To justify the quasidynamic approximation for shear flow, we show that
for moderate values of $|w_x|$ solutions of
the homogeneous system
\begin{equation}\label{eqn:qd-6}
\partial_t S + A \partial_x S - L S = 0,
\end{equation}
with $A$ and $L$ as described in (\ref{eqn:qd-4}) decay to zero for
large times.
We look for solutions of the form
\begin{equation}\label{eqn:qd-5}
S(x,t) = r \exp(\lambda t + i k x),
\end{equation}
with $\lambda, k \in \mathbb{R}-\{0\}$.
Computing such solutions amounts to finding eigenvalues $\lambda$ and
eigenvectors $r$ for the problem
$$
(\lambda I + i k A -L) r = 0.
$$
The solutions $S(x,t)$ of (\ref{eqn:qd-6}) decay to zero, 
if $\mbox{Re}(\lambda) < 0$ for
all possible solutions $\lambda$, $r$ of (\ref{eqn:qd-5}).   
We find that $\lambda$ has the form
\begin{equation}\label{eqn:shear-1}
\begin{split}
\lambda_{1,2} & = - 6 \pm \frac{1}{21} \kappa, \quad \mbox{with } \kappa:=\sqrt{36 i k w_x - 12 k^2
  -414w_x^2}\\
\lambda_{3,4} & = -6.
\end{split}
\end{equation}
Corresponding eigenvectors are given by
\begin{equation}
r_1 = \frac{1}{\kappa} \left( \begin{array}{c}
2(i k - 9w_x)\\ -2(2 i k + 3 w_x)\\ 2(i k + 12 w_x) \\ \kappa
\end{array}\right), \ r_2 = \frac{1}{\kappa} \left( \begin{array}{c} -2(i
  k - 9 w_x) \\ 2(2 i k + 3 w_x) \\ - 2(i k + 12 w_x) \\
  \kappa\end{array}\right),
r_3 = \left(\begin{array}{c}0\\0\\1\\0\end{array}\right), r_4 =
\left(\begin{array}{c}
\frac{i k - 2 w_x}{7 w_x}\\1 \\0 \\0\end{array}\right)
\end{equation}

Now we use the relation
$$
Re(\sqrt{a + ib}) = \sqrt{\frac{a+\sqrt{a^2 + b^2}}{2}}
$$
in order to compute the real part of $\kappa$.
\begin{equation*}
\begin{split}
\mbox{Re} (\kappa) & = \mbox{Re}\left( \sqrt{-12 k^2 -414w_x^2 + 36 i k w_x} \right) \\
& = \sqrt{\frac{-12 k^2 - 414 w_x^2 + \sqrt{12^2 k^4 + 11232 k^2 w_x^2
    + 414^2 w_x^4}}{2}}\\
& < \sqrt{\frac{-12 k^2 - 414 w_x^2 + \sqrt{(12 k^2 + 468
      w_x^2)^2}}{2}}\\
& = \sqrt{27} |w_x|
\end{split}
\end{equation*}
Thus, $\mbox{Re}(\lambda) < -6+\frac{\sqrt{27}}{21}|w_x| < 0$ whenever
 $|w_x| < 14 \sqrt{3} $.

Finally, we  present numerical results for the shear flow
problem, which confirm that the quasidynamic approximation leads to an
accurate representation of the solution structure. 
We use the
parameter values $D_r = \delta = Re =1$ and $\gamma = 0$.  

For shear flow, the full
model has the form
\begin{equation}\label{eqn:numerics1}
\begin{split}
& \partial_t f  
 +\nabla_{\vec{n}} \cdot \left( P_{\vec{n}^\bot} (0,0,n_1 w_x)^T
  f\right) 
- \partial_{x} \left(n_1 n_2 f \right)   = \Delta_{\vec{n}}f\\
& \partial_{t} w  = \frac{\partial^2 w}{\partial x^2} + 
  \left( m -\int_{S^2} f d\vec{n} \right).
\end{split}
\end{equation}
The effective equation for the evolution of $\rho$ has the form
\begin{equation}\label{eqn:numerics2}
\begin{split}
\partial_t \rho & = \partial_x \left( \frac{\kappa}{90}
  \frac{1}{\kappa + 46 w_x^2}
  (3 \rho w_x + \rho_x) \right) \quad (\kappa = 42^2)\\
v_t & = v_{xx} +  (m-\rho).
\end{split}
\end{equation}
Furthermore, we also compare the simulations of the full model (\ref{eqn:numerics1})
with results for the linear model 
\begin{equation}\label{eqn:numerics3}
\begin{split}
\partial_t \rho - \partial_x S_{13}& = 0  \\
\partial_t S_{11} - \frac{2}{21} \partial_x S_{13} & = -6S_{11}\\
\partial_t S_{22} + \frac{4}{7} \partial_x S_{13}  & = -6 S_{22}\\
\partial_t S_{33} - \frac{2}{21} \partial_x S_{13} & = -6 S_{33}\\
\partial_t S_{13} + \frac{1}{7} \partial_x S_{22} -
\frac{1}{15} \partial_x \rho & = - 6S_{13} + \frac{1}{5} \partial_x
w\\
\partial_t w & = \partial_{xx} w + (m-\rho),
\end{split}
\end{equation}
which is obtained by linearizing the nonlinear moment closure model
(\ref{eqn:mc-shear}) around the state $\rho = 1$ and $w=0$.

\begin{figure}[h!]
\includegraphics[width=0.45\textwidth]{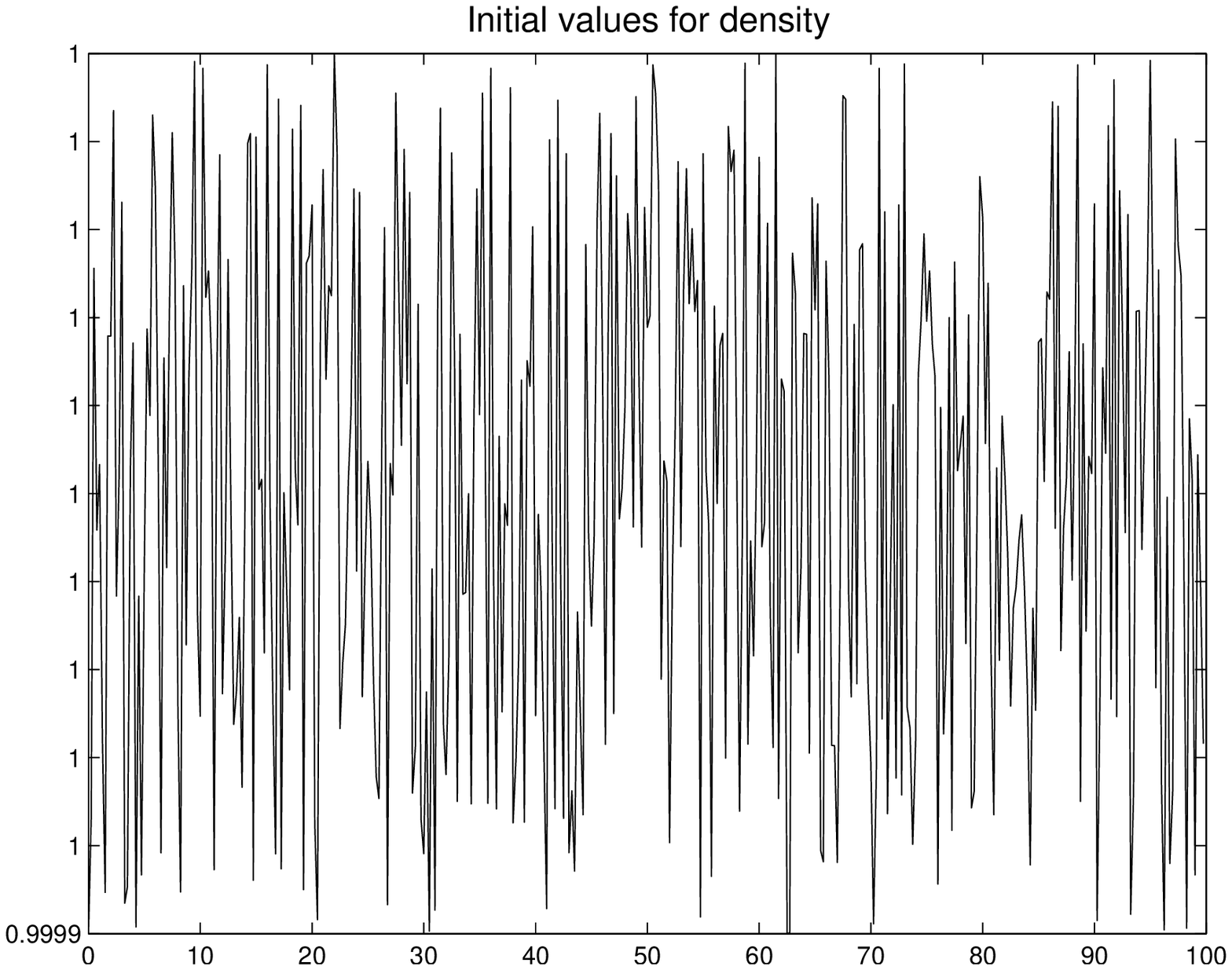}\hfill
\includegraphics[width=0.45\textwidth]{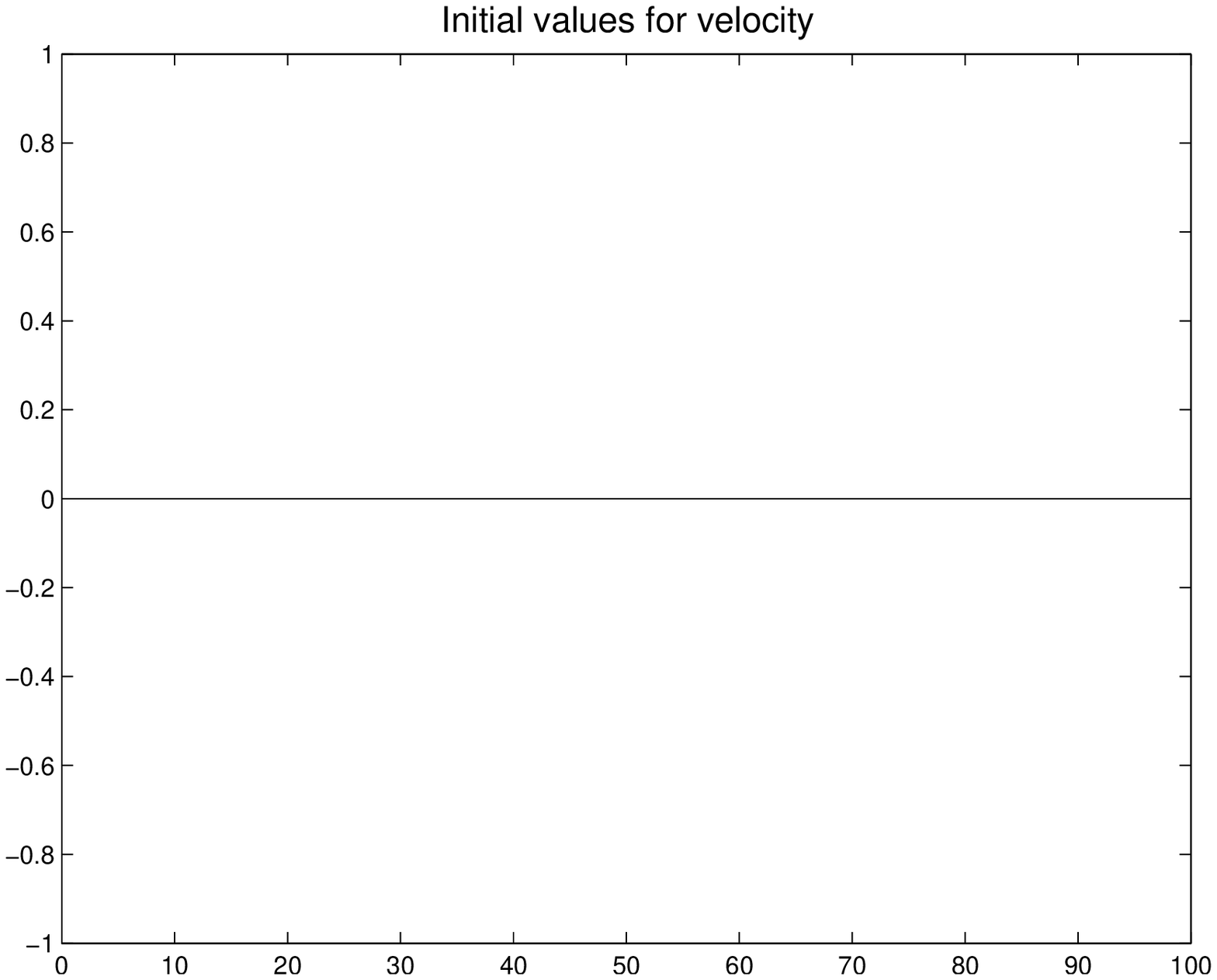}
\caption{\label{fig:numerics-0}
Initial values for $\rho$ and $w$.}
\end{figure}

We compute periodic solutions on the intervall $0 \le x \le 100$. The
computational domain in the $x$-direction is discretized with 400 grid
cells. The initial values are set to be
\begin{equation*}
\rho(x_k,0)  = 1 + 10^{-4} \left( \epsilon(x_k) - \frac{1}{2} \right), 
\end{equation*}
where $\epsilon(x_k)$ is a random number between 0 and 1 and $k = 1,
\ldots, 400$.
The initial values for the velocity are set to zero, i.e.\ $w(x,0) =
0$. Figure \ref{fig:numerics-0} shows plots of the initial values.

In our simulations of the full model, the initial values for $f$  are
set to be
$$
f(x_k,0,\vec{n}) = \frac{\rho(x_k)}{4 \pi}, \quad k=1, \ldots, 400,
\vec{n}\in S^2
$$
In the simulations of the linear model (\ref{eqn:numerics3}), the
initial values for $S_{11}, S_{22}, S_{33}, S_{13}$ are computed via
the relation
$$
S_{ij}(x_k,0) = \int_{S^2} \left( n_i n_j - \frac{1}{3} \delta_{ij}
\right) f(x_k, 0, \vec{n}) d\vec{n}, \quad k=1, \ldots, 400.
$$

\begin{figure}[htbp!]
\includegraphics[width=0.45\textwidth]{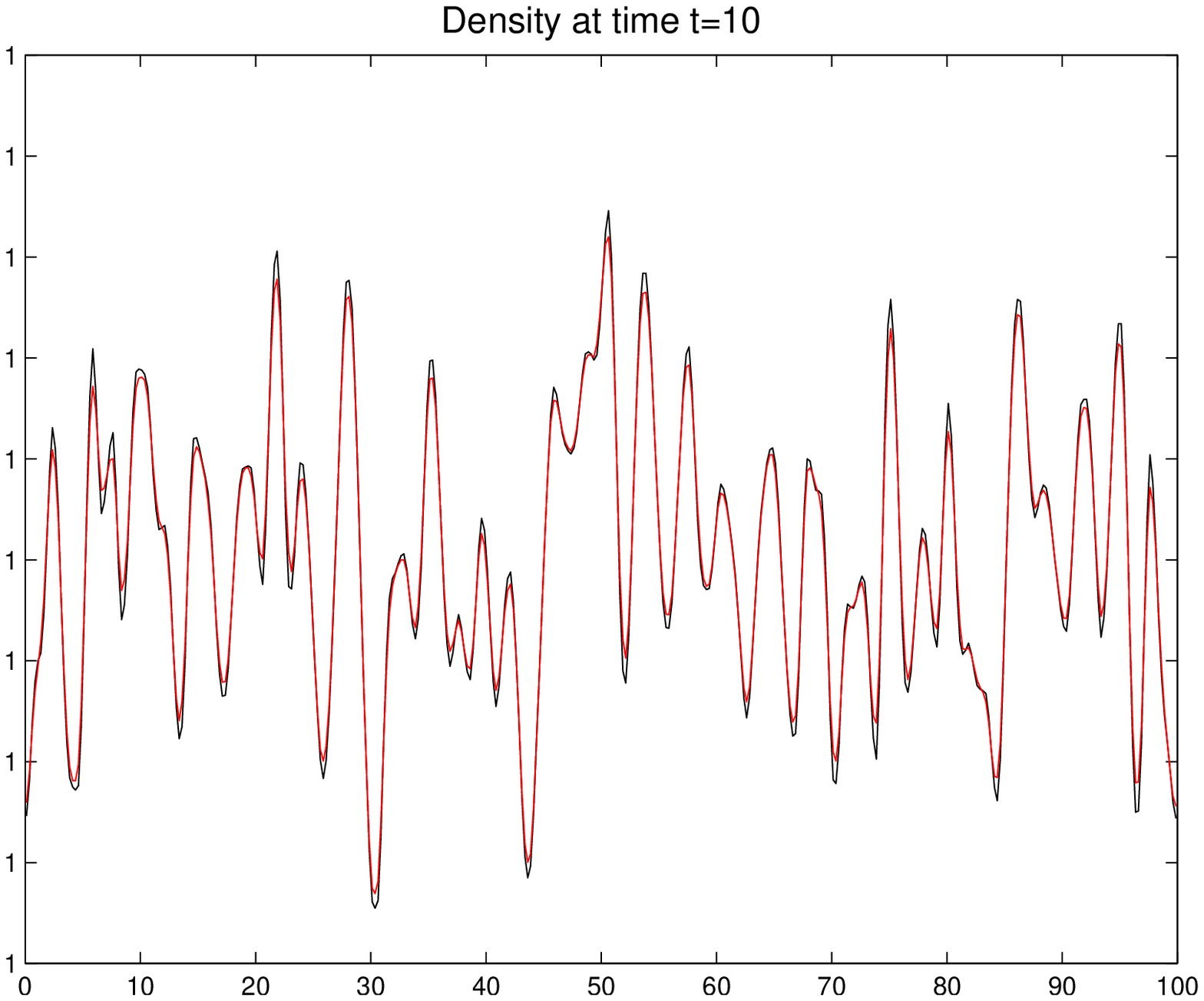}\hfill
\includegraphics[width=0.45\textwidth]{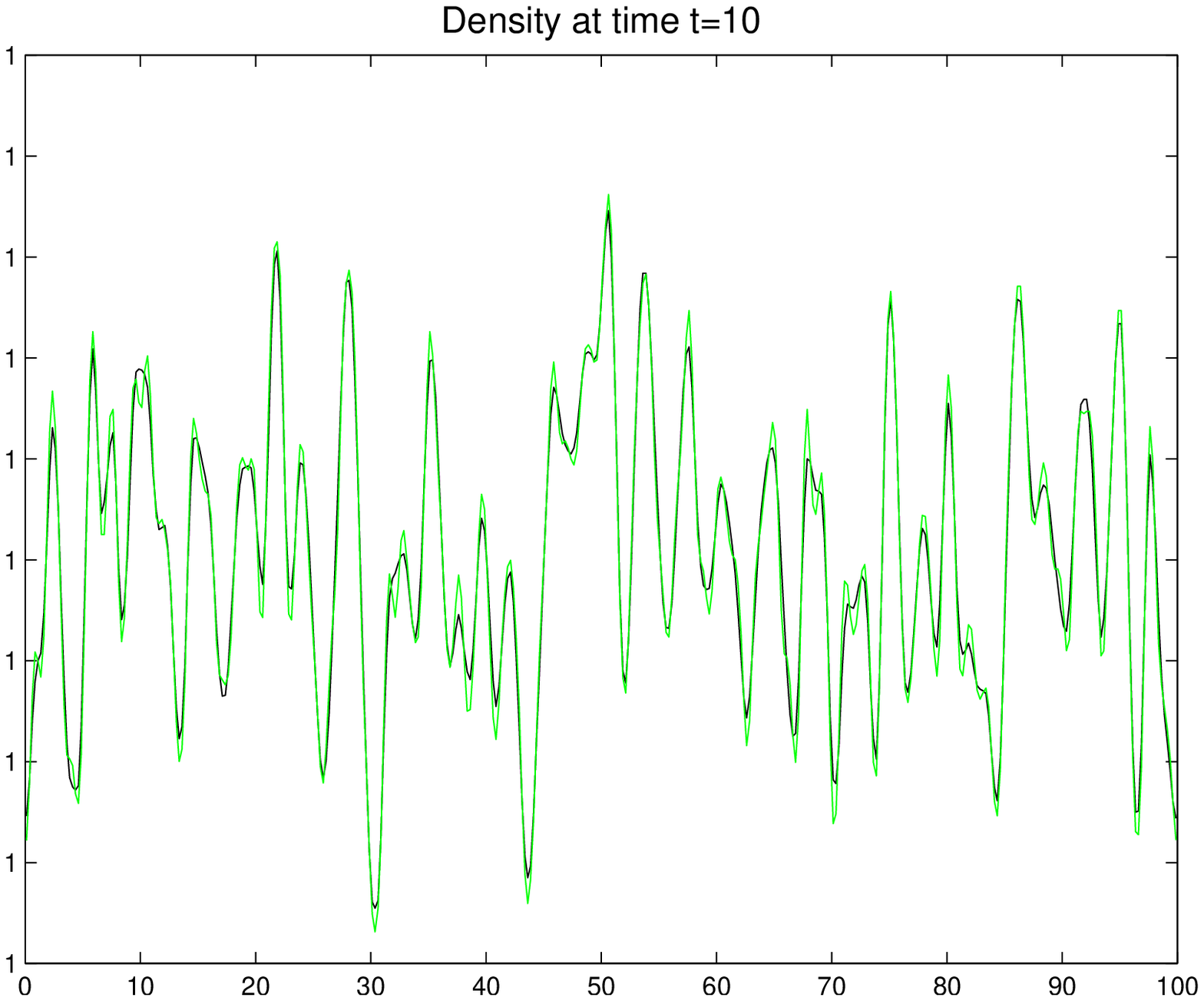}

\includegraphics[width=0.45\textwidth]{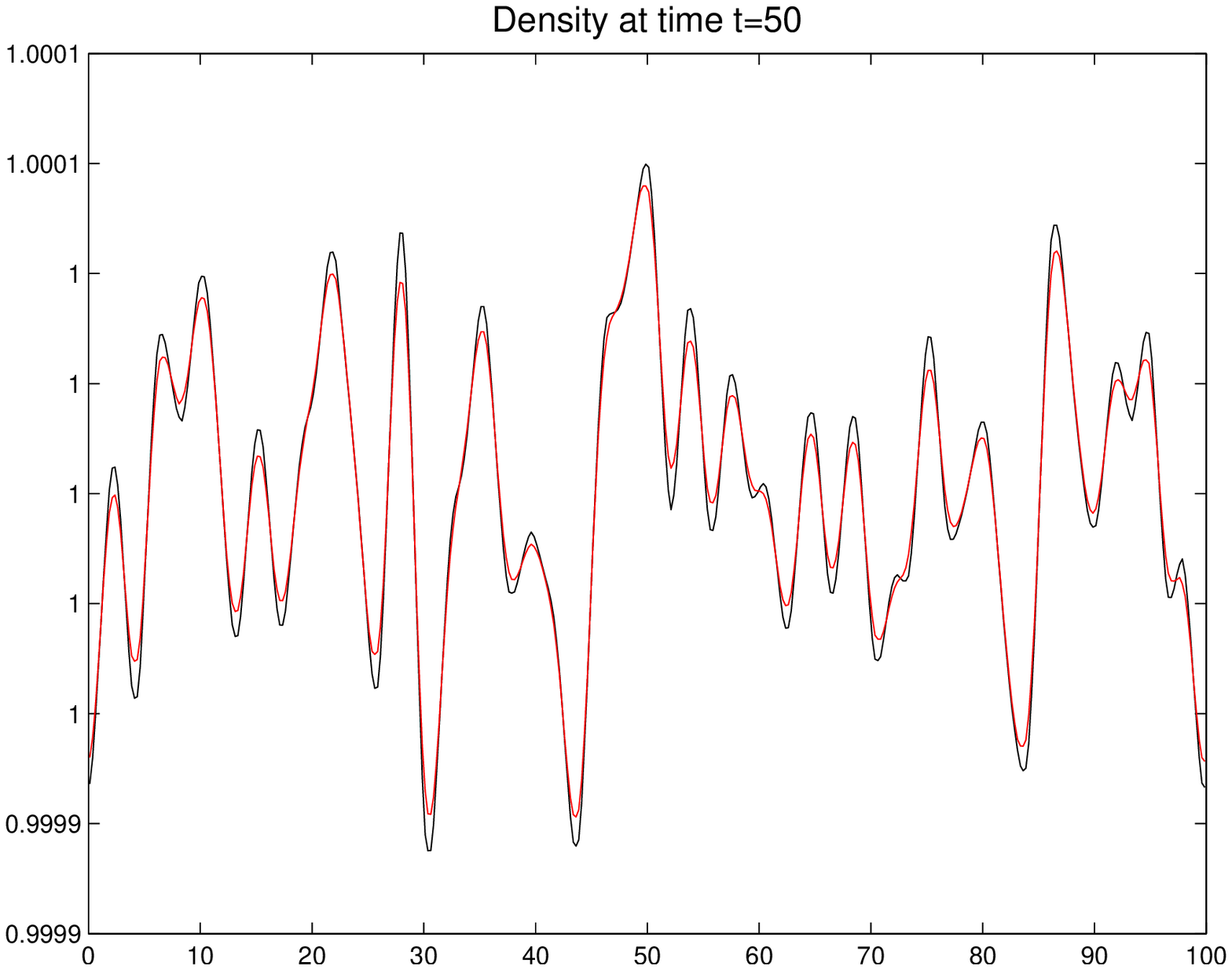}\hfill
\includegraphics[width=0.45\textwidth]{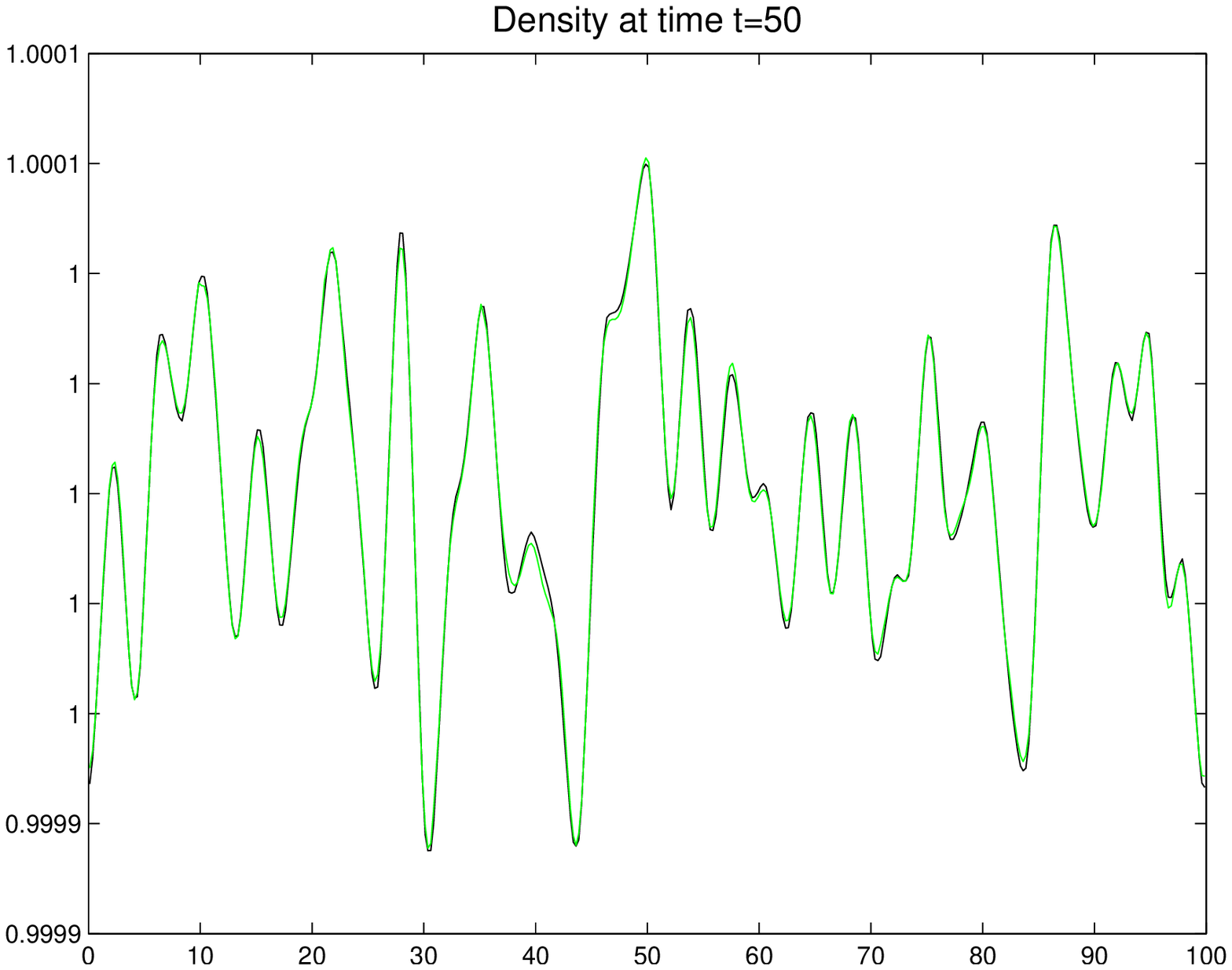}

\includegraphics[width=0.45\textwidth]{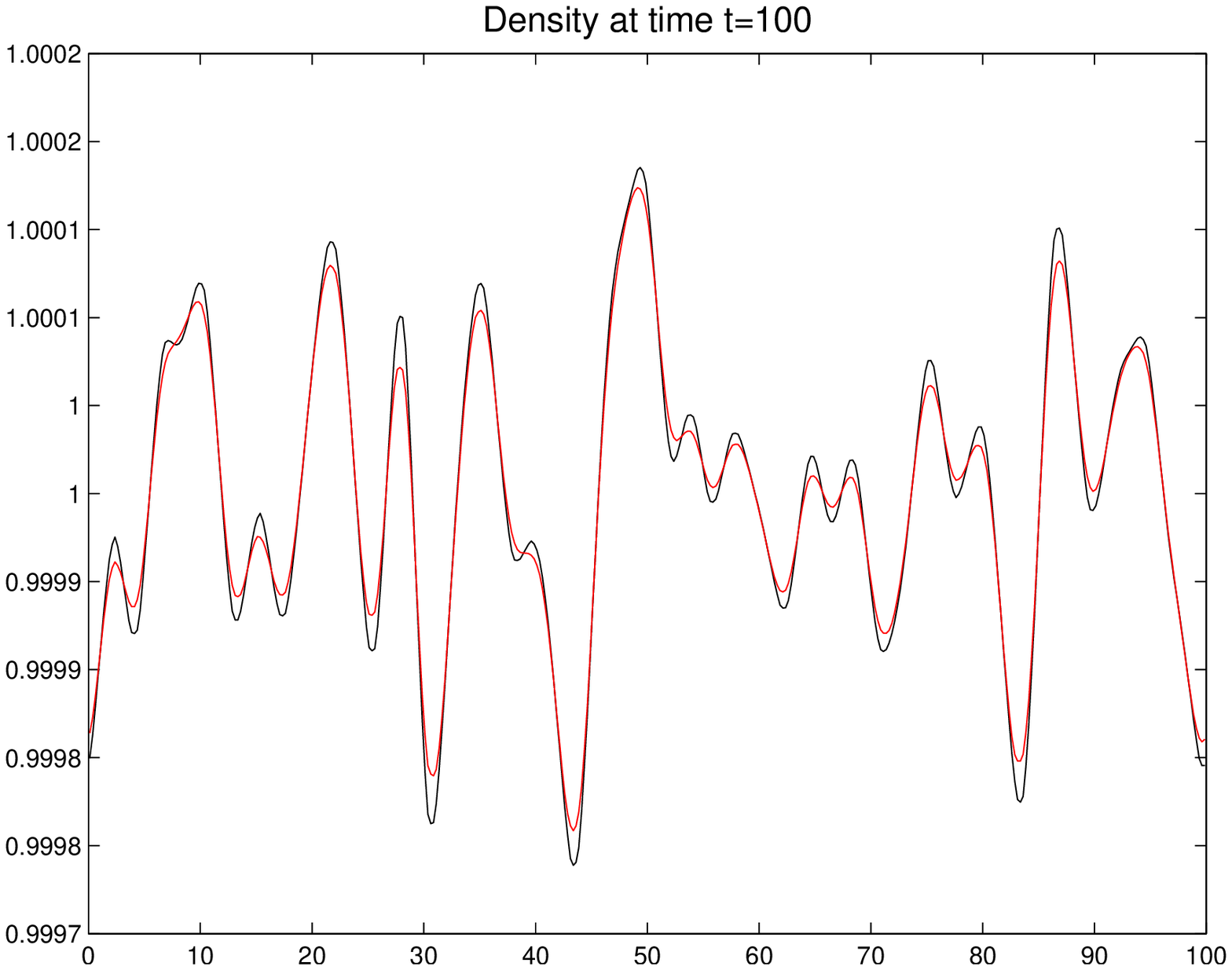}\hfill
\includegraphics[width=0.45\textwidth]{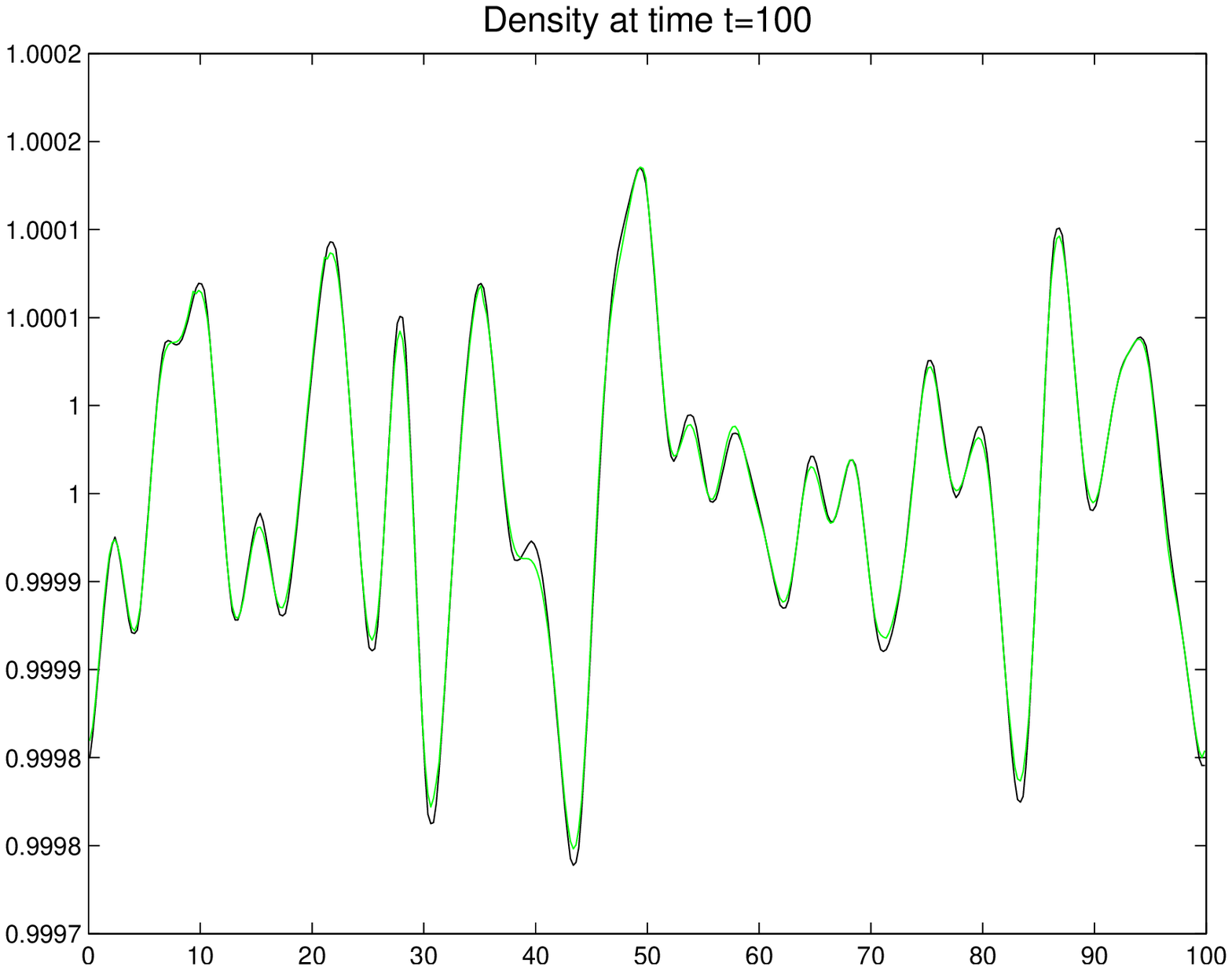}
\caption{\label{fig:numerics-1}
Left plots: Comparison of the linear model (red line) with
  the full model (black line). Right plots: Comparison of the
  quasidynamic approximation (green line) with the full model (black line).}
\end{figure}

\begin{figure}[htbp!]
\includegraphics[width=0.45\textwidth]{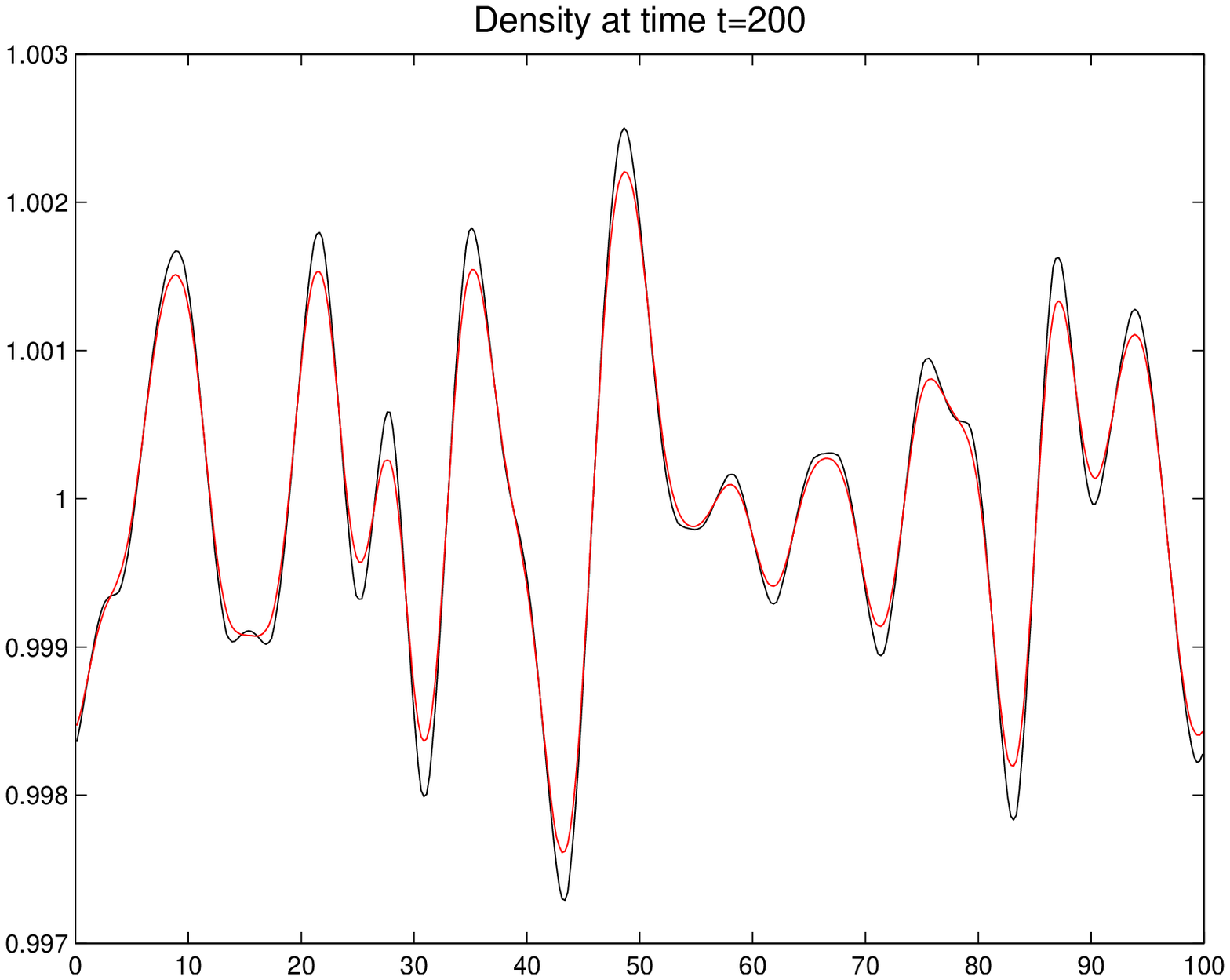}\hfill
\includegraphics[width=0.45\textwidth]{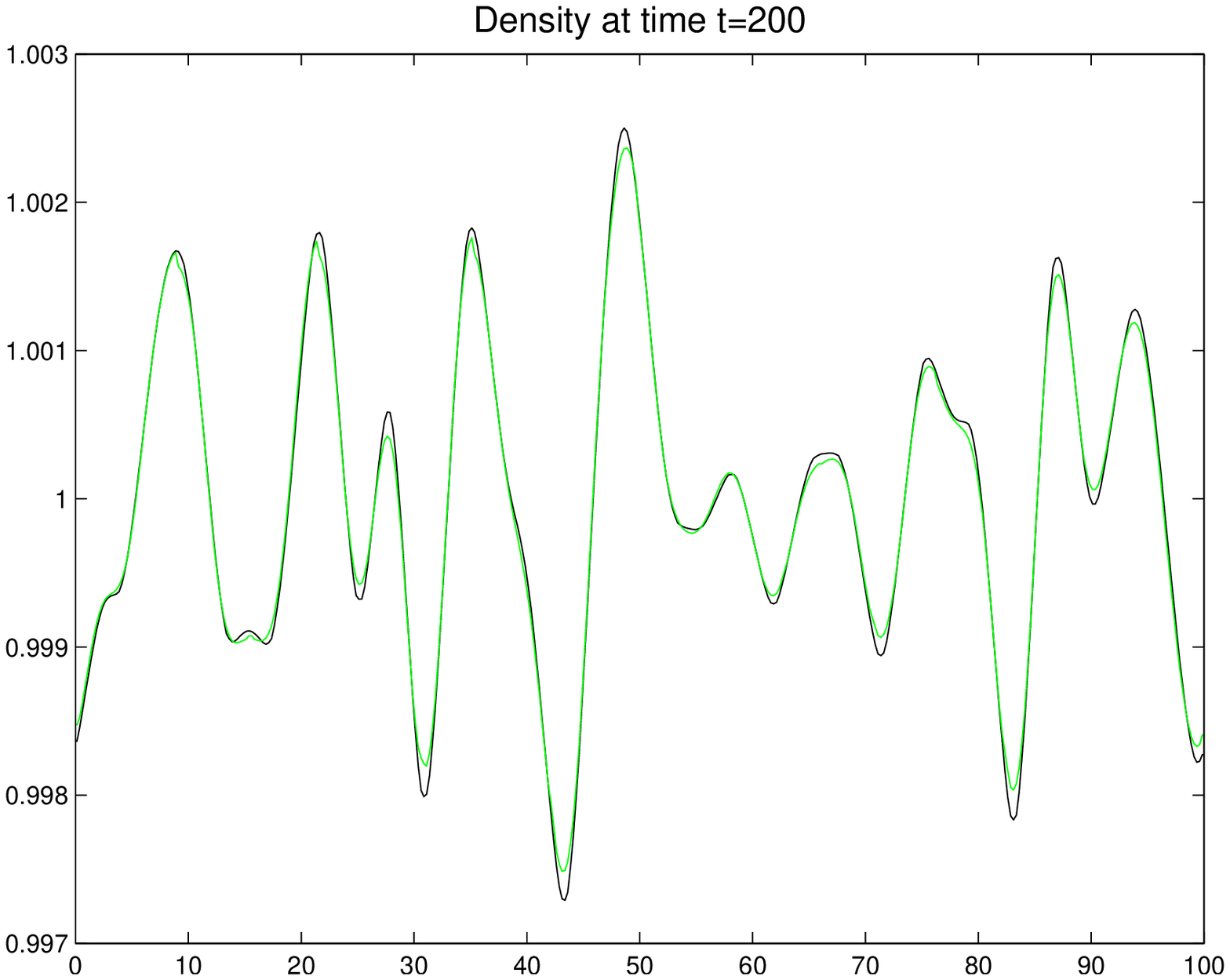}

\includegraphics[width=0.45\textwidth]{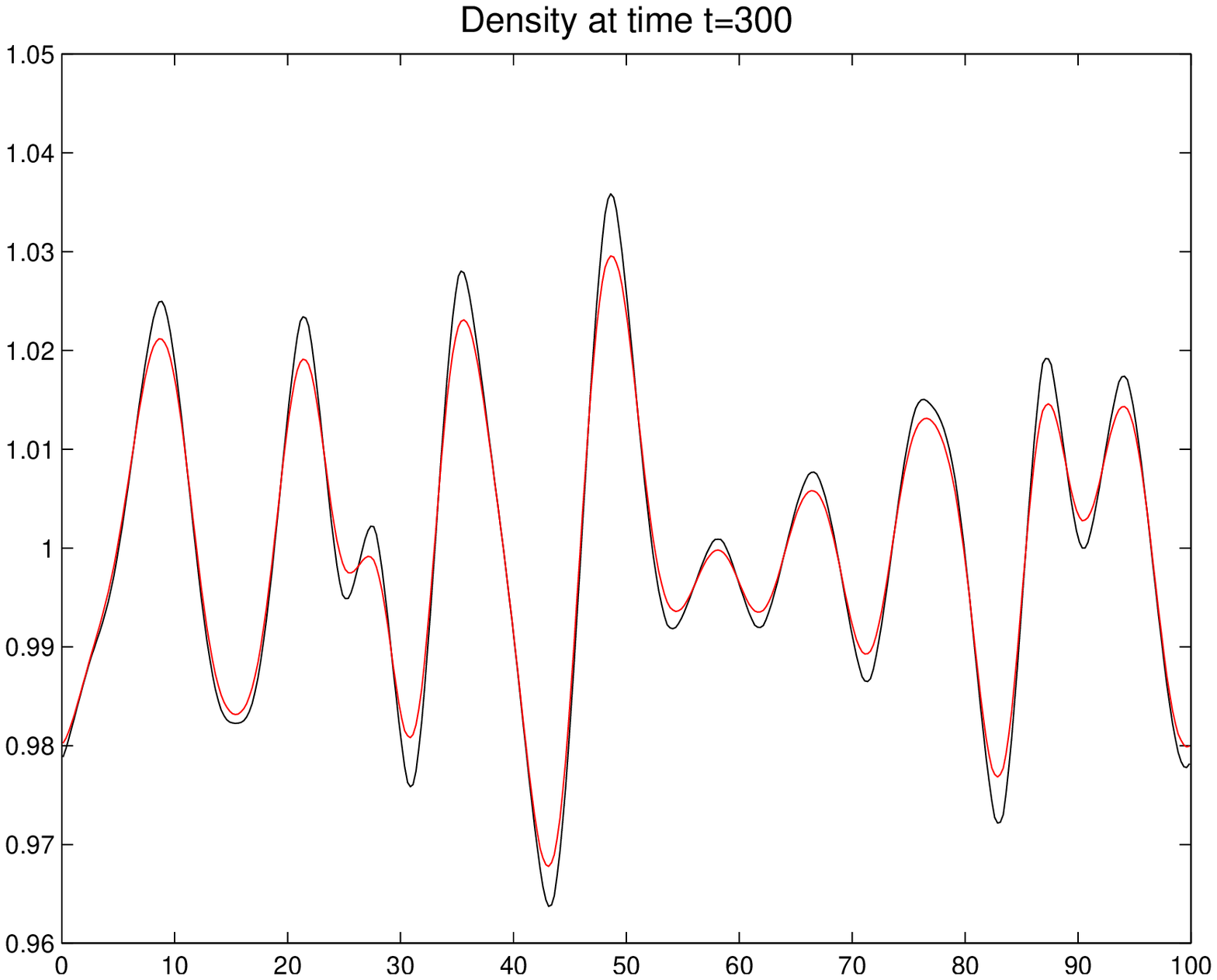}\hfill
\includegraphics[width=0.45\textwidth]{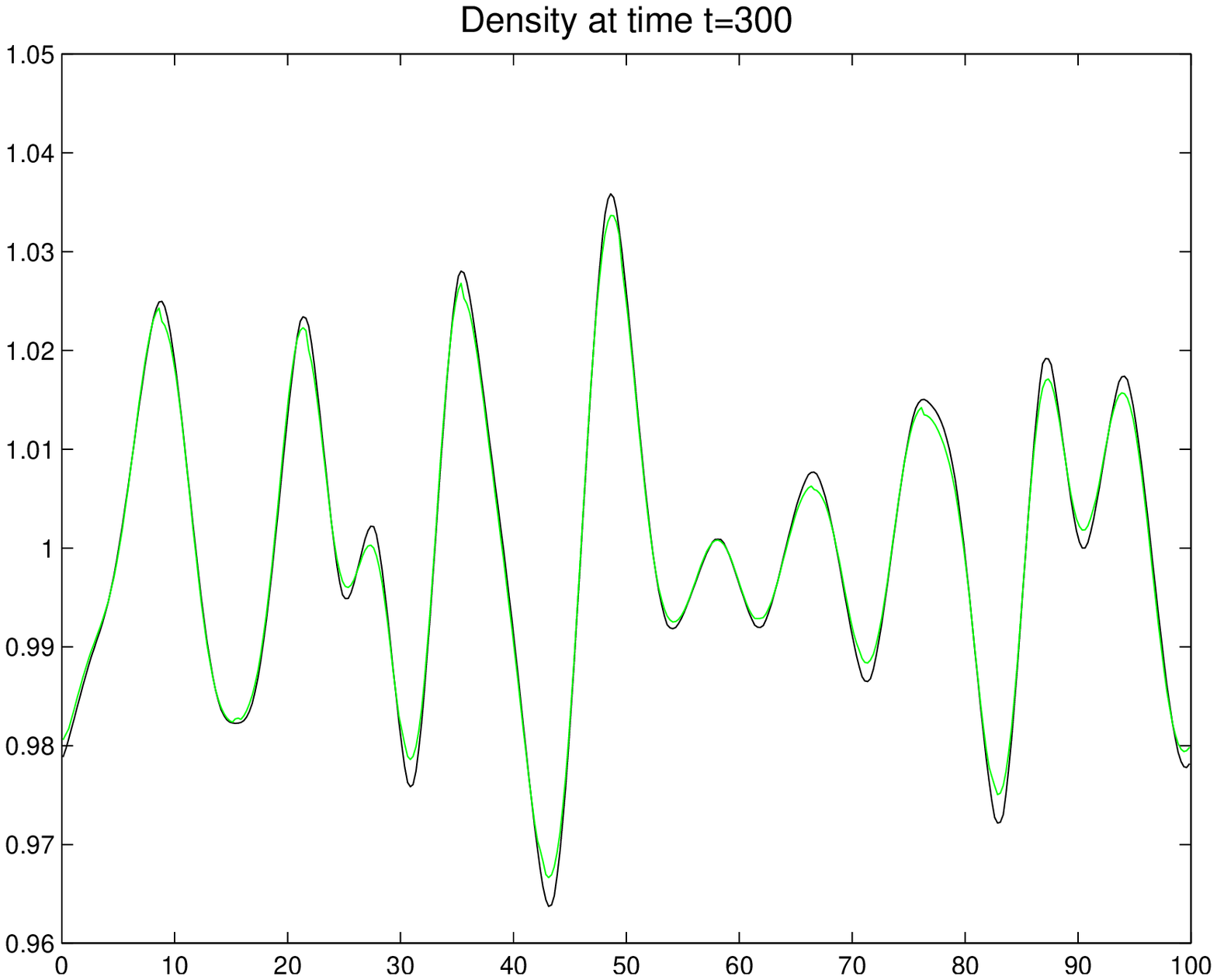}

\includegraphics[width=0.45\textwidth]{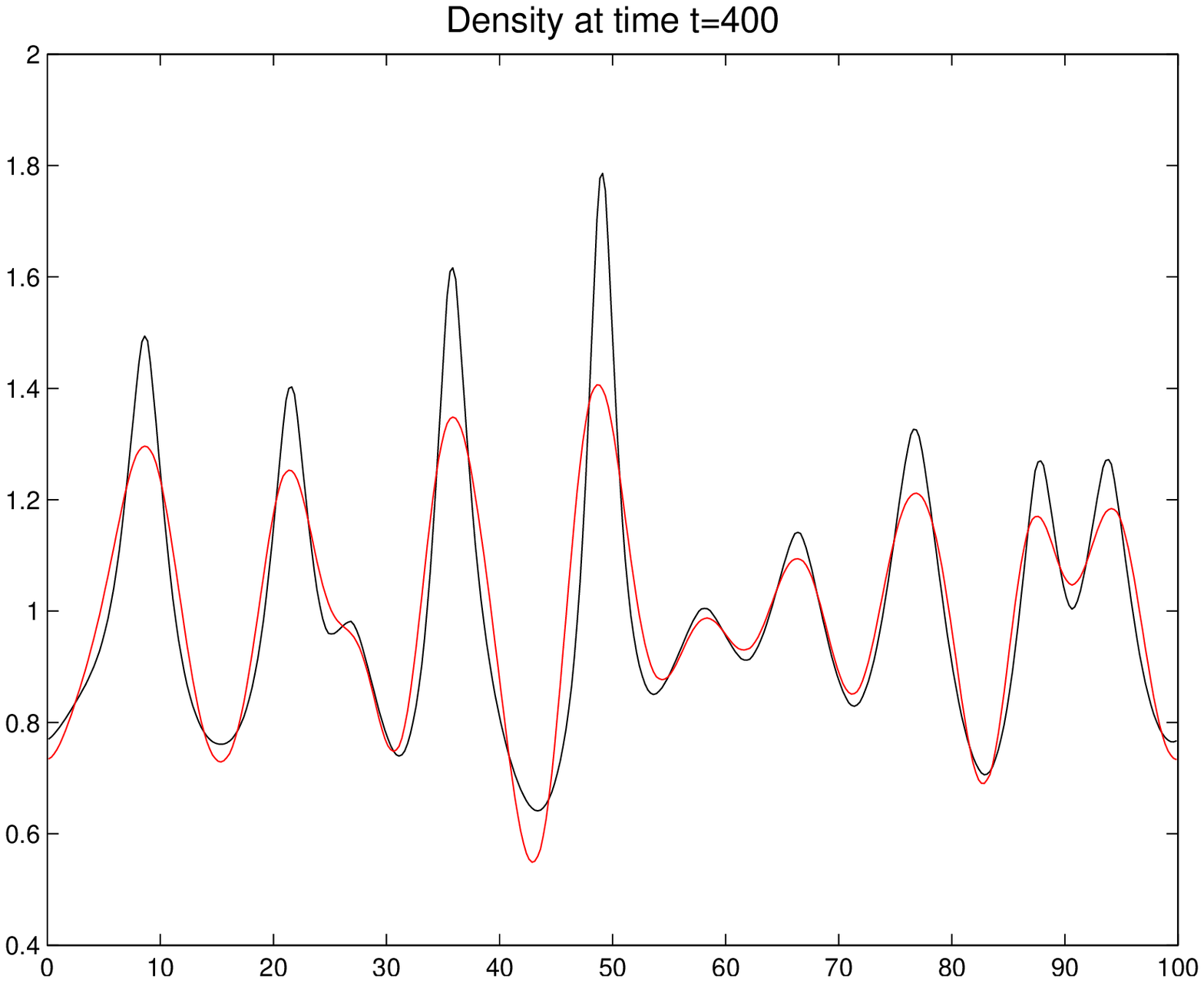}\hfill
\includegraphics[width=0.45\textwidth]{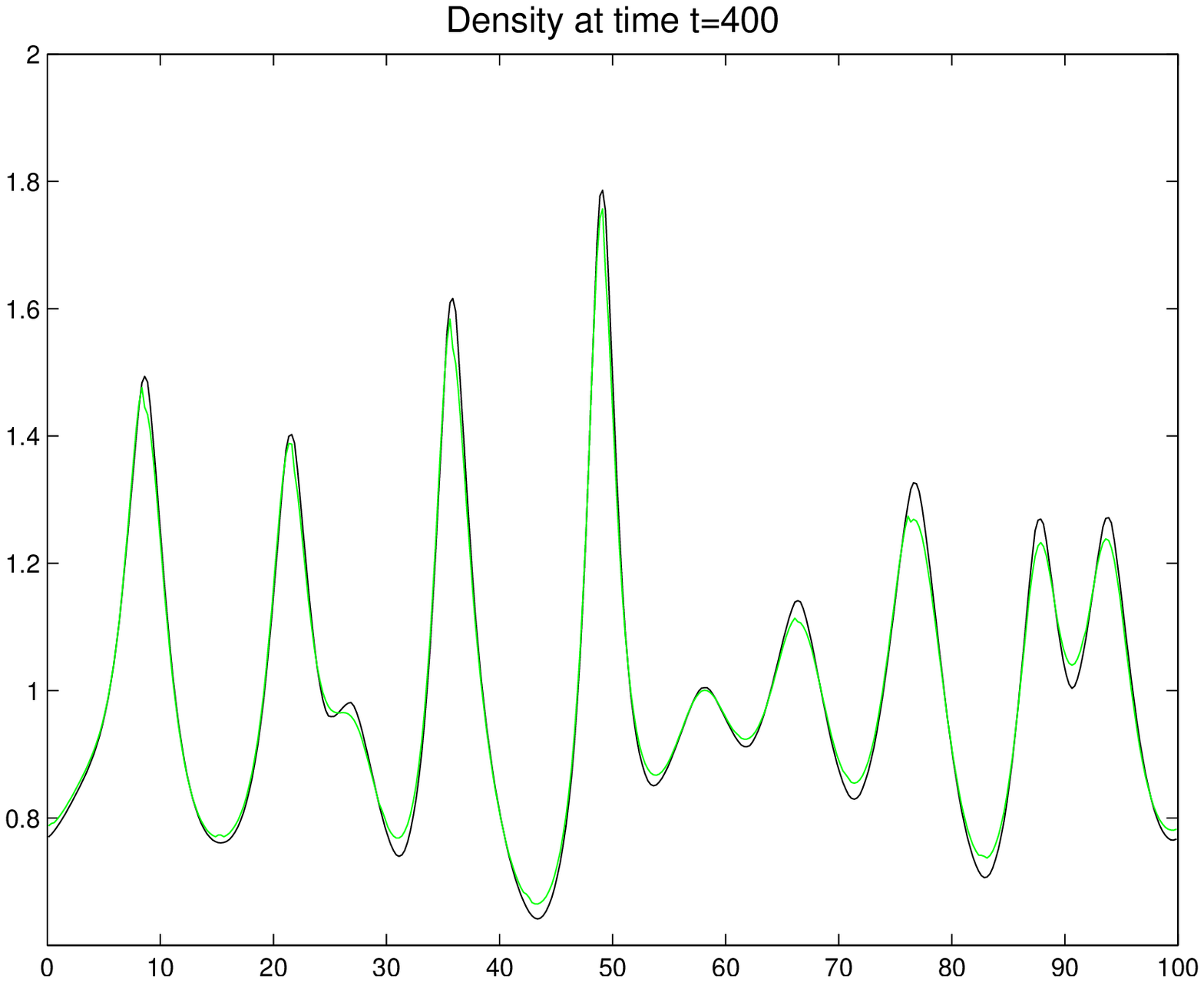}
\caption{\label{fig:numerics-2}
Left plots: Comparison of the linear model (red line) with
  the full model (black line). Right plots: Comparison of the
  quasidynamic approximation (green line) with the full model (black line).}
\end{figure}

\begin{figure}[htb!]
\includegraphics[width=0.45\textwidth]{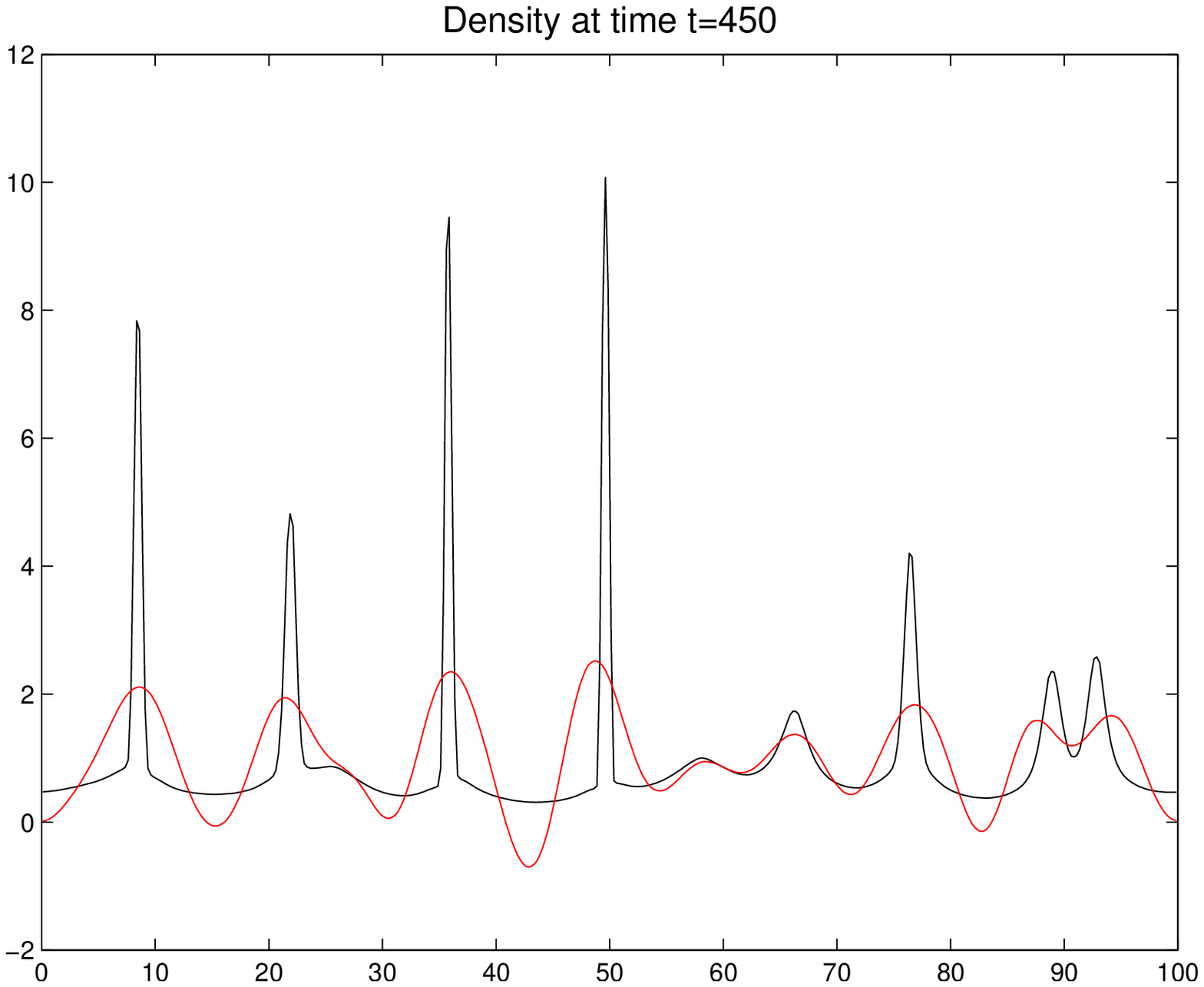}\hfill
\includegraphics[width=0.45\textwidth]{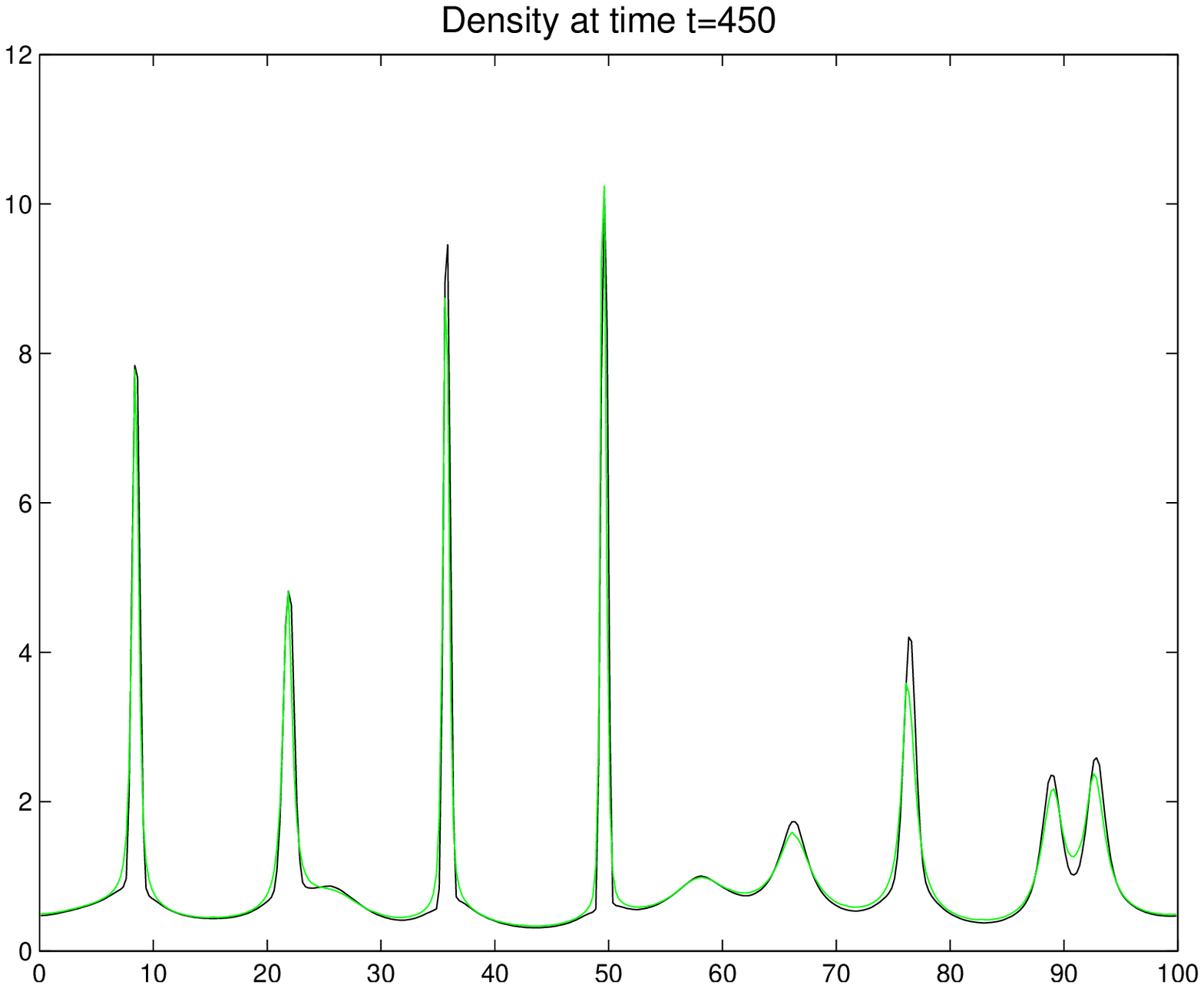}

\includegraphics[width=0.45\textwidth]{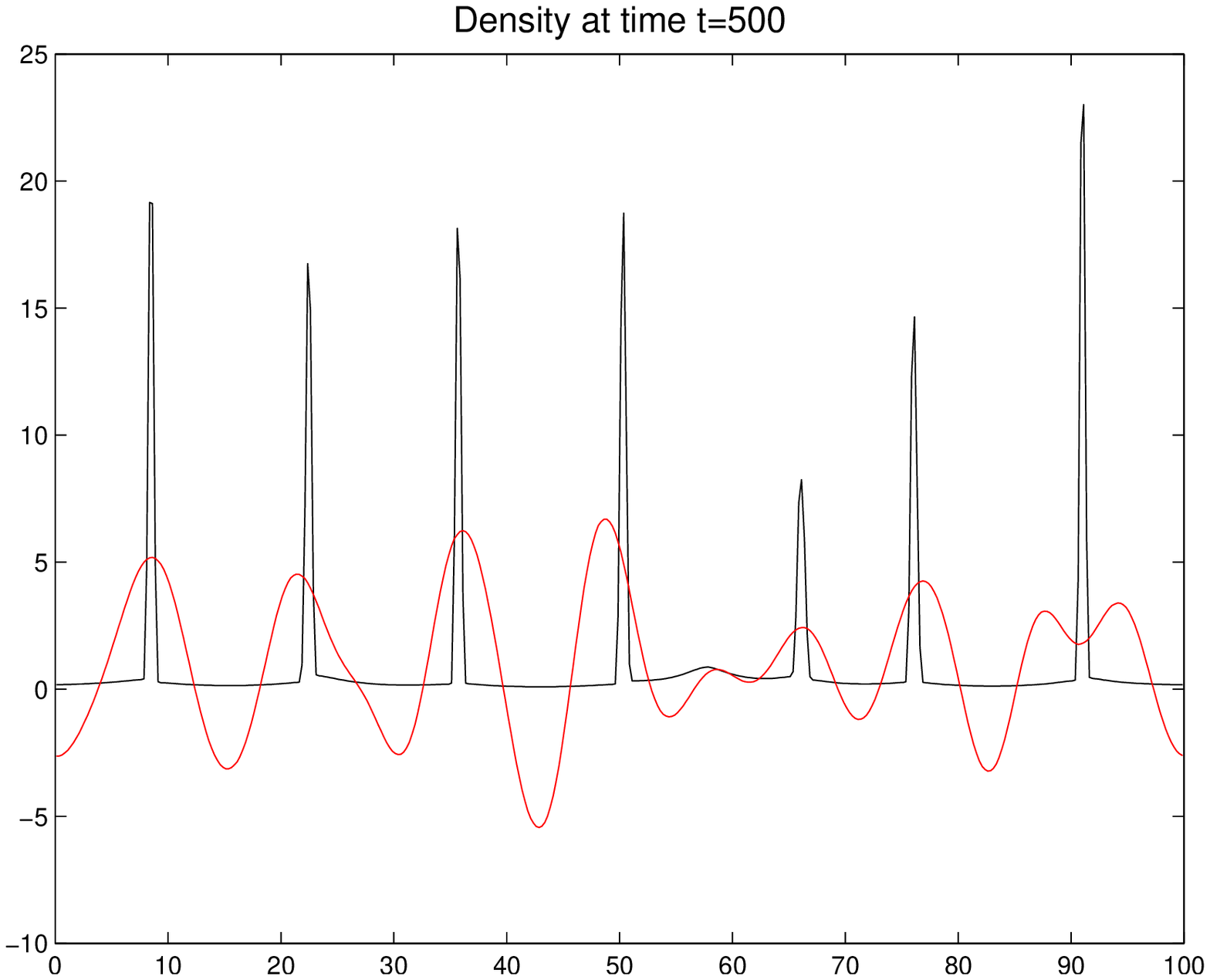}\hfill
\includegraphics[width=0.45\textwidth]{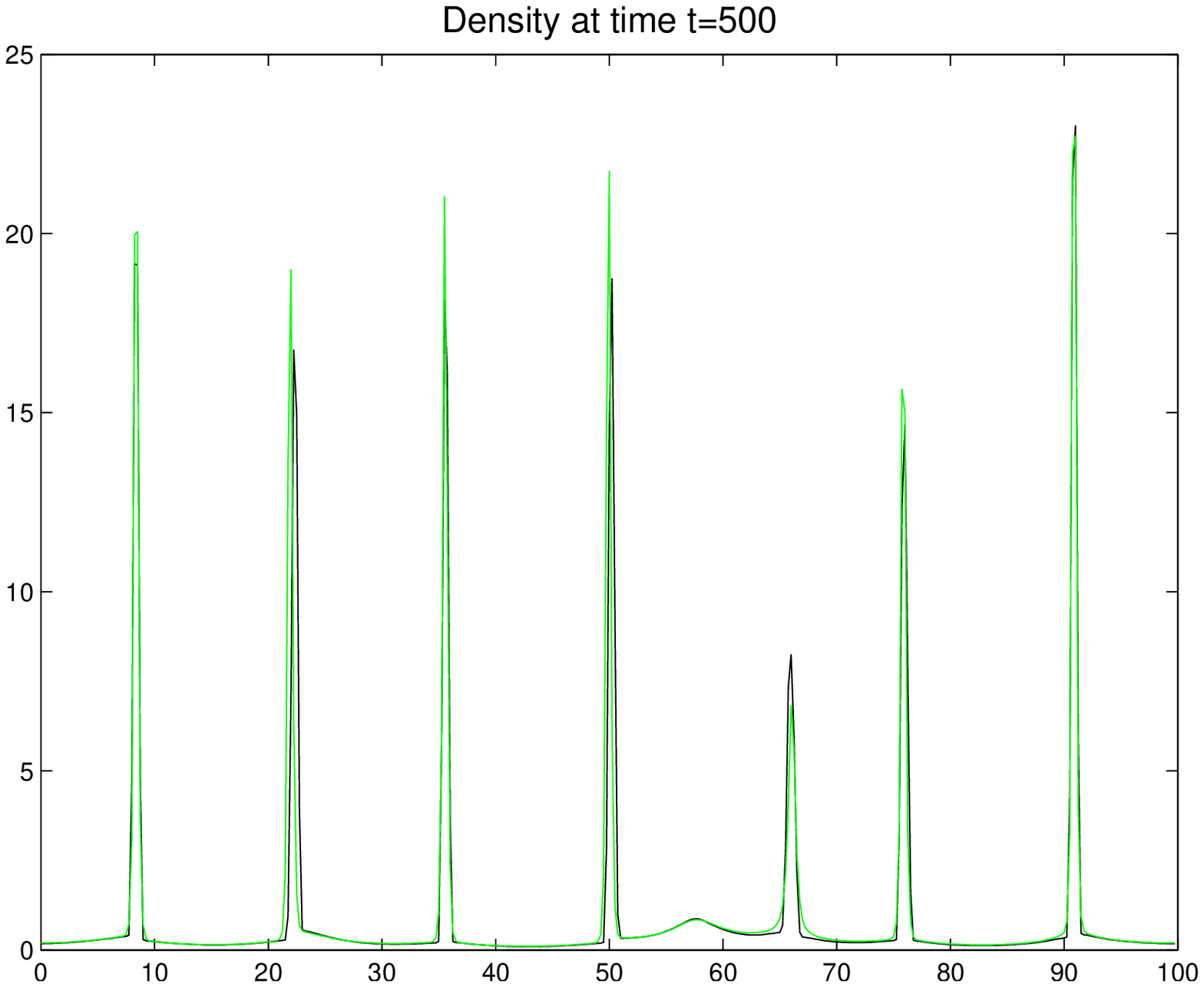}
\caption{\label{fig:numerics-3}
Left plots: Comparison of the linear model (red line) with
  the full model (black line). Right plots: Comparison of the
  quasidynamic approximation (green line) with the full model (black line).}
\end{figure}

Figures \ref{fig:numerics-1}-\ref{fig:numerics-3} present the results of
our numerical simulations. In all these plots, the black line is the
density $\rho$ as computed from the full model. This
solution is compared with results of the linear model (red curves) as
well as with results of the quasidynamic approximation (green curves).   

For small times, we see that both the linear as well as the nonlinear
quasidynamic approximation produce results which compare well with the
results of the full model. For the time $t=10$ the results of the
linear model appear to be even more accurate than the results of the
quasidynamic approximation. For large times
the predictions of the solution structure obtained by the quasidynamic 
approximation are much
more accurate than the predictions 
obtained from the linear model.

\subsection{Some remarks about the numerical methods}
We finish this section with some remarks about the numerical methods,
which were used to simulate the three different models. For each
model, the  method is based on an
operator splitting approach, i.e.\ during each time step we
successively approximate the different components of the coupled system.

We use a staggered grid and discretize the velocity $w$ at the nodes of
the grid, i.e.$\ w_{i+\frac{1}{2}}^n \approx w(x_{i+\frac{1}{2}},t_n)$
for $i=0, \ldots, m$
and  the density $\rho$ at midpoints of a grid cell, i.e.\
$$
\rho_i^n \approx \frac{1}{\Delta x}
\int_{x_{i-\frac{1}{2}}}^{x_{i+\frac{1}{2}}} \rho(x, t_n) dx \approx
\rho(x_i, t_n) \quad i=1, \ldots, m.
$$
Each subproblem is discretized using either a second order accurate
finite volume or a second order accurate finite difference
method. This means in particular, that we don't need to distinguish
between cell averaged values and cell centered point values of $\rho$.
The update of $w$ is computed using the Crank-Nicolson method for
periodic solutions.

For the full model (\ref{eqn:numerics1}), the evolution of $f$ is
split into the subproblems
\begin{equation}\label{eqn:numerics-f1}
\partial_t f + \nabla_{\vec{n}} \cdot 
\left( P_{\vec{n}^\bot} (0, 0 , n_1 w_x)^T f \right)  = \Delta_{\vec{n}} f
\end{equation}
and
\begin{equation}\label{eqn:numerics-f2}
\partial_t f - \partial_x (n_1 n_3 f) = 0.
\end{equation}
Recall that $f = f(x,t,\vec{n})$. Thus, subproblem (\ref{eqn:numerics-f1})
is a drift diffusion equation on the sphere, which needs to be solved
at all the discrete positions $x_1, \ldots, x_m.$ In our example, we solve during
each time step 400 times this drift diffusion equation on the
sphere ($m=400$). In each case we need to use another value for 
$w_x$, i.e.\ 
$$w_x(x_i,t_n) = 
\frac{w_{i+\frac{1}{2}}^n-w_{i-\frac{1}{2}}^n}{\Delta x}, \quad i=1,
\ldots, m.$$  

To discretize (\ref{eqn:numerics-f1}) at a point $x_i$, we use the
sphere grid from \cite{CHL2008}. In this approach, a single rectangular
computational domain (discretized with an equidistant Cartesian mesh)
is mapped to the sphere. In our simulations, the sphere was
discretized by mapping a rectangular Cartesian mesh with $40 \times
20$ grid cells to the sphere. This means that the position vector
$\vec{n}$ is discretized at 800 discrete points on the sphere, which are
denoted by ${\bf n}_{j,k}$, $j=1, \ldots, 40$, $k=1, \ldots, 20$.

In \cite{CHL2008}, it was  shown how 
LeVeque's wave propagation method \cite{book:RJL2002} can be used to discretized hyperbolic
pdes on the sphere. A version of the wave propagation algorithm was here
used to discretize 
$$
\partial_t f + \nabla_{\vec{n}} \cdot 
\left( P_{\vec{n}^\bot} (0, 0 , n_1 w_x)^T f \right)  = 0.
$$   Diffusion on the sphere, i.e.\ the subproblem
$$
\partial_t f = \Delta_{\vec{n}} f,
$$
was approximated using a finite volume method for parabolic problems
on surfaces, see \cite{CH2009}.

Finally, we discretize the transport equation (\ref{eqn:numerics-f2})
for all discrete representations of the position vector $\vec{n}$ on
the sphere. Note that for fixed values of $j$ and $k$, 
$n_1 n_2$ in (\ref{eqn:numerics-f2}) is constant. To discretize
(\ref{eqn:numerics-f2}), we solve an advection equation for each
choice of $j$ and $k$, i.e.\ in our test simulations  we solved during
each time step 800 advection equations, each with a different
advection speed. Each of these advection equations was discretized on
a grid with 400 grid cells (the number of grid cells used to
discretize the macroscopic space). We used again the high-resolution wave
propagation algorithm to approximate these advection problems.
  
Compared to the full system, the discretization of the effective
equation (\ref{eqn:numerics2}) as well as the discretization of the linear
model (\ref{eqn:numerics3}) is simple and computationally much less
expensive.
The evolution equation for $\rho$ in the effective equation
(\ref{eqn:numerics2}) is a combination of the wave propagation
algorithm for an advection problem with spatially varying advection
speed and a finite difference discretization for the nonlinear
diffusion term.
To discretize the linear moment closure model (\ref{eqn:numerics3}),
we note that the evolution of $\rho, S_{11}, S_{22}, S_{33}, S_{13}$
can be formulated as a linear hyperbolic system with
source term. For its discretization, we use the wave propagation algorithm for linear
hyperbolic systems together with an ODE solver for the source term.



 
\section*{Appendices}
\begin{appendix}

\section{Properties of the differential operators in spherical coordinates}
\label{section:appendix2}

The operator $\nabla_{\vec{n}}$ satisfies certain elementary
properties that are extensively used in this article:
Let $\vec{F}$ be a vector-valued function and $f$, $g$ be scalar-valued
functions, then
\begin{align}
\label{A.1}
\int_{S^{2}} ( \nabla_{\vec{n}} \cdot \vec{F} ) f d\vec{n} 
&= 
- \int_{S^{2}} \vec{F} \cdot (\nabla_{\vec{n}} f - 2 \vec{n} f ) d\vec{n} 
\\
\label{A.2}
\int_{S^{2}} (\nabla_{\vec{n}} \cdot \nabla_{\vec{n}} f) g d\vec{n} 
&= \int_{S^{2}} (\nabla_{\vec{n}} \cdot \nabla_{\vec{n}} g) f d\vec{n} 
\\
\label{A.3}
\int_{S^{2}} \vec{n} \otimes \nabla_{\vec{n}} f  d\vec{n} 
= \int_{S^{2}}  \nabla_{\vec{n}} f \otimes \vec{n}  \, d\vec{n}
&= \int_{S^{2}} (3 \vec{n} \otimes \vec{n} - {\rm id} ) f d\vec{n} 
\end{align}
The  components of the tensor $3\vec{n}\otimes \vec{n}-{\rm id}$
are the surface spherical harmonics of order 2. That is they are
harmonic polynomials on $\mathbb{R}^3$ of order 2, restricted
to $S^2$. The surface spherical harmonics
are eigenfunctions of the Laplacian on $S^2$ with corresponding
eigenvalue $-\ell(\ell+1)$, where $\ell$ is the order
\cite[App. E]{book:BCAH87}. Hence
\begin{equation}\label{surfhar}
\triangle_{\vec{n}} (3\,n_i\,n_j-\delta_{ij})\;=\;-6\,(3\,n_i\,n_j-\delta_{ij}).
\end{equation}

It is convenient to  use spherical coordinates in proving such formulas, 
see \cite[App. A.6 and E.6]{book:BCAH87}. 
A point 
$P$ with Cartesian coordinates $(n_{1}, n_{2}, n_{3})$ is expressed in spherical
coordinates via
$$n_{1} = r \sin \theta \, \cos \varphi \, , \;
n_{2} = r \sin \theta \, \sin \varphi \, , \;
n_{3} = r \cos \theta
$$
where $0<\theta < \pi$, $0 \le \varphi < 2\pi$. Let
$\vec{e}_{r}$, $\vec{e}_{\theta}$, $\vec{e}_{\varphi}$
be the orthonormal coordinate system associated to spherical
coordinates and attached at $P$. It satisfies the derivative formulas
\begin{equation}
\label{A.4}
\begin{array}{ccc}
\frac{\partial \vec{e}_{r} }{\partial r} = 0 \, , 
&
\frac{\partial \vec{e}_{r} }{\partial \theta} = \vec{e}_{\theta} \, , 
&
\frac{\partial \vec{e}_{r} }{\partial \varphi} = 
\vec{e}_{\varphi} \, \sin \theta  \, ,
\\
\frac{\partial \vec{e}_{\theta} }{\partial r} = 0 \, , 
&
\frac{\partial \vec{e}_{\theta} }{\partial \theta} = - \vec{e}_{r} \, , 
&
\frac{\partial \vec{e}_{\theta} }{\partial \varphi} = 
\vec{e}_{\varphi} \, \cos \theta \,,
\\
\frac{\partial \vec{e}_{\varphi} }{\partial r} = 0 \, , 
&
\frac{\partial \vec{e}_{\varphi} }{\partial \theta} = 0 \, , 
&
\frac{\partial \vec{e}_{\varphi} }{\partial \varphi} = 
-  \vec{e}_{r}  \, \sin \theta-  \vec{e}_{\theta} \,\cos \theta \, .
\end{array}
\end{equation}

We visualize the sphere $S^{2}$ as embedded in the Euclidean space.
The surface gradient $\nabla_{\vec{n}}$ is related to the (ambient) gradient operator $\nabla$
through
$$
\nabla_{\vec{n}} = r ({\rm id} - \vec{n} \otimes \vec{n} ) \cdot \nabla 
= \vec{e}_{\theta} \frac{\partial}{\partial \theta} + 
\vec{e}_{\varphi} \frac{1}{\sin \theta} \frac{\partial}{\partial \varphi}
$$
and, for a scalar-valued function $f$, the surface gradient and the Laplace-Beltrami
operator are given respectively by
$$
\begin{aligned}
\nabla_{\vec{n}} f &= 
 \vec{e}_{\theta} \frac{\partial f}{\partial \theta} + 
\vec{e}_{\varphi} \frac{1}{\sin \theta} \frac{\partial f}{\partial \varphi}
\\
\triangle_{\vec{n}} f = \nabla_{\vec{n}} \cdot \nabla_{\vec{n}} f &=
\frac{1}{\sin \theta} \frac{\partial}{\partial \theta}
\left ( \sin \theta \frac{\partial f}{\partial \theta} \right )
+ 
\frac{1}{\sin^{2} \theta} \frac{\partial^{2} f}{\partial \varphi^{2}}
\end{aligned}
$$
The divergence of  a vector-valued function 
$\vec{F} = F_{r} \vec{e}_{r} + F_{\theta} \vec{e}_{\theta} + F_{\varphi} 
\vec{e}_{\varphi}$ has the form
$$
\begin{aligned}
\nabla_{\vec{n}} \cdot \vec{F} &= 
\left ( \vec{e}_{\theta} \frac{\partial}{\partial \theta} + 
\vec{e}_{\varphi} \frac{1}{\sin \theta} \frac{\partial}{\partial \varphi} \right )
\cdot
	\big ( \vec{F}_{r} \vec{e}_{r} + F_{\theta} \vec{e}_{\theta} + F_{\varphi} 
\vec{e}_{\varphi} \big )
\\
&\stackrel{\eqref{A.4}}{=} \frac{1}{\sin \theta} \frac{\partial}{\partial \theta} 
\left ( \sin \theta F_{\theta} \right ) + 
\frac{1}{\sin \theta} \frac{\partial F_{\varphi}}{\partial \varphi}
+ 2 F_{r}
\end{aligned}
$$

It is now easy to compute \eqref{A.1}-\eqref{A.3}.
Observe that
\begin{align*}
\int_{S^{2}} ( \nabla_{\vec{n}} \cdot \vec{F} ) f d\vec{n}
 &=
\iint \left (
\frac{1}{\sin \theta} \frac{\partial}{\partial \theta} 
\left ( \sin \theta F_{\theta} \right ) + 
\frac{1}{\sin \theta} \frac{\partial F_{\varphi}}{\partial \varphi}
+ 2 F_{r} \right ) f \sin \theta \, d\theta d\varphi
\nonumber\\
&=
- \iint \left ( -2 F_{r} f + F_{\theta} \frac{\partial f}{\partial 
\theta} + \frac{1}{\sin \theta} F_{\varphi} 
\frac{\partial f}{\partial \varphi} \right ) \sin \theta \, d\theta d\varphi
\nonumber \\
&=
- \int_{S^{2}} \vec{F} \cdot (\nabla_{\vec{n}} f - 2 \vec{n} f ) d\vec{n}
\nonumber
\end{align*}
which gives \eqref{A.1}, and  \eqref{A.2} follows by applying  \eqref{A.1} twice:
$$
\begin{aligned}
\int_{S^{2}} (\nabla_{\vec{n}} \cdot \nabla_{\vec{n}} f) g d\vec{n}
&= - \int_{S^{2}} \nabla_{\vec{n}} f \cdot ( \nabla_{\vec{n}} g - 2 \vec{n} g) \, d\vec{n}
\\
&= \int_{S^{2}} f (\nabla_{\vec{n}} \cdot \nabla_{\vec{n}} g)  d\vec{n} \, .
\end{aligned}
$$

Using integration by parts, we obtain the chain of identities
\begin{eqnarray}\label{A.15}
\lefteqn{\int_{S^{2}} \vec{n} \otimes \nabla_{\vec{n}} f  d\vec{n} 
= \iint \vec{e}_{r} \otimes \Big ( 
\vec{e}_{\theta} \frac{\partial f}{\partial \theta} 
+ \vec{e}_{\varphi} \frac{1}{\sin \theta} \frac{\partial f}{\partial \varphi}
\Big ) \, \sin \theta d\theta d\varphi
}\nonumber\\
&=&
- \iint \Big [ 
\frac{\partial}{\partial \theta} (\vec{e}_{r} \otimes \vec{e}_{\theta}) 
+ \frac{\cos \theta}{\sin \theta} \vec{e}_{r} \otimes \vec{e}_{\theta}
+ \frac{1}{\sin \theta} \frac{\partial }{\partial \varphi}
    (\vec{e}_{r} \otimes \vec{e}_{\varphi}) 
\Big ] \, f \, \sin \theta d\theta d\varphi
\nonumber \\
&\stackrel{(\ref{A.4})}{=}&
- \iint \Big [
\vec{e}_{\theta} \otimes \vec{e}_{\theta} +  \vec{e}_{\varphi} \otimes \vec{e}_{\varphi}
- 2 \vec{e}_{r} \otimes \vec{e}_{r}
\Big ] \, f \, \sin \theta d\theta d\varphi
\nonumber\\
&=&
\int_{S^{2}} (3 \vec{n} \otimes \vec{n} - {\rm id} ) f d\vec{n}
\end{eqnarray}
Since the final equation in 
\eqref{A.15} is a symmetric tensor, \eqref{A.3} follows.

Note that, for any $3\times 3$ matrix $\kappa$, the equation holds
\begin{equation}\label{A.5}
\nabla_{ \vec{n}} \cdot \left( P_{\vec{n}^\bot} \kappa \vec{n} \right) 
= \mbox{tr } \kappa - 3 \vec{n} \cdot \kappa \vec{n} \, ,
\end{equation} 
where tr stands for the trace operator. Indeed, using \eqref{A.4}
we obtain
\begin{align}
\nabla_{ \vec{n}} \cdot \left( P_{\vec{n}^\bot} \kappa \vec{n} \right)  
&=
\left ( \vec{e}_{\theta} \frac{\partial}{\partial \theta} + 
\vec{e}_{\varphi} \frac{1}{\sin \theta} \frac{\partial}{\partial \varphi} \right )
\cdot
\big ( \kappa \vec{e}_r - (\vec{e}_r \cdot \kappa \vec{e}_r) \vec{e}_r \big ) 
\nonumber\\
&\stackrel{(\ref{A.4})}{=}
\vec{e}_\theta \cdot \kappa \vec{e}_\theta + \vec{e}_\varphi \cdot \kappa \vec{e}_\varphi
- 2 \vec{e}_r \cdot \kappa \vec{e}_r
\nonumber\\
&=
\mbox{tr } \kappa - 3 \vec{n} \cdot \kappa \vec{n}
\end{align}


\section{Harmonic Polynomials and Spherical Harmonics}\label{app3}

For the reader's convenience we review some material on harmonic polynomials and spherical harmonics that is used 
in the text. We refer to Stein and Weiss \cite[Ch 4]{SW} and Helgason \cite[Intro Thm 3.1]{H} for further details.

Let $\cP_k (\R^d)$ be the space of homogeneous polynomials of degree $k$ on $\R^d$. We denote by $\cP_k (S^{d-1})$  the restrictions
of polynomials $P \in \cP_k (\R^d)$ on the sphere $S^{d-1}$. The dimension $dim \cP_k (\R^d) = \binom{d + k -1}{k}$.
\medskip

Let $\cH_k (\R^d)$ be the space of homogeneaous polynomials of degree $k$ on $\R^d$ that are harmonic,
$$
\cH_k (\R^d) = \{ P \in \cP_k (\R^d) : \triangle P = 0 \} .
$$
The elements of  $\cH_k (\R^d)$ are called solid harmonics. The spherical harmonics $\cH_k (S^{d-1})$ are defined as the restrictions of 
$P \in \cH_k (\R^d)$ on the sphere.
The map $\rho : \cH_k (\R^d) \to \cH_k (S^{d-1})$ that maps the above polynomial to its restriction is a bijection.

The relation between these spaces of polynomials is given by the following proposition.

\begin{theorem} The map $\triangle : \cP_k (\R^d) \to \cP_{k-2} (\R^d)$ is onto for all $d$ and $k \ge 2$. Furthermore, we have the
orthogonal direct sum decomposition
$$
\cP_k (\R^d) = \cH_k (\R^d) \oplus |x|^2 \, \cP_k (\R^d)  \quad \mbox{for $k \ge 2$}.
$$
\end{theorem}
This proposition allows to compute the dimension of  $ \cH_k (S^{d-1})$ as
$$
\dim \cH_k (\R^d) =  \cH_k (S^{d-1}) = \binom{d+k-1}{k} - \binom{d+k-3}{k-2} \quad \mbox{for $k \ge 2$}.
$$
Note also that $\dim \cH_0 (\R^d)  =1$ and $\dim \cH_0 (\R^d)  = d$. Moreover, proceeding via 
induction we deduce  the direct sum decompositions
\begin{align}
\label{ds1}
\cP_k (\R^d ) &= \;  \cH_k ( \R^d ) \oplus |x|^2  \cH_{k-2} ( \R^d )  \oplus ... \oplus |x|^{2 \left[ \frac{k}{2} \right ] }  \cH_{ k - 2 \left[ \frac{k}{2} \right ] } ( \R^d )
\\
\label{ds2}
\cP_k (S^{d-1}) &= \cH_k ( S^{d-1} ) \oplus \cH_{k-2} ( S^{d-1} )  \oplus ... \oplus \cH_{ k - 2 \left[ \frac{k}{2} \right ] } ( S^{d-1} ) 
\end{align}

The importance of spherical harmonics stems from the property that  for $P \in \cH_k (\R^d)$ the restriction
 $H = P \Big |_{S^{d-1}} \in \cH_k ( S^{d-1} )$
is an  eigenfunction of the Laplace-Beltrami operator $\triangle_{S^{d-1}}$. 
To see that  for  $P \in \cH_k (\R^d)$ we write $x = r n$, with $r = |x|$ and $n = \frac{x}{|x|}$, and e write  $P(x) = P(r n) = r^d H(n)$.
Recall that for $f \in C^\infty (\R^d)$ the Laplacian is expressed as
$$
\triangle f = \frac{1}{r^{d-1}} \frac{\del}{\del r} \left ( r^{d-1} \frac{\del f}{\del r} \right ) + \frac{1}{r^2} \triangle_{S^{d-1}} f 
$$
Using this formula we compute
$$
\triangle P = 0 \quad  \Longleftrightarrow \quad  \triangle_{S^{d-1}} H = - k (d + k -2) H
$$
Moreover, one has the following theorem (see \cite[Intro Thm 3.1]{H}:
\begin{theorem}
The eigenspaces of the Laplace-Beltrami operator $\triangle_{S^{d-1}}$ are the spaces of spherical harmonics $\cH_k (S^{d-1})$
with associated eigenvalue -k (d+k-2). The eigenspaces are ortogonal and $L^2(S^{d-1})$ admits the direct sum decomposition
\begin{equation}
\label{directsum}
L^2 (S^{d-1}) = \bigoplus_{k=0}^{\infty} \cH_k (S^{d-1})
\end{equation}
\end{theorem}

In the present context we are interested in functions on the sphere $S^{d-1}$ that are even, that is $f(-n) = f(n)$.
In this case the basis will only involve the even spherical harmonics and the direct sum in \eqref{directsum} will extend over
the even integers. On $S^2$ the spherical harmonics are computed by using the Legendre and associated Legendre polynomials.
In Table \ref{tableharmonic} we list the harmonic polynomial basis functions (up to order $4$) that are used
in the calculations of the present article.

\begin{table}
\label{tableharmonic}
\begin{center}
\begin{tabular}{c|c|c}
$0th$ order & $2nd$ order & $4th$ order\\
\hline \hline
& & $P_4^{-4} =
\sin^4 \theta \cos 4 \phi$\\ 
& & $P_4^{-3} = \sin^3
\theta \cos \theta \cos 3 \phi$\\
& $P_2^{-2} = \sin^2 \theta \cos 2 \phi$ & 
$P_4^{-2} = \sin^2 \theta (7
\cos^2 \theta - 1) \cos 2 \phi$\\ 
&  $P_2^{-1} = \sin \theta \cos \theta \cos \phi$ &
$P_4^{-1} = \sin
\theta (7 \cos^3 \theta - 3 \cos \theta ) \cos \phi$\\
$P_0^0 = 1$ &  $P_2^0 = \cos^2 \theta - \frac{1}{3}$ &
$P_4^0 = 35 \cos^4 \theta - 30
\cos^2 \theta + 3$\\
& $P_2^1 = \sin \theta \cos \theta \sin \phi$ &
$P_4^1 = \sin \theta (7 \cos^3 \theta - 3 \cos \theta) \sin
\phi$\\
& $P_2^2 = \sin^2 \theta \sin 2\phi$ &
$P_4^2 = \sin^2 \theta (7 \cos^2 \theta - 1) \sin 2 \phi$ \\
& &  $P_4^3 = \sin^3 \theta \cos \theta \sin 3 \phi$ \\
& & $P_4^4 = \sin^4 \theta \sin 4 \phi$ \\
\end{tabular}
\end{center}
\vspace*{0.25cm}

\caption{Harmonic polynomial basis functions.}
\end{table}


\section{Linear stability analysis}\label{appendix:linstab}
Experimental studies for the sedimentation of suspensions with
rod-like particles \cite{HG96, HG99, MBG07a} reveal the formation of
packets of particles which seem to have a mesoscopic equilibrium
width. Our goal  is to give an explanation of this wave length
selection mechanism based on linear stability theory.

We consider the linear pde for shear flow
\begin{equation}\label{eqn:linear-moment-closure-shear}
\begin{split}
\partial_t \rho & = \partial_x S_{13} \\
\partial_t S_{11} & = \frac{2}{21} \partial_x S_{13} - 6 D_r S_{11} \\
\partial_t S_{22} & = - \frac{4}{7} \partial_x S_{13} - 6 D_r S_{22}\\
\partial_t S_{33} & = \frac{2}{21} \partial_x S_{13} - 6 D_r S_{33}\\
\partial_t S_{13} & = - \frac{1}{7} \partial_x S_{22} +
\frac{1}{15} \partial_x \rho - 6 D_r S_{13} + \frac{1}{5} \partial_x
w\\
Re \, \partial_t w & = \partial_{xx} w + \delta (m-\rho),
\end{split}
\end{equation}
which is obtained by linearizing the
nonlinear moment closure system (\ref{eqn:mc-shear})  around the state 
$\rho = 1$ and $\vec{u}=0$. 

Fourier transformation of the first five equations of
(\ref{eqn:linear-moment-closure-shear}) leads to the linear system of
ordinary differential equations
\begin{equation}\label{eqn:stability1}
\begin{split}
\hat{\rho}'(\xi,t) & = i \xi \hat{S}_{13}(\xi,t) \\
\hat{S}_{11}'(\xi,t) & = \frac{2}{21} i \xi \hat{S}_{13}(\xi,t) - 6 D_r
\hat{S}_{11}(\xi,t) \\
\hat{S}_{22}'(\xi,t) & = -\frac{4}{7} i \xi \hat{S}_{13}(\xi,t) - 6 D_r
\hat{S}_{22}(\xi,t)\\
\hat{S}_{33}'(\xi,t) & = \frac{2}{21} i \xi \hat{S}_{13}(\xi,t) - 6 D_r
\hat{S}_{33}(\xi,t)\\
\hat{S}_{13}'(\xi,t) & = -\frac{1}{7} i \xi \hat{S}_{22}(\xi,t) + \frac{1}{15}
i \xi \hat{\rho}(\xi,t) - 6 D_r \hat{S}_{13}(\xi,t) + \frac{1}{5} i \xi \hat{w}(\xi,t). 
\end{split}
\end{equation}
\subsection*{The case $Re=0$:}
First we consider the case $Re = 0$. In this case, Fourier
transformation of the last equation of
(\ref{eqn:linear-moment-closure-shear})
leads to the relation
$$
0 = - \xi^2 \hat{w}(\xi) + \delta \left( m \delta_\xi - \hat{\rho}(\xi) \right),
$$
where $\delta_\xi$ is the delta function.
Thus we can replace $\hat{w}$ in the last equation of (\ref{eqn:stability1}) 
by 
$$
\hat{w}(\xi) = \frac{1}{\xi^2} \delta \left( m \delta_\xi - \hat{\rho}(\xi) \right).
$$
We obtain a linear ode system of the form 
$$
\partial_t \vec{U}(\xi,t) = A(\xi,\delta,D_r) \vec{U}(\xi,t) +
\delta m \delta_\xi \vec{e}_5,
$$
with $\vec{U} = (\hat{\rho}, \hat{S}_{11}, \hat{S}_{22}, \hat{S}_{33},
\hat{S}_{13})^T$ and 
\begin{equation}\label{eqn:stability2}
A = \left( \begin{array}{ccccc}
0 & 0 & 0 & 0 & i \xi \\
0 & -6Dr & 0 & 0 & \frac{2}{21} i \xi \\
0 & 0 & -6 D_r & 0 & -\frac{4}{7} i \xi \\
0 & 0 & 0 & -6 D_r & \frac{2}{21} i \xi \\
-\frac{1}{5 \xi} i \delta   + \frac{1}{15} i \xi & 0 & -\frac{1}{7} i \xi & 0 &
-6D_r
\end{array}\right). 
\end{equation}
Let $\alpha(A)$ denote the spectral abscissa of the matrix $A$, i.e.
$$
\alpha(A) := \max_j \mbox{Re} (\lambda_j),
$$
where $\lambda_1,\ldots, \lambda_5$ are the eigenvalues of $A$. The
solution $\vec{U}(\xi,t)$ of the linear system remains bounded
provided $\alpha(A) \le 0$ and $\vec{U}(\xi,t) \rightarrow 0$ as
$t\rightarrow \infty$ if $\alpha(A) < 0$. 

For $D_r = 0$ the eigenvalues of (\ref{eqn:stability2}) are
$$
\lambda_{1,2,3} = 0, \quad \lambda_{4,5} = \pm \sqrt{-1635 \xi^2 + 2205 \delta}.
$$
This predicts the instability of density modulations for horizontal
waves with a sufficiently small wavenumber but not a wavelength
selection mechanism. This agrees with previous findings of Koch and
Shaqfeh \cite{KS89}. We also see that the problem becomes more unstable
if we increase $\delta$.

\begin{figure}[htb]
(a)\includegraphics[width=0.3\textwidth]{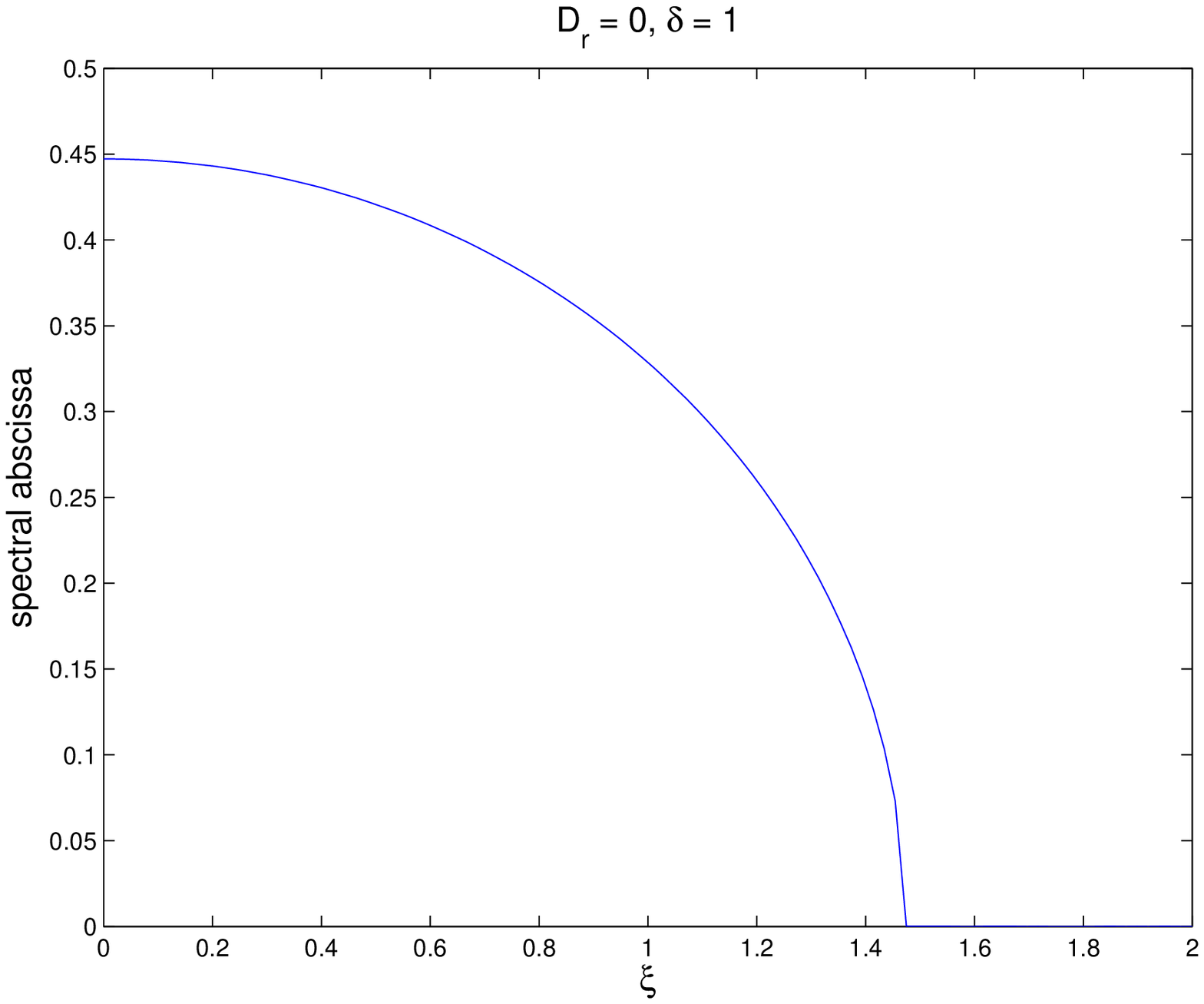}\hfil
(b)\includegraphics[width=0.3\textwidth]{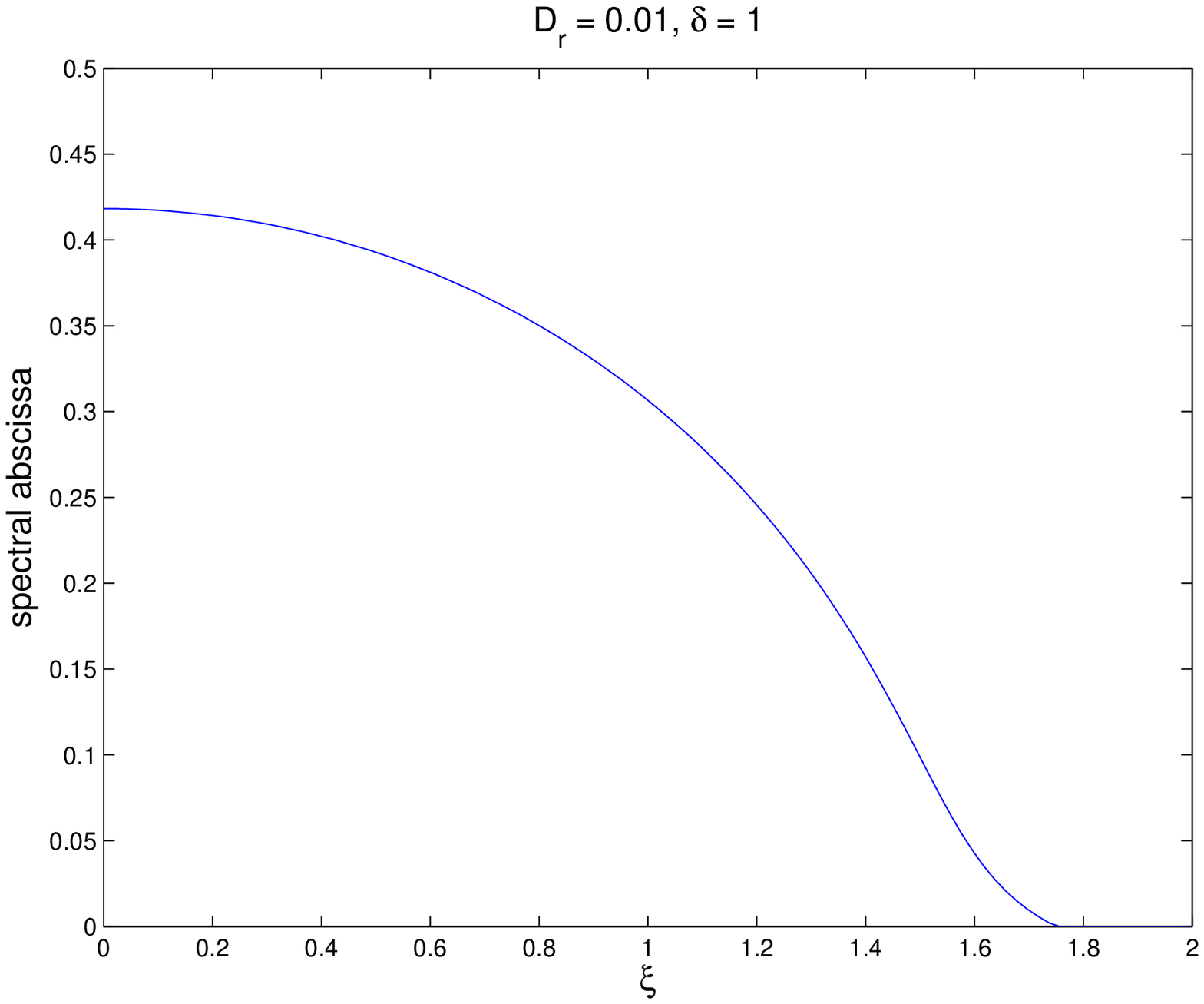}\hfil
(c)\includegraphics[width=0.3\textwidth]{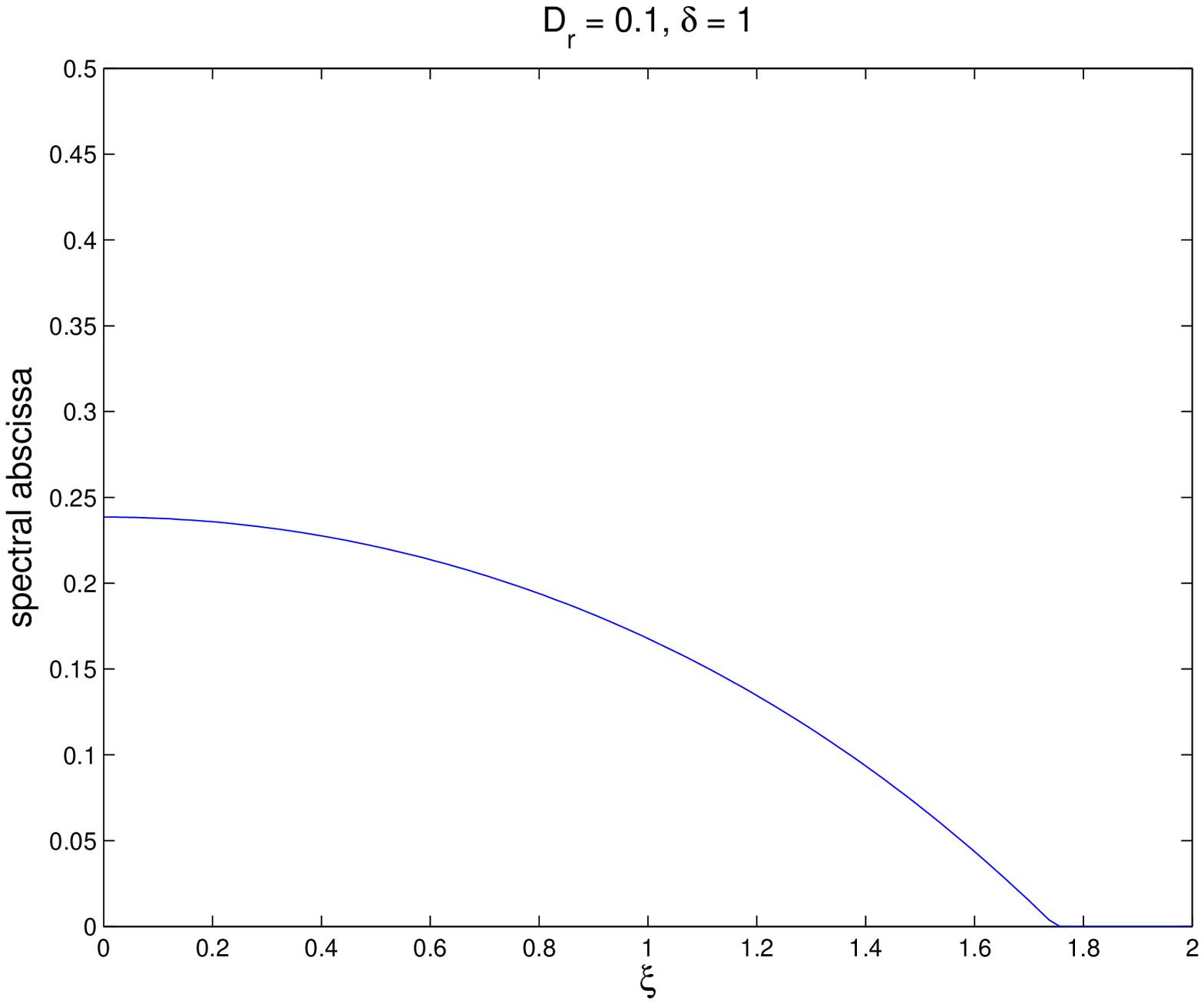}
\caption{\label{fig:stability3}Spectral abscissa as a function of wavenumber $\xi$ for
  $\delta = 1$, $Re=0$ and different values of $D_r$.}
\end{figure}

The introduction of Brownian effects in terms of translational
difusion (a case which was excluded here by setting $\gamma =0$)  does
also not provide a wavelength selection mechanism at the level of
linear stability analysis, see Saintillan \cite{SSD06b}.
If we include Brownian effects in terms or rotational diffusion, i.e.\
in the case $D_r > 0$, then we don't have such a simple formula for
the eigenvalues. However, we computed the spectral abscissa as a
function of $\xi$. The results of these computations are shown in
Figure \ref{fig:stability3}. In all three plots we set $\delta =1$ and
varied the value of $D_r$.  
By increasing the value of $D_r$ the problem becomes less
unstable. There is no wavelength selection mechanism.

\subsection*{The case $Re > 0$:} 
\begin{figure}[htb]
(a)\includegraphics[width=0.3\textwidth]{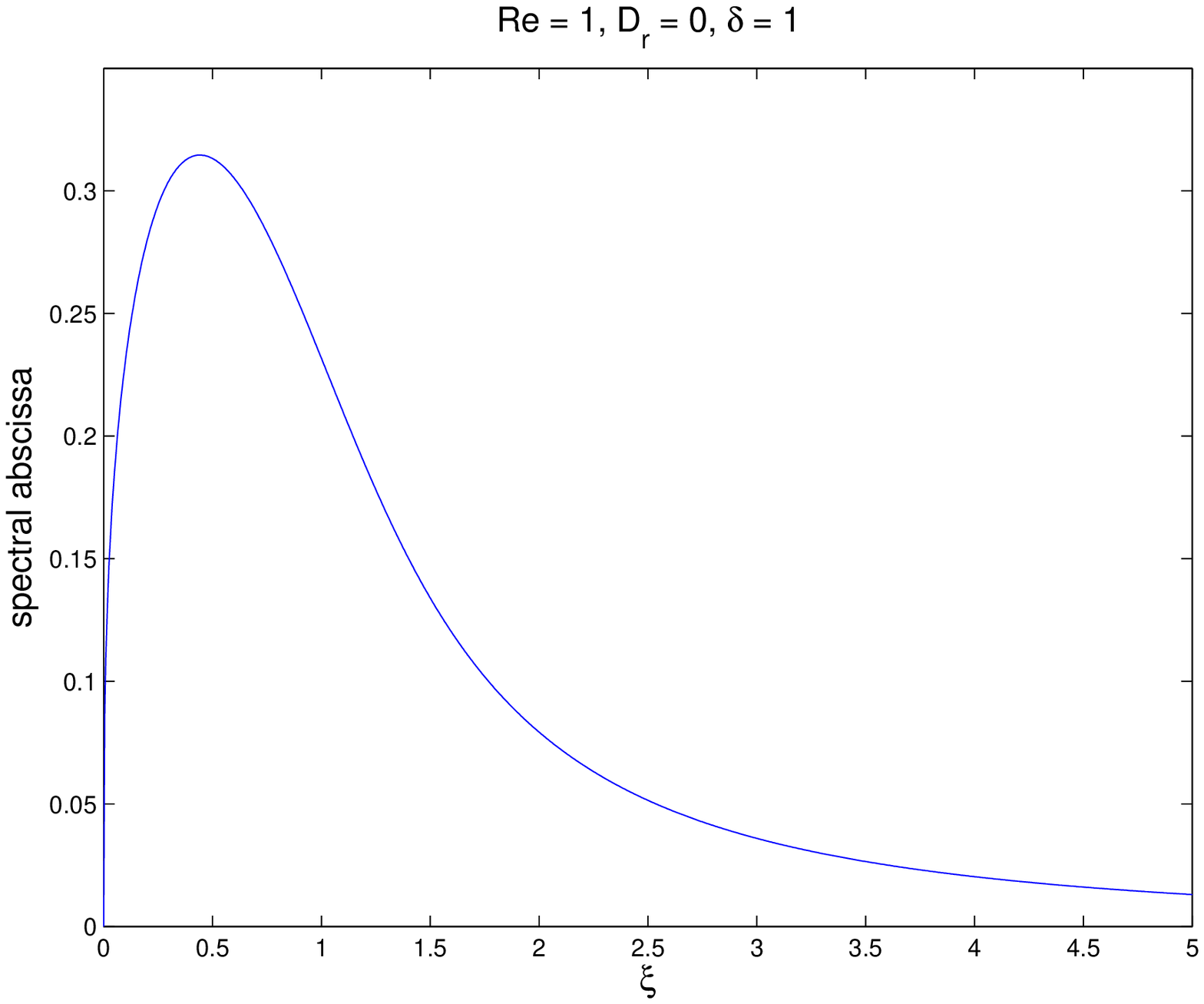}\hfil
(b)\includegraphics[width=0.3\textwidth]{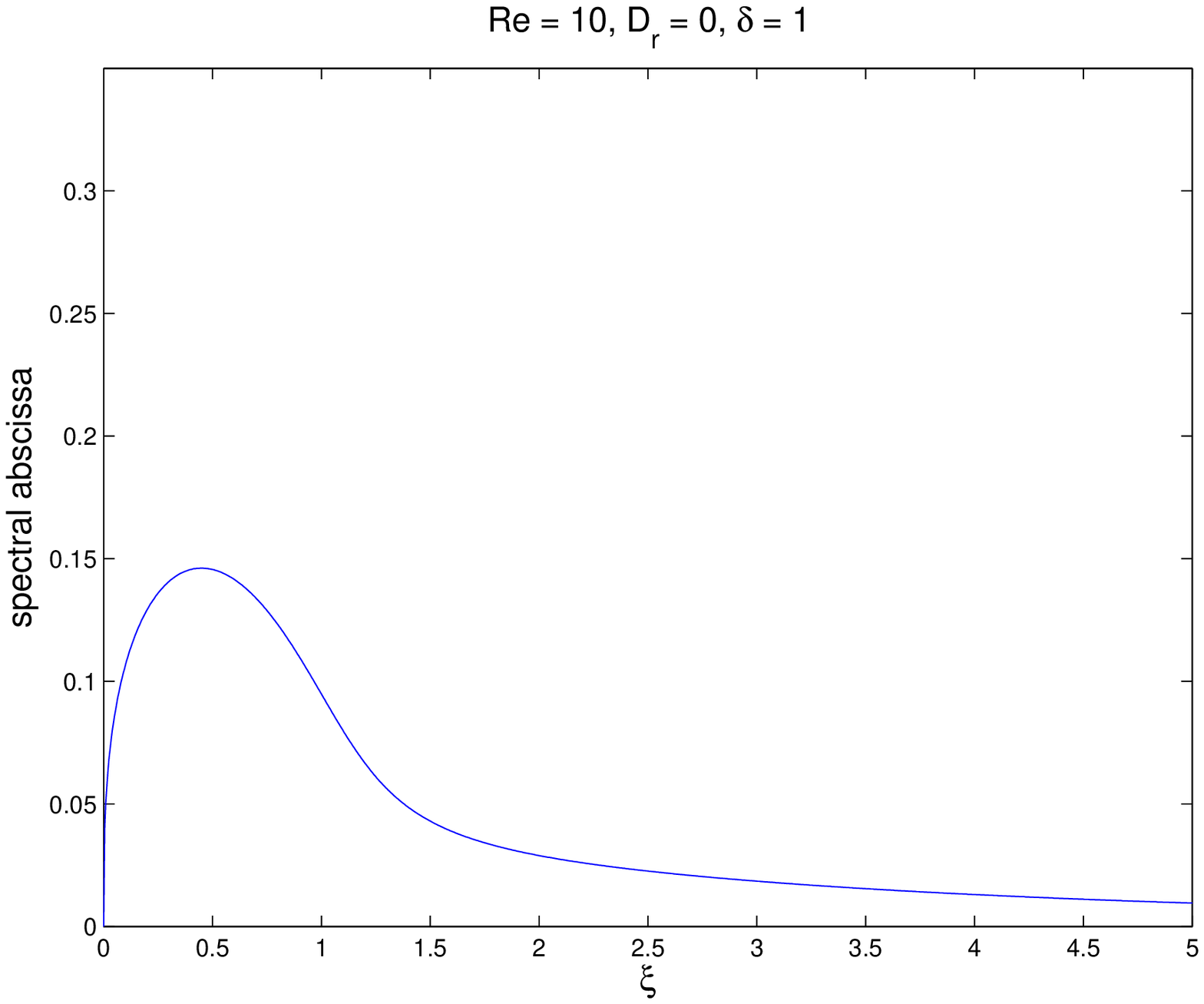}\hfil
(c)\includegraphics[width=0.3\textwidth]{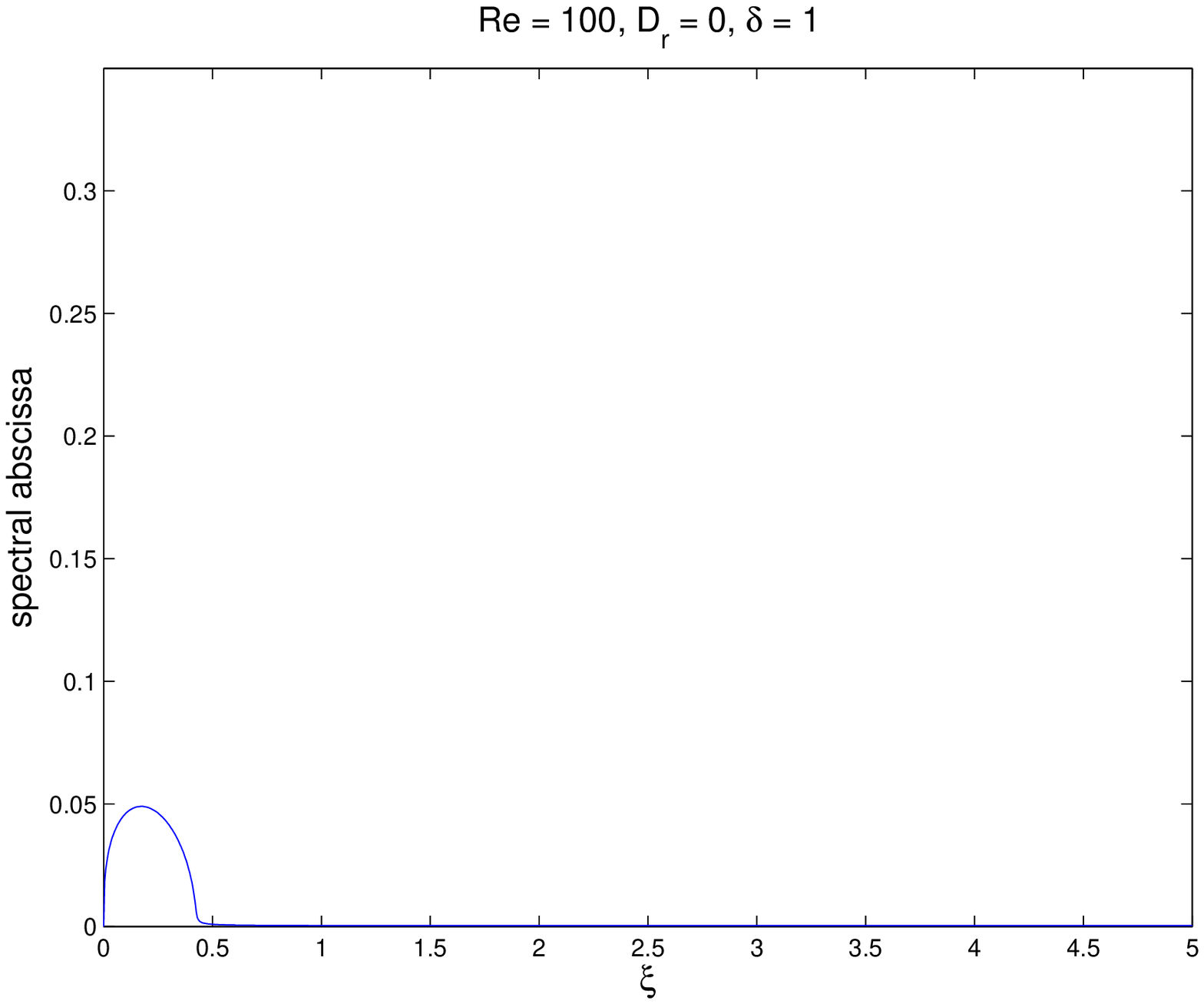}
\caption{\label{fig:stability4}Spectral abscissa as a function of wavenumber $\xi$ for
  $\delta = 1$, $D_r=0$ and different values of $Re$.}
\end{figure}

Now we consider the case
$Re>0$. Fourier transformation of the linearized equation for $w$
gives
\begin{equation}\label{eqn:stability4}
Re \partial_t \hat{w}(\xi,t) = -\xi^2 \hat{w}(\xi,t)+ \delta(m
\delta_\xi - \hat{\rho}(\xi,t)).
\end{equation}
We consider the linear stability of the system (\ref{eqn:stability1})
together with (\ref{eqn:stability4}).
This system has the form 
$$
\partial_t \vec{U}(\xi,t) = A(\xi,\delta,D_r,Re) \vec{U}(\xi,t) +
\delta m \delta_\xi \vec{e}_6,
$$
with $\vec{U} = (\hat{\rho}, \hat{S}_{11}, \hat{S}_{22}, \hat{S}_{33},
\hat{S}_{13},\hat{w})^T$ and
\begin{equation}
A = \left( \begin{array}{cccccc}
0 & 0 & 0 & 0 & i \xi & 0\\
0 & -6Dr & 0 & 0 & \frac{2}{21} i \xi & 0\\
0 & 0 & -6 D_r & 0 & -\frac{4}{7} i \xi & 0\\
0 & 0 & 0 & -6 D_r & \frac{2}{21} i \xi & 0\\
\frac{1}{15} i \xi & 0 & -\frac{1}{7} i \xi & 0 &
-6D_r & \frac{1}{5} i \xi \\
-\frac{\delta}{Re} & 0 & 0 & 0 & 0 & - \frac{1}{Re} \xi^2
\end{array}\right).
\end{equation}
For $\delta = 0$ and $D_r=0$, the eigenvalues of $A$ are 
$$
\lambda_{1,2,3} = 0, \quad \lambda_{4,5} = \pm \frac{1}{105} i
\sqrt{1635} \xi, \quad \lambda_6 = - \frac{\xi^2}{Re},
$$
and thus the linear system is stable. 
For $\delta > 0$,  we observe instability as well as a wavelength
selection.  In Figure \ref{fig:stability4}, we plot the spectral
abscissa as a function of $\xi$ for $\delta = 1$, $D_r = 0$ and
different values of $Re$.
An increase of $Re$ reduces the spectral abscissa, i.e.\ it makes the
system less unstable. The wavelength of the most unstable wave
increases (the wavenumber decreases).  

\begin{figure}[htb]
(a)\includegraphics[width=0.3\textwidth]{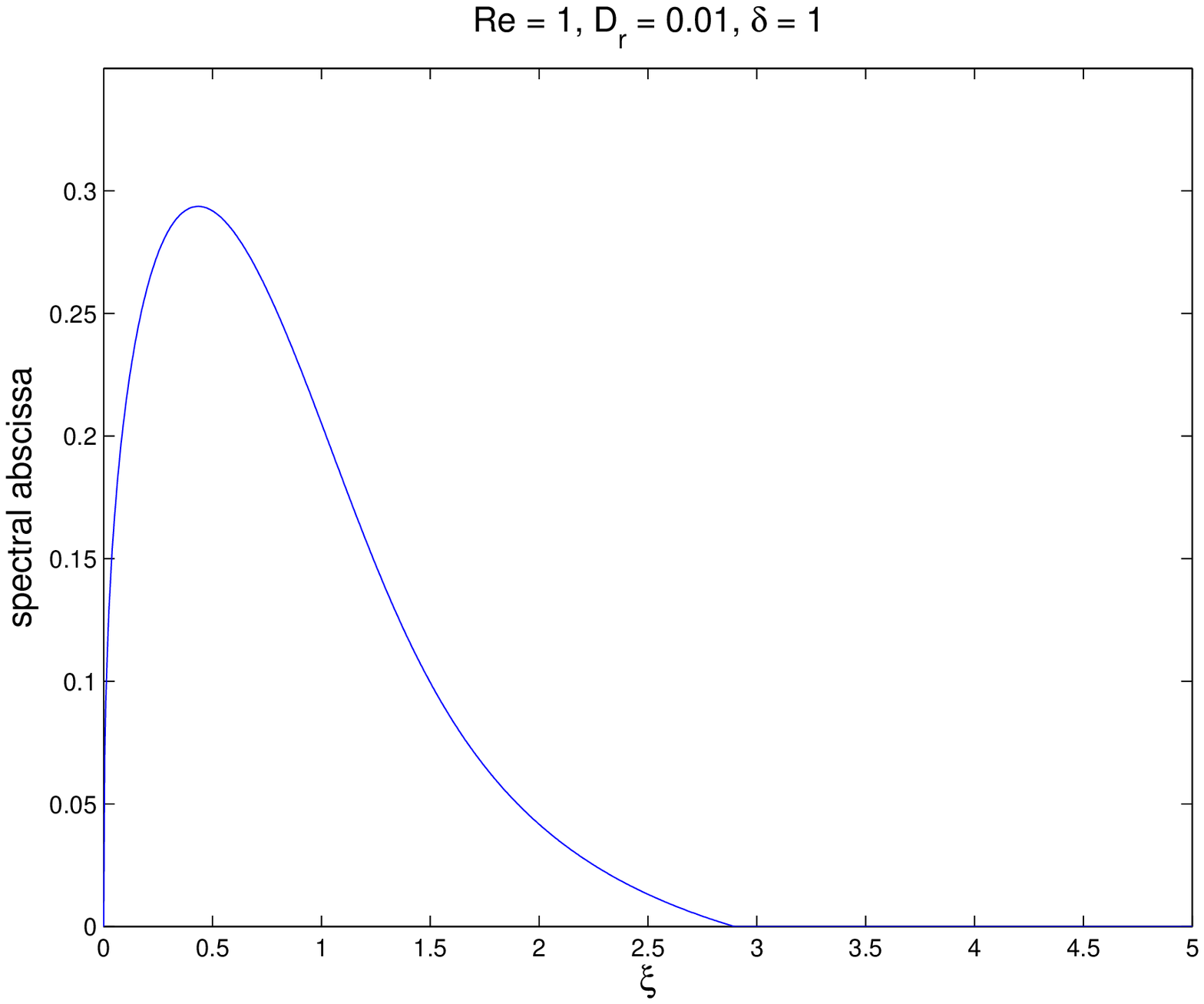}\hfil
(b)\includegraphics[width=0.3\textwidth]{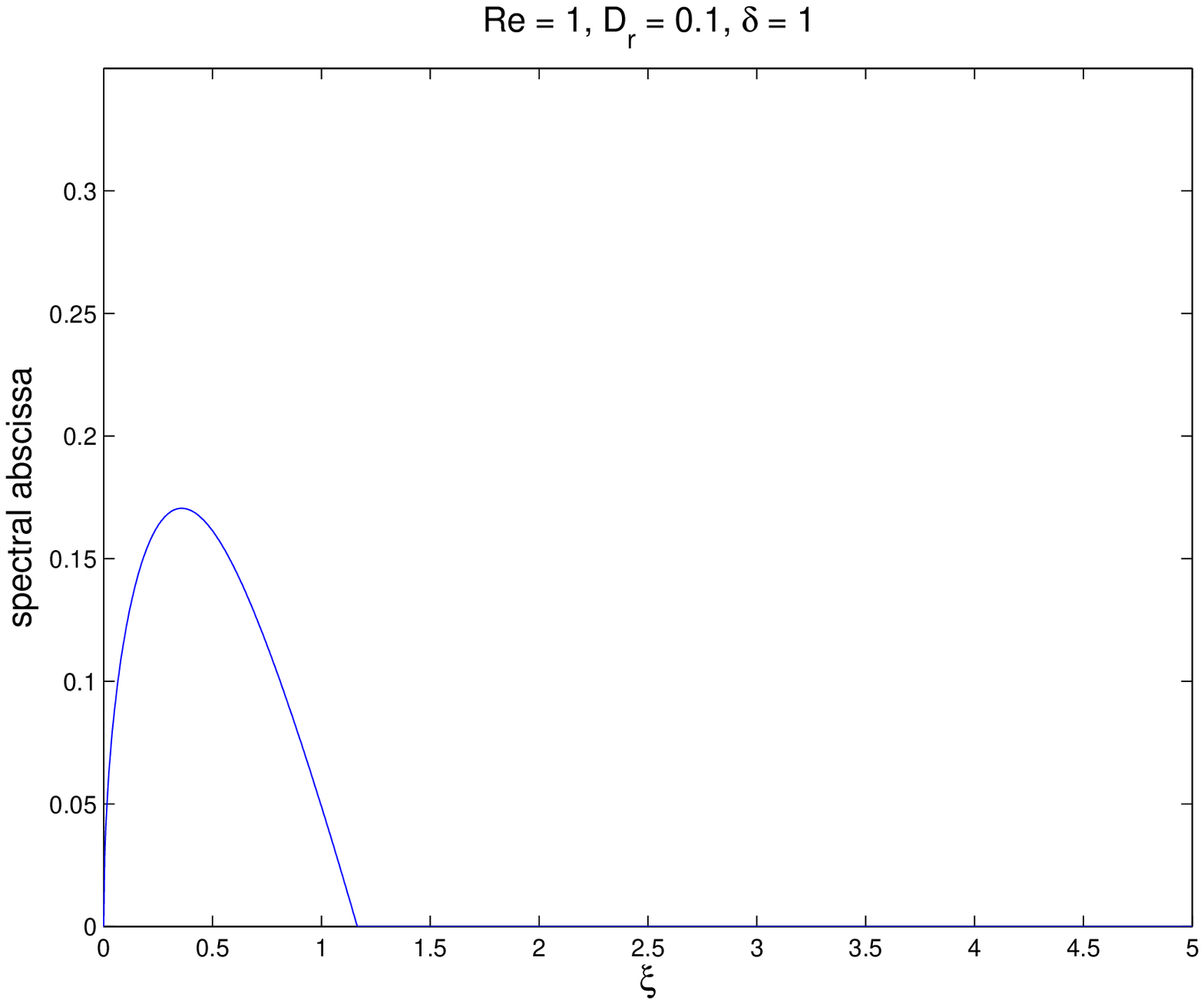}\hfil
(c)\includegraphics[width=0.3\textwidth]{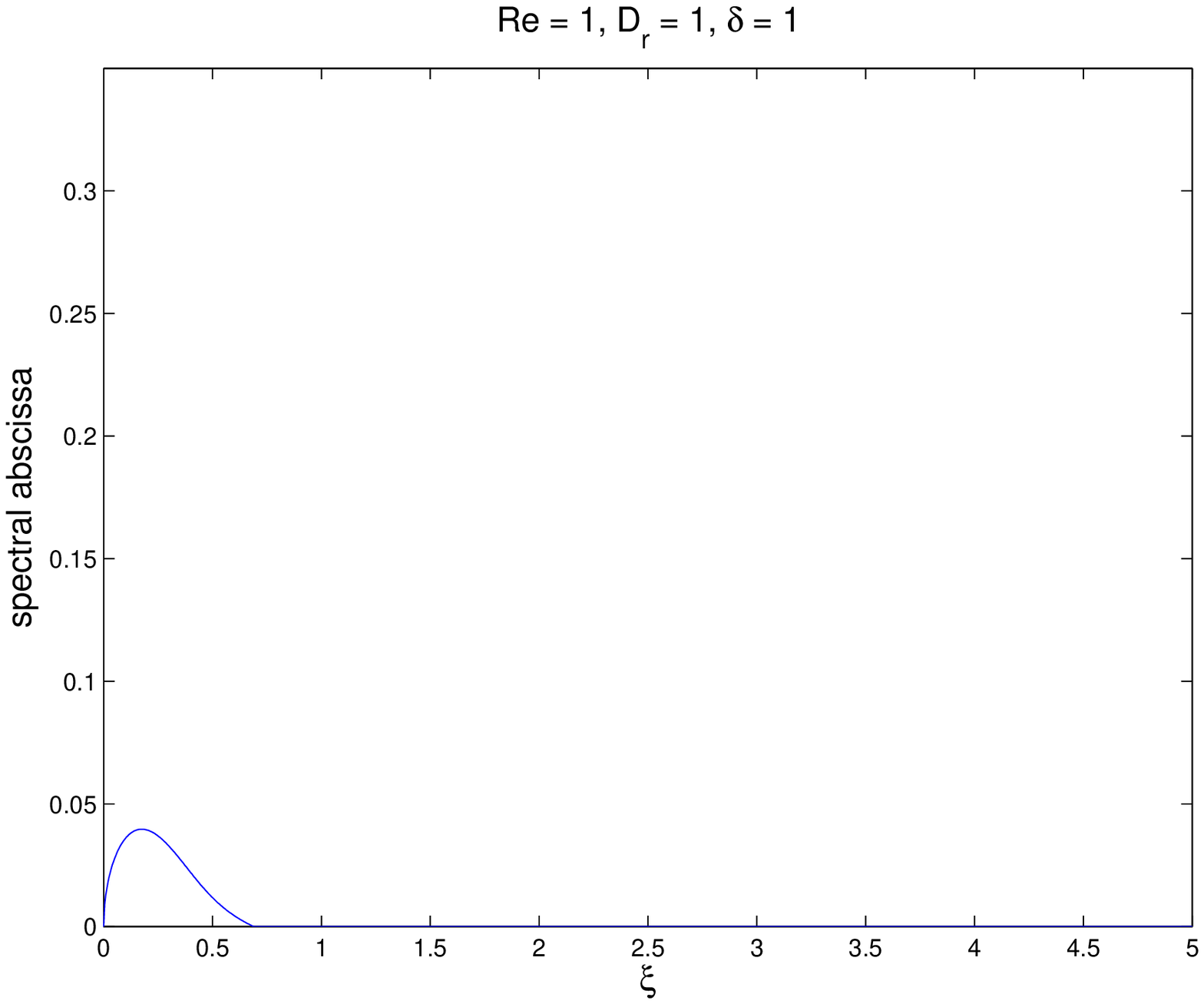}
\caption{\label{fig:stability5}Spectral abscissa as a function of wavenumber $\xi$ for
  $\delta = 1$, $Re=1$ and different values of $D_r$.}
\end{figure}

As in the case $Re=0$, an increase of $D_r$ has a stabilizing
effect, while an increase of $\delta$  has a destabilizing effect.
In addition, an increase of $D_r$ leads to the selection of a longer
wavelength, while an increase of $\delta$ leads to the selection of a
shorter wavelength.    
In Figure \ref{fig:stability5} we show plots of the spectral abscissa
vs.\ the wavenumber
for $Re=1$, $\delta = 1$ and different values of $D_r$.

\end{appendix}

\section*{Acknowledgements}
The authors express their thanks to Professor {\sc Felix Otto} for his valuable comments
and his involvement in several discussions concerning this work.
Research supported in part by the King Abdullah University of Science and Technology (KAUST), 
by the  project ACMAC of the FP7-REGPOT-2009-1 program of the European Commission, 
and by the Aristeia program of the Greek Secretariat for Research.

\bibliographystyle{plain}
\bibliography{references}

\end{document}